%% file: main.tex
\documentclass[journal]{IEEEtran}
\usepackage[table,xcdraw]{xcolor}
\usepackage{amsmath,amsfonts}
\usepackage{algorithmic}
\usepackage{algorithm}
\usepackage{array}
\usepackage[caption=false,font=normalsize,labelfont=sf,textfont=sf]{subfig}
\usepackage{textcomp}
\usepackage{stfloats}
\usepackage{url}
\usepackage{verbatim}
\usepackage{graphicx}
\usepackage{cite}
\usepackage{algorithmic}
\usepackage{algorithm}
\usepackage{amssymb}
\usepackage{multirow}
\usepackage{adjustbox}
\usepackage{bookmark}
\usepackage{orcidlink}
\usepackage[acronym]{glossaries}
\usepackage[normalem]{ulem}
\usepackage{microtype}
\usepackage{hyperref}
\useunder{\uline}{\ul}{}

\newacronym{hsi}{HSI}{HyperSpectral Imaging}
\newacronym{hs}{HS}{HyperSpectral}

\begin{document}

\newcommand{\warn}[1]{{\color{red}#1}}
\newcommand{\real}{\mathbb{R}}
\newcommand{\tth}{\text{th}}

\title{Preprocessing Algorithm Leveraging Geometric Modeling for Scale Correction in Hyperspectral Images for Improved Unmixing Performance}

\author{
	Praveen~Sumanasekara\(^{\dagger}\) ,~\IEEEmembership{Student Member,~IEEE}\orcidlink{0009-0000-7044-9944},
	Athulya~Ratnayake,~\IEEEmembership{Student Member,~IEEE}\orcidlink{0009-0008-9582-4606},
	Buddhi~Wijenayake,~\IEEEmembership{Student Member,~IEEE}\orcidlink{0009-0001-2624-0251},
	Keshawa~Ratnayake,~\IEEEmembership{Student Member,~IEEE}\orcidlink{0009-0004-8232-1365},
	Roshan~Godaliyadda,~\IEEEmembership{Senior Member,~IEEE}\orcidlink{0000-0002-3495-481X},
	Parakrama~Ekanayake,~\IEEEmembership{Senior Member,~IEEE}\orcidlink{0000-0002-5639-8105},
	and~Vijitha~Herath,~\IEEEmembership{Senior Member,~IEEE}\orcidlink{0000-0002-2094-0716}%
	\thanks{Praveen Sumanasekara, Athulya Ratnayake, Buddhi Wijenayake, Roshan Godaliyadda,
		Parakrama Ekanayake, and Vijitha Herath are with the Department of Electrical and Electronic Engineering,
		University of Peradeniya, Sri Lanka.
		E-mails: e19391@eng.pdn.ac.lk,  e19328@eng.pdn.ac.lk, e19445@eng.pdn.ac.lk, roshangodd@ee.pdn.ac.lk, mpb.ekanayake@ee.pdn.ac.lk, vijitha@ee.pdn.ac.lk.}%
	\thanks{Keshawa Ratnayake is with the School of Electrical and Computer Engineering, Purdue University, West Lafayette, IN, USA.
		E-mail: Ratnayar@purdue.edu.}
	\thanks{\(\dagger\)Corresponding Author: Praveen Sumanasekara, email: e19391@eng.pdn.ac.lk}
}

\IEEEpubid{}

\maketitle

\begin{abstract}
	Spectral variability significantly impacts the accuracy and convergence of hyperspectral unmixing algorithms. Many methods address complex spectral variability; yet large-scale distortions to the scale of the observed pixel signatures due to topography, illumination, and shadowing remain a major challenge. These variations often degrade unmixing performance and complicate model fitting. Because of this, correcting these variations can offer significant advantages in real-world GIS applications. In this paper, we propose a novel preprocessing algorithm that corrects scale-induced spectral variability prior to unmixing.
	By estimating and correcting these distortions to the scale of the pixel signatures, the algorithm produces pixel signatures with minimal distortions in scale. Since these distortions in scale (which hinder the performance of many unmixing methods) are greatly minimized in the output provided by the proposed method, the abundance estimation of the unmixing algorithms is significantly improved.
	We present a rigorous mathematical framework to describe and correct for scale variability and provide extensive experimental validation of the proposed algorithm. Furthermore, the algorithm’s impact is evaluated across a wide range of state-of-the-art unmixing methods on two synthetic and two real hyperspectral datasets. The proposed preprocessing step consistently improves the performance of these algorithms, achieving error reductions of around 50\%, even for algorithms specifically designed to handle spectral variability. This demonstrates that scale correction acts as a complementary step, facilitating more accurate unmixing with existing methods. The algorithm’s generality, consistent impact, and significant influence highlight its potential as a key component in practical hyperspectral unmixing pipelines. The implementation code is available at \url{https://github.com/DMUPraveen/Perspecitve_Transform.git}.

\end{abstract}


\input{Chapters/introduction}

\input{Chapters/relatedwork}

\input{Figures_Latex/scaling_no_scaling}
\input{Figures_Latex/Main_Algo_Figure}

\input{Chapters/methodology}
\input{Chapters/algorithm}

\input{Chapters/Experiments/experiments}
\input{Chapters/discussion}
\input{Chapters/Conclusion}

\section*{Acknowledgments}
The authors acknowledge the support received from the LK Domain Registry in publishing this paper .

\input{Chapters/appendix}

\bibliographystyle{IEEEtran}
\bibliography{Perspective_Transform_HSI}

\input{Chapters/Author_Biography.tex}

\end{document}

%% file: Chapters/introduction.tex
\section{Introduction}

Hyperspectral Imaging (\acrshort{hsi}) provides high spectral resolution, capturing rich spectral information and enabling solutions beyond the capabilities of conventional RGB or multispectral imaging \cite{HSIBible}. As a result, \acrshort{hsi} finds applications in a wide array of fields such as mineral mapping \cite{Mineral1,Mineral2}, agriculture \cite{Agriculture1,Agriculture2}, military applications \cite{Military1}, and environmental monitoring \cite{Environ1,Environ2}. Each pixel of a hyperspectral image records reflectance values across numerous narrow spectral bands, providing detailed insights into material composition \cite{KeshawaHSI}.

However, hyperspectral (HS) images typically suffer from limited spatial resolution, causing many pixels to contain mixtures of multiple materials. The process of identifying the constituent materials within each pixel is known as Hyperspectral Unmixing (HU) \cite{HSIBible}. The pure materials within each pixel are called the endmembers, and their proportional contributions are called the abundances.

For the development and analysis of HU algorithms, the mixing phenomenon is often represented mathematically. One of the most widely used models is the Linear Mixture Model (LMM) \cite{KeshawaHSI}, which models each pixel as a linear combination of endmember signatures weighted by their abundances. Building upon the LMM, a vast array of unmixing algorithms have been proposed. Traditional unsupervised approaches heavily utilize the LMM \cite{FCLSU,SUnSAL,ICAHU}, while other methods have been developed to account for complex multiple scattering and nonlinear interactions between endmembers \cite{NonLinearMM, Yang2018, GLMM}. In recent years, the state-of-the-art (SOTA) has increasingly shifted toward deep learning architectures. Notable examples include convolutional and autoencoder-based networks such as CyCU-Net \cite{CYCU} and PGMSU \cite{PGMSU}, Siamese networks \cite{SiameseHSU}, and transformer-based models \cite{Deeptrans,USTNet}, as well as algorithms such as SSAF-Net \cite{SSAFNet} designed to capture global spatial–spectral dependencies. Beyond standard network architectures, recent work has explored sophisticated optimization paradigms and temporal dependencies, including gradient-based multiobjective methods with greedy hashing (GMOGH) \cite{GMOGH} and copula-guided temporal dependency methods (Cog-TD) \cite{CogTD} for modeling dynamic endmembers over time.

While many traditional unmixing algorithms rely heavily on the assumption of consistent spectral signatures across the image, real hyperspectral data frequently violates this assumption through pixel-wise variations known as Spectral Variability (SV) \cite{SVBible}. Arising from differences in illumination, topography, and atmospheric conditions, SV can severely distort both the scale and shape of spectral signatures. To address this, several of the advanced models mentioned above, such as those using spectral libraries or complex parametric models—explicitly incorporate SV into their architectures. However, these variability-aware approaches often introduce severe computational complexity, optimization challenges, and difficult hyperparameter tuning \cite{SVBible}. Furthermore, while they effectively model complex, small-scale, or frequency-dependent variations, their performance is frequently hindered by large-scale magnitude variations. Massive distortions in scale can overwhelm optimization processes and regularization mechanisms, causing even SOTA SV-aware methods to converge to sub-optimal scaling parameters, ultimately leading to severe inaccuracies in abundance estimation.

Fortunately, in many practical scenarios, variations in scale are the predominant form of SV \cite{MerconScale,ScalingSpectralImaging}. These scale variations, primarily caused by differences in illumination and surface topography, can frequently be modeled using a simple, wavelength-independent scaling factor applied to the pixel’s spectral signature \cite{ScalingFacIndependent}. Because these dominant scale distortions are highly amenable to correction prior to the unmixing stage, applying transformations or preprocessing algorithms is a highly desirable alternative to overly complex unmixing networks.

Standard normalization methods can mitigate scaling effects but often distort the intrinsic simplex geometry of the data cloud \cite{Normalization}. Perspective Projection \cite{HSIBible} provides a more geometrically meaningful approach by rescaling pixel signatures so that they lie on a hyperplane consistent with the LMM. Despite this property, perspective projection remains underexplored as a preprocessing technique for abundance estimation. Existing methods that utilize it, such as Vertex Component Analysis (VCA) \cite{VCA}, project onto arbitrary hyperplanes (e.g., using the mean of all signatures). While acceptable for extracting endmembers, arbitrary projections scale pixels unevenly, introducing severe distortions in abundance estimates \cite{HSIBible}.

Currently, the literature lacks a robust, unsupervised method for determining optimal projection parameters that correct scale distortions while strictly preserving abundance relationships. In this work, we demonstrate that uncorrected large-scale variations significantly hinder the performance of state-of-the-art (SOTA) unmixing algorithms, even those explicitly designed to handle general spectral variability.

To bridge this gap, we propose a mathematically justified perspective projection algorithm that corrects large-scale variations without distorting abundance fractions. By optimizing the projection parameters, the proposed algorithm restores the simplex geometry, providing downstream unmixing algorithms with a more suitable input for downstream unmixing.

In summary, our main contributions are:
\begin{enumerate}
    \item Development of a mathematically justified theory for estimating optimal projection parameters to correct scale distortions arising from topographical and illumination effects.
    \item Development of an unsupervised algorithm based on this mathematical formulation, demonstrating its ability to robustly estimate and correct scale distortions.
    \item Extensive experimental validation showing that the proposed preprocessing algorithm reduces abundance estimation errors by as much as 50\% across a wide range of SOTA HU algorithms, including deep learning and variability-aware models.
\end{enumerate}

The main notation used throughout the paper is summarized in Table~\ref{tab:notation}.
\input{Chapters/notation}

%% file: Chapters/notation.tex

\newcommand{\img}{Y_0}
\newcommand{\pix}{Y_0}
\newcommand{\npix}{N}
\newcommand{\nend}{K}
\newcommand{\nband}{L}
\newcommand{\matsig}{M_0}
\newcommand{\matabd}{A}
\newcommand{\noise}{E}
\newcommand{\sigo}[1]{(\mathbf{m}_0)_{#1}}
\newcommand{\abdo}{\mathbf{a}}
\newcommand{\noiso}{\mathbf{e}}
\newcommand{\pixlo}[1]{(\mathbf{y}_0)_{#1}}
\newcommand{\onemat}{\mathbf{1}}
\newcommand{\scalo}{\mu}
\newcommand{\unpix}{Y_0^{*}}
\newcommand{\corpix}{\widehat{Y}_0}
\newcommand{\unpixlo}[1]{(\mathbf{y}_0^*)_{#1}}
\newcommand{\corpixlo}[1]{(\mathbf{\widehat{y}}_0)_{#1}}
\newcommand{\estscalo}[1]{{\widehat{\mu}}_{#1}}
\newcommand{\yhat}[1]{(\mathbf{\widehat{y}}_0)_{#1}}
\newcommand{\muhat}[1]{{\widehat{\mu}}_{#1}}
\newcommand{\Ystar}{Y^{*}}
\newcommand{\KK}{K}
\newcommand{\LL}{L}
\newcommand{\yystar}[1]{\mathbf{y}^{*}_{#1}}

\newcommand{\yy}[1]{\mathbf{y}_{#1}}
\newcommand{\MM}{M}
\newcommand{\aai}[1]{\mathbf{a}_{#1}}
\newcommand{\nnstar}{\mathbf{n}^*}
\newcommand{\ccstar}{\mathbf{c}^*}
\newcommand{\mean}[1]{\left<{#1}\right>}

\newcommand{\ee}[1]{\text{E}\left[{#1}\right]}
\newcommand{\Nsum}[1]{\sum_{i=1}^N{#1}}
\newcommand{\norm}[1]{\left\lVert#1\right\rVert}
\newcommand{\yyhat}[1]{\mathbf{\widehat{y}}_{#1}}
\newcommand{\nhat}{\hat{\mathbf{n}}}
\newcommand{\nn}{{\mathbf{n}}}
\newcommand{\gammai}[2]{\Gamma_{#1}(#2)}

\newcommand{\maxvar}{\sigma_{max}^2}
\newcommand{\minvar}{\sigma_{min}^2}

\newcommand{\reconpixlo}[1]{(\tilde{\mathbf{y}}_0)_{#1}}
\newcommand{\reconpix}{\tilde{Y}_0}
\newcommand{\Var}[1]{\text{Var}\left[#1\right]}

\begin{table*}[t]
    \caption{Main notation used throughout the paper. All vectors are column vectors.}
    \label{tab:notation}
    \centering
    \small
    \renewcommand{\arraystretch}{1.12}
    \begin{tabular}{l|l}
        \hline
        \textbf{Symbol}                                                         & \textbf{Meaning} \\
        \hline
        \(\npix,\nend,\nband\)                                                  &
        Number of pixels, endmembers, and spectral bands, respectively.                            \\

        \(\img=[\pixlo{1},\dots,\pixlo{\npix}] \in \real^{\nband\times\npix}\)  &
        Observed hyperspectral image.                                                              \\

        \(\pixlo{i}\in\real^{\nband}\)                                          &
        Observed spectral signature of the \(i^{\tth}\) pixel.                                     \\

        \(\matsig=[\sigo{1},\dots,\sigo{\nend}] \in \real^{\nband\times\nend}\) &
        Endmember signature matrix in the original spectral space.                                 \\

        \(\sigo{j}\in\real^{\nband}\)                                           &
        Spectral signature of the \(j^{\tth}\) endmember.                                          \\

        \(\abdo_i\in\real^{\nend}\)                                             &
        Abundance vector of the \(i^{\tth}\) pixel.                                                \\

        \(\matabd=[\abdo_1,\dots,\abdo_{\npix}] \in \real^{\nend\times\npix}\)  &
        Abundance matrix.                                                                          \\

        \(\noiso_i\in\real^{\nband},\quad \noise\in\real^{\nband\times\npix}\)  &
        Noise vector of the \(i^{\tth}\) pixel and noise matrix.                                   \\

        \(\onemat\in\real^{\nend}\)                                             &
        All-ones vector used in the abundance sum-to-one constraint, \(\onemat^\top \abdo_i = 1\). \\

        \(\scalo_i\)                                                            &
        True pixel-wise wavelength-independent scaling factor.                                     \\

        \(\unpix=[\unpixlo{1},\dots,\unpixlo{\npix}]\)                          &
        Hypothetical image without scale distortion.                                               \\

        \(\corpix=[\corpixlo{1},\dots,\corpixlo{\npix}]\)                       &
        Scale-corrected image produced by the proposed preprocessing.                              \\

        \(\mathbb{T}:\real^{\nband}\rightarrow\real^{\nend}\)                   &
        Dimensionality-reduction map obtained using SVD.                                           \\

        \(\yy{i}\in\real^{\nend}\)                                              &
        Dimensionally reduced version of \(\pixlo{i}\).                                            \\

        \(\yystar{i}\in\real^{\nend}\)                                          &
        Dimensionally reduced version of the unscaled pixel \(\unpixlo{i}\).                       \\

        \(\MM\in\real^{\nend\times\nend}\)                                      &
        Endmember matrix in the reduced domain.                                                    \\

        \(\ccstar\)                                                             &
        A point on the desired hyperplane in the reduced domain.                                   \\

        \(\nnstar\)                                                             &
        Normal vector of the desired hyperplane.                                                   \\

        \(\nhat\)                                                               &
        Estimate of the hyperplane normal.                                                         \\

        \(\muhat{i}\)                                                           &
        Estimated scaling factor of the \(i^{\tth}\) pixel.                                        \\

        \(\yyhat{i}=\yy{i}/\muhat{i}\)                                          &
        Corrected reduced-domain pixel used in the optimization.                                   \\
        \hline
    \end{tabular}
\end{table*}

%% file: Chapters/relatedwork.tex
\section{Related Work}

\subsection{Spectral Variability and Current Mitigation Strategies}
While standard mixture models assume fixed endmember signatures, real-world hyperspectral data exhibits significant pixel-wise spectral variability (SV). The main factors contributing to these variations include atmospheric conditions, illumination, topographic effects, and intrinsic material differences \cite{SVPhysics,SVBible}. To account for SV, numerous variability-aware strategies have been developed. Spectral library-based methods address SV by selecting representative signatures from pre-existing collections \cite{Mesma,SparseUnmixing1}. Alternatively, probabilistic and parametric models extend the linear mixture model to explicitly include variability parameters, such as per-pixel scaling factors, endmember-wise scaling, or band-wise variations \cite{DeepGenerativeModelling,PerturbedLMM,ELMM}.

Recent studies have also highlighted an important connection between spectral variability and nonlinearity. Although these are physically distinct phenomena, certain nonlinear mixing effects can be approximated by space-varying linear models under suitable conditions \cite{NonToSV}. In particular, scaling-based variability models such as the Extended Linear Mixing Model (ELMM) \cite{SVtoNon} can account for part of the distortion induced by some nonlinear mixtures.

Despite their theoretical robustness, incorporating SV directly into the unmixing optimization often increases computational complexity and makes convergence challenging \cite{SVBible}. Consequently, spectral preprocessing algorithms, such as spectral transformations and library pruning, are frequently employed to project data into minimally variable subspaces \cite{MinVolTrans,Fisher,LibraryPruning}. However, many of these transformations require prior topographic or atmospheric data, or they fail to remove scale variations in an unsupervised manner \cite{TopoNorm,Topocompare}. Correcting these variations as a preprocessing step allows unmixing algorithms to converge properly and focus their attention on the unmixing problem and minor frequency-dependent as well endmember dependent variations.

\subsection{Scale-Induced Variability and Loss Function Limitations}
\label{sec:scale-sv}
Among the various causes of SV, topographic and illumination-induced variations primarily affect the scale of spectral signatures \cite{ScalingSpectralImaging}. Even small variations in topography can result in substantial spectral distortions \cite{SmallTopo,Hapke}. In many practical scenarios, these dominant effects can be modeled by a wavelength-independent scaling factor that varies from pixel to pixel \cite{ScalingFacIndependent}. Large-scale variations in this scaling factor can severely hinder the performance of algorithms, especially those relying on deep learning and gradient descent.

To mitigate these large-scale magnitude variations, many contemporary unsupervised algorithms replace or complement the highly scale-sensitive Mean Square Error (MSE) reconstruction loss with Spectral Angle Distance (SAD) loss \cite{SSABN,DAAN}. Because SAD compares only the direction of spectral signatures, it is inherently invariant to scale. However, completely ignoring spectral magnitude introduces a critical flaw: spectra with identical directions but differing magnitudes can yield substantially different abundance values \cite{Deeptrans,USTNet}.

\input{Figures_Latex/sad_problem}

As illustrated in Figure \ref{fig:sad-problem}, if the predicted endmember signatures (e.g., \(e_1'\) and \(e_2'\)) have the same direction as the true signatures (\(e_1\) and \(e_2\)) but different magnitudes, the SAD loss will be zero. However, the predicted abundance fractions for the pixel \(p'\) will be highly biased compared to the true fractions for pixel \(p\). This issue is especially pronounced in the absence of accurate endmember initialization \cite{ICIISEEAimpact}. If predicted magnitudes deviate from the true ones, resulting abundance estimates are biased even if spectral angles are correct.

While fusing MSE and SAD losses can mitigate these effects by balancing scale invariance with magnitude sensitivity, this introduces additional hyperparameters that are difficult to tune in fully unsupervised settings without access to ground truth data. Other methods attempt simple normalization of signatures \cite{Normalization}, but this distorts the geometry of the data points, losing the simplex geometry imposed by the LMM.

%% file: Figures_Latex/sad_problem.tex
\begin{figure}[!t]
    \centering
    \includegraphics[width=\linewidth]{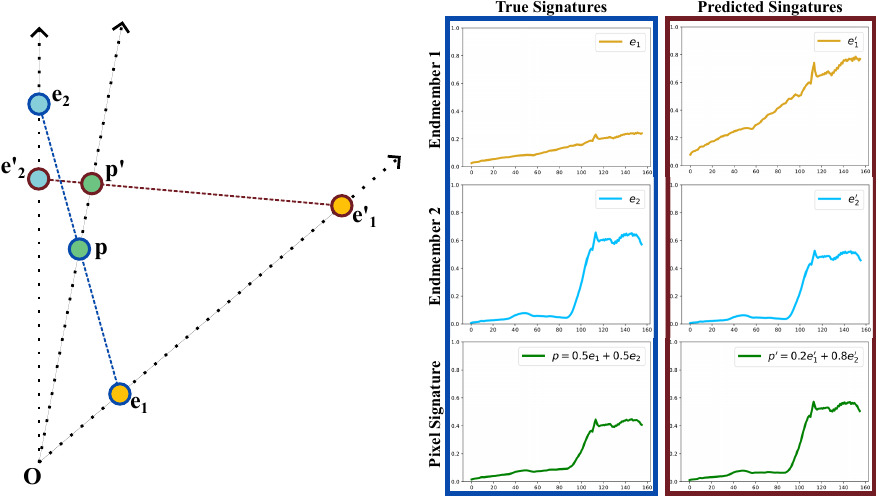}
    \caption{Illustration of the limitations of using only SAD loss. The left diagram shows endmember signatures as vectors in an L-dimensional space, with endpoints marked by circles. The corresponding signatures for each of these vectors are shown on the diagram on the right. $e_1$, $e_2$, and $p$ (outlined in blue) are the true signatures of endmember 1, endmember 2, and the pixel, respectively. The quantities $e_1',e_2',$ and $p'$ are the predicted counterparts of these quantities.} 
    \label{fig:sad-problem}
\end{figure}

%% file: Figures_Latex/scaling_no_scaling.tex
\begin{figure}[!t]
    \centering
    \subfloat[]{\includegraphics[width=0.5\linewidth]
        {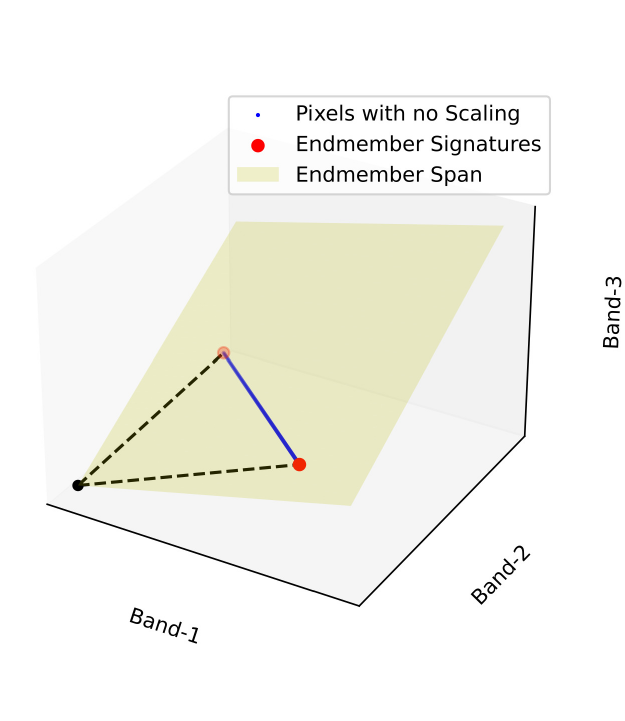}%
        \label{fig:Perfect_LMM}}
    \hfil
    \subfloat[]{\includegraphics[width=0.5\linewidth]
        {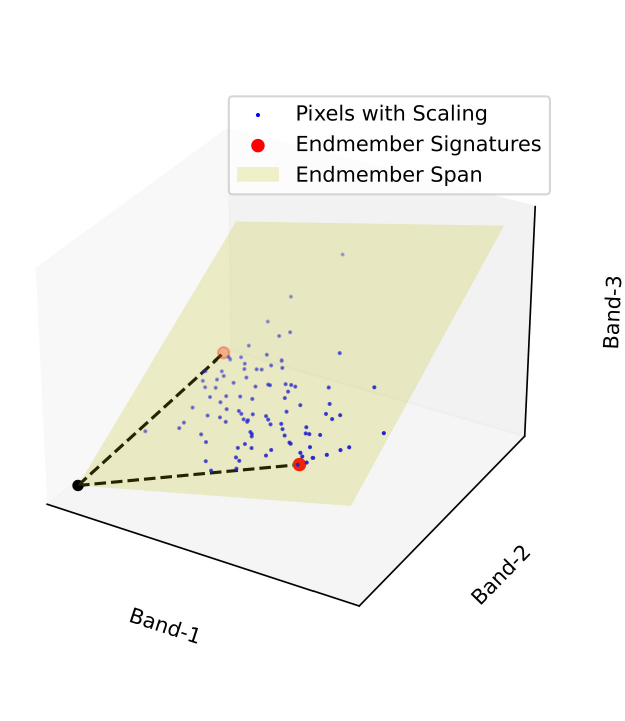}%
        \label{fig:LMM_Scaling}}
    \caption{Geometry of a case where there are two endmembers (represented as red dots) and the signatures have three bands for purposes of visualization. (a) The signatures adhere to LMM perfectly. (b) Signatures are scaled by random scaling factors \(\mu_i,\) distorting the original simplex geometry. The origin in both cases is represented by a black dot. Notice that in both cases the signatures occupy the span of the endmembers given in light yellow. However, in case (a), ASC additionally constrains the pixels to be on a hyperplane (for two endmembers this is a line going through the endmember signatures) and further restricted to a simplex (for two endmembers this is a line segment joining the two signatures) by ANC.}
    \label{fig:Pixel_Geometry}
\end{figure}

%% file: Figures_Latex/Main_Algo_Figure.tex
\begin{figure*}[!t]
    \centering
    \includegraphics[width=0.8\linewidth]{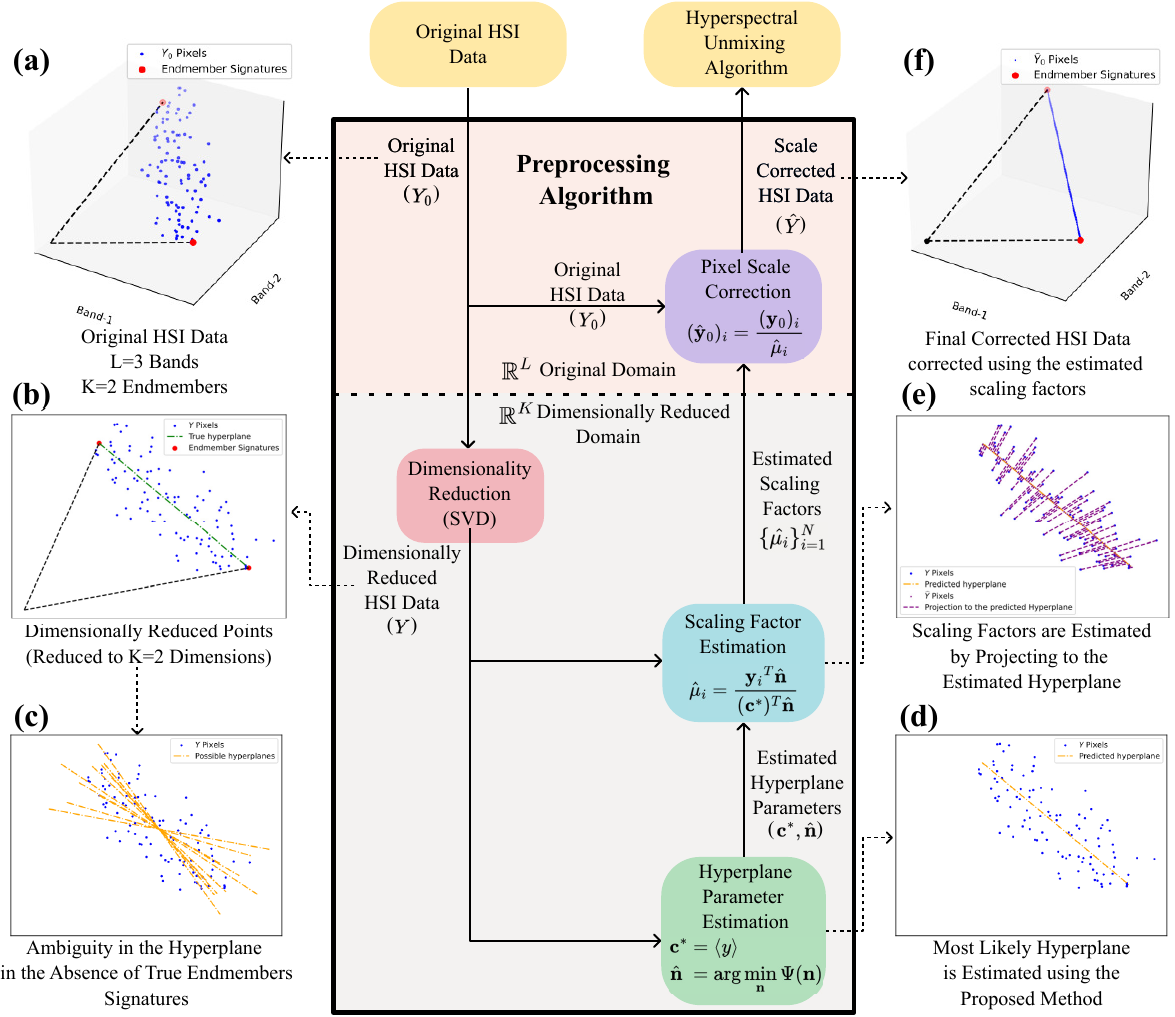}
    \caption{Overview of the Signature Scale Correction Pre-Processing Algorithm. The Figure illustrates the main steps and the data flow of the algorithm. As the input Raw HSI Data is fed to the algorithm and it outputs corrected data which can be used by a HU algorithm in order to improve the HU algorithm's performance when compared with using raw data. At the margins of the algorithm geometric visualization of the data is given to give a visual understanding of the steps of the algorithm. For ease of visualization, this is illustrated for the case where there are 3 channels in the signatures ($L$=3) and two endmembers ($K$=2).}
    \label{fig:main_algo}
\end{figure*}

%% file: Chapters/methodology.tex
\section{Methodology}

\input{Chapters/problem_setup}

\input{Chapters/motivation}

\input{Chapters/method_overview}

\input{Chapters/dimensionality_reduction}

\input{Chapters/hyperplane}

\input{Chapters/DOF_mu}

\input{Chapters/estimating_nstar}

%% file: Chapters/problem_setup.tex
\subsection{Problem Setup}

Let \(\img = [\pixlo{1},\pixlo{2},\dots,\pixlo{\npix}] \in \real^{\nband \times \npix}\) denote the observed hyperspectral image, where \(\npix\) is the number of pixels and \(\nband\) is the number of spectral bands. Let \(\matsig = [\sigo{1},\sigo{2},\dots,\sigo{\nend}] \in \real^{\nband \times \nend}\) denote the endmember signature matrix, where \(\nend\) is the number of endmembers. The abundance vectors are denoted by \(\abdo_i \in \real^{\nend}\) and collected into the abundance matrix \(\matabd = [\abdo_1,\abdo_2,\dots,\abdo_{\npix}]\).

Under the standard Linear Mixture Model (LMM), each pixel is modeled as,

\begin{align}
    \pixlo{i} = \matsig \abdo_i + \noiso_i ,
\end{align}
where \(\noiso_i \in \real^{\nband}\) represents noise and modelling errors. The abundance vectors satisfy the abundance non-negativity (ANC) and abundance sum-to-one (ASC) constraints
\begin{align}
    \abdo_i \ge 0, \qquad \onemat^T \abdo_i = 1,
    \label{eq:asc}
\end{align}
where \(\onemat\) denotes the all-ones vector.

In this work, we focus on the dominant wavelength-independent component of spectral variability, which can often be approximated as a per-pixel scaling effect. Accordingly, the observed pixels are modeled as
\begin{align}
    \pixlo{i} = \scalo_i \matsig \abdo_i + \noiso_i \;; \scalo_i > 0
    \label{LMM-scaling}
\end{align}

where \(\scalo_i\) denotes the scaling factor (which is positive) associated with the \(i^{\text{th}}\) pixel and \(\noiso_i\) represents the noise and modelling errors such as wavelength/endmember dependent SV and nonlinear mixing effects.

Let \(\unpixlo{i}\) denote the hypothetical pixel that would be observed if this scale distortion were removed. Then

\begin{align}
    \label{eq:unscaled-def}
    \unpixlo{i} = \matsig \abdo_i
\end{align}

and therefore

\begin{align}
    \label{eq:pixel-unscaled}
    \pixlo{i} = \scalo_i \unpixlo{i} + \noiso_i.
\end{align}

The goal of the proposed preprocessing framework is to estimate \(\scalo_i\) for each pixel and use it to obtain corrected pixels that better satisfy the simplex geometry implied by the LMM.

%% file: Chapters/motivation.tex
\subsection{Motivation}
\label{sec:motivation}
Although the standard LMM does not fully capture the true complexity of most hyperspectral datasets, it imposes a fundamental simplex geometry on the data that numerous algorithms exploit. In this framework, the vertices of the simplex correspond to the endmember signatures. Notably, if there are \(K\) endmembers, the resulting simplex is \((K-1)\)-dimensional.

Figure \ref{fig:Perfect_LMM} represents an ideal case where the HSI data consists of \(L=3\) bands with \(K=2\) endmembers for ease of visualization. Under the LMM, pixel signatures are linear combinations of the endmembers, restricting them to the \(K\)-dimensional span of the endmember signatures (depicted in yellow). Furthermore, the ASC constrains the pixels to a \((K-1)\)-dimensional hyperplane, which happens to be a line in this example. Finally, the ANC restricts the space further to a \((K-1)\)-dimensional simplex \cite{HSIBible}. In Figure \ref{fig:Perfect_LMM}, this is the blue line segment joining the red endmember signatures.

Variations in the scale of pixel signatures severely deform this usual simplex geometry. As illustrated in Figure \ref{fig:LMM_Scaling}, scale distortions cause signatures to spread toward or away from the origin. Mathematically, this corresponds to scaling each pixel signature by a scalar factor. As a result, the pixels are no longer restricted to the hyperplane imposed by the ASC. However, the pixel signatures remain within the \(K\)-dimensional span of the endmembers because scalar multiplication does not alter the underlying subspace.

If unaccounted for, these massive scaling variations force SOTA HU algorithms to converge to sub-optimal solutions with highly inaccurate abundance estimates. This raises a critical question: can we isolate and estimate these scaling factors by analyzing the distorted geometry?

By projecting the dimensionally reduced pixel signatures back onto a \((K-1)\)-dimensional hyperplane, we can find the ratio between the observed and projected signatures to estimate the scaling factors.

Once estimated, these scaling factors are used to inversely scale the original high-dimensional pixels. This corrects the massive scale deformations and aligns the data much more closely with the fundamental simplex geometry, while carefully preserving the smaller, frequency-dependent spectral variabilities and nonlinearities. Consequently, downstream HU algorithms, especially SV-aware networks, are provided with an improved input state where they can focus entirely on the unmixing problem and exploiting residual variations.

However, extracting these scaling factors reveals an inherent ambiguity regarding which hyperplane the data should be projected onto. As illustrated in Figure \ref{fig:main_algo}(c), projecting onto different candidate hyperplanes results in different abundance estimates for the exact same pixel signature. Since accurate abundance estimation is central to the HU problem, establishing a mathematical framework to select the optimal hyperplane is paramount.

%% file: Chapters/method_overview.tex
\subsection{Methodology Overview}
\label{sec:methodology_overview}

The goal of the proposed preprocessing framework is to correct large-scale magnitude variations, producing corrected pixels \(\corpix\) that closely adhere to the LMM. By correcting these dominant frequency-independent distortions, the preprocessed data are placed in a much more suitable state, allowing downstream HU algorithms to focus their optimization on the actual unmixing problem and minor frequency-dependent variations.

An overview of the entire preprocessing algorithm is presented in Figure \ref{fig:main_algo}. The procedure consists of four primary steps:

\begin{enumerate}
    \item \textbf{Dimensionality Reduction:} As established in the geometric motivation, even with the presence of scale variations, the pixel signatures lie in a \(K\)-dimensional subspace. Therefore, Singular Value Decomposition (SVD) is first used to project the original \(L\)-dimensional data into a \(K\)-dimensional subspace. This step hopes to attenuate minor modelling errors and noise, isolating the geometric scaling problem.

    \item \textbf{Hyperplane Parameter Estimation:} To project the reduced pixels back into a proper simplex, the algorithm must identify the optimal \((K-1)\)-dimensional hyperplane. This requires estimating two parameters: a position vector (\(c^*\)) that lies on the hyperplane, and the normal vector (\(\hat{n}\)) orthogonal to it.

    \item \textbf{Scale Factor Estimation:} Once the optimal hyperplane is identified, the pixels are projected onto it using perspective projection. The ratio between the observed pixels and their projected counterparts yields the estimated per-pixel scaling factors, denoted as \(\hat{\mu}_i\).

    \item \textbf{Pixel Scale Correction:} Finally, the estimated scaling factors are applied to the original, \(L\)-dimensional hyperspectral data. The \(i^{\text{th}}\) corrected pixel, \(\corpixlo{i}\), is obtained by dividing the original pixel by its corresponding estimated scaling factor as given in Equation \eqref{eq-Correction-Equation}.
\end{enumerate}

\begin{align}
    \label{eq-Correction-Equation}
    \corpixlo{i}  = \frac{\pixlo{i}}{\estscalo{i}}
\end{align}

Note that scale correction is performed in the original \(\real^L\) dimensional space rather than the reduced space. This ensures that no unintended information loss occurs, preserving the minor frequency-dependent spectral variability that advanced SV-aware algorithms might still need to model.

The primary mathematical challenge in this pipeline lies in Step 2, specifically in finding a robust estimate for the optimal normal vector \(\hat{n}\). As will be shown in Section \ref{sec:estimate-n}, determining this normal can be formulated as an optimization problem. However, the resulting objective function contains numerous local minima. Therefore, the proposed algorithm utilizes a hybrid optimization strategy. First, Particle Swarm Optimization (PSO) \cite{PSOoriginal,PSOGD} initialized with specially generated candidate normals is used to find a global region of optimality. This is followed by Gradient Descent (GD) fine-tuning to secure the precise normal vector.

The following sections provide the rigorous mathematical derivations and proofs supporting these algorithmic steps.

%% file: Chapters/dimensionality_reduction.tex
\subsection{Dimensionality Reduction}
\label{sec:dimred}
This section will demonstrate that performing the dimensionality reduction, which is essential for subsequent steps of the algorithm, preserves the geometry of the problem. Specifically, it will be demonstrated that scaling factors remain identical to the ones in the original L-dimensional space, and the mathematical structure imposed by Equation \eqref{LMM-scaling} remains unaltered.

According to the LMM with Scaling Effects, if the effect of noise is not considered, the pixels occupy a K-dimensional subspace, which happens to be the span of the endmember signatures. This is because, according to this model, the pixels are a linear combination of the endmember signatures (see Equation \eqref{LMM-scaling}).

Therefore, the dimensionality of pixel signatures can be reduced to a \(\KK\)-dimensional subspace. In theory, any linear transformation that preserves the linear independence of the spectral signatures is valid. However, due to the presence of nonlinearities, spectral variability, and noise in real hyperspectral data, Singular Value Decomposition (SVD) is commonly employed to identify the dominant subspace spanned by the endmembers \cite{UnDIP,KeshawaHSI}. This estimated subspace can then be used to effectively reduce the dimensionality of the signatures while retaining the most relevant information.

Let this Linear Transformation be denoted by \(\mathbb{T}: \real^\nband \rightarrow \real^\nend\). Then, let the dimensionally reduced version of \(x_0\) be \(x\) i.e. \(\mathbb{T}(x_0) = x \). For example, the dimensionally reduced version of the \(i^\tth\) pixel signature \(\pixlo{i}\in \real^\LL\) is \(\yy{i}\in \real^K\), the dimensionality reduced version of the endmember matrix \(\matsig \in \real^{\LL \times \KK}\) is  \(\MM \in \real^{\KK \times \KK}\), etc. This transformation and the preservation of the geometry of the problem can be seen visually by comparing Figure \ref{fig:main_algo}(a) and Figure \ref{fig:main_algo}(b).

Therefore, from the linearity property, it can be easily seen that the LMM with Scaling is preserved under this transformation, as can be observed by applying \(\mathbb{T}\) to \eqref{LMM-scaling} to get \eqref{LMM-scaling-dimred}.

\begin{align}
     \mathbb{T}(\pixlo{i}) & = \mathbb{T}\left(\scalo_i \matsig\abdo_i + \noiso_i \right) \nonumber                                              \\
     \implies \yy{i}       & = \scalo_i \MM \abdo_i+ \mathbb{T}(\noiso_i) \nonumber                                                              \\
     \implies \yy{i}       & \approx \scalo_i \MM \abdo_i \; \text{(Assuming \(\mathbb{T}(\noiso_i)\) is negligible)} \label{LMM-scaling-dimred}
\end{align}
Here it is assumed that \(\noiso_i\) is small enough such that SVD effectively identifies the subspace spanned by the endmembers, and therefore \(\mathbb{T}(\noiso_i)\) is negligible \cite{UnDIP,HSIBible}.

If there were no scaling factors, the pixels would lie in a (K-1) dimensional simplex, which in turn resides in a (K-1) dimensional hyperplane. Once the dimensionality reduction is done, the reduced pixels \(\yy{i}\) can be projected onto the hyperplane by scaling them. This is known as Perspective Projection. This can then be used to predict the scaling factors, which are then used to correct the pixel signatures.

Neglecting the nonlinear and noise terms is an approximation adopted to estimate the dominant scaling component. The SVD-based projection does not, in general, remove nonlinear effects; rather, it identifies the dominant \(K\)-dimensional subspace and attenuates low-energy residual components due to noise and model mismatch, while preserving the scaled simplex geometry under the ideal model.

This dimensionality reduction is only done to estimate the scaling factors as can be seen in Figure \ref{fig:main_algo}. Once the scaling factors are estimated, the scaling effect is removed by dividing the original observed pixels in the original dimensional space \(\pixlo{i} \in \real^L\) by the estimated scaling factors \(\muhat{i}\) as given in \eqref{eq-Correction-Equation} (See Figure \ref{fig:main_algo}). The scale correction is done in the original dimensional space instead of projecting the dimensionally reduced pixels because this minimizes unintended information loss and distortions.

%% file: Chapters/hyperplane.tex
\subsection{Estimating \texorpdfstring{\(\muhat{i}\)}{muhat i}  using Hyperplane Parameters}
\label{sec:estimating-mu}

Under the Dimensionality reduction \eqref{eq:unscaled-def} and \eqref{eq:pixel-unscaled} become,
\begin{align}
    \label{eq:unscaled-def-dimred}
    \yystar{i} & =  \MM\aai{i}         \\
    \label{eq:pixel-unscaled-dimred}
    \yy{i}     & = \scalo_i \yystar{i}
\end{align}
Then, the set of all \(\yystar{i}\) lie in a \((\KK-1)\) dimensional simplex in \(\real^\KK\) because of ASC and ANC. This also means that the set of \(\yystar{i}\) lie in a \((\KK-1)\) dimensional hyperplane (because of ASC). Therefore, \(\forall i, \yystar{i}\) satisfy the hyperplane equation,
\begin{align}
    \label{eq:hyperplane}
    (\yystar{i}-\ccstar)^T \nnstar = 0
\end{align}
where \(\ccstar\) is any position vector on the desired hyperplane and \(\nnstar\) is the normal vector to the desired hyperplane.

This can be easily shown using \eqref{eq:asc} and \eqref{eq:unscaled-def-dimred} (see Appendix).

Using \eqref{eq:pixel-unscaled-dimred} and \eqref{eq:hyperplane} it can be shown that,
\begin{align}
    \label{eq:mu}
    \scalo_i = \frac{\yy{i}^T\nnstar}{(\ccstar)^T\nnstar}
\end{align}

However, since \(\ccstar\) and \(\nnstar\) are unknown, they need to be estimated.

%% file: Chapters/DOF_mu.tex
\subsection{Degree of Freedom (DOF) in \texorpdfstring{\(\scalo\)}{mu} and expression for \texorpdfstring{\(\ccstar\)}{cstar}}
\label{sec:dof-mu}

As mentioned in section \ref{sec:methodology_overview}, there is an inherent DOF in the scaling factors \(\scalo_i\) since we can scale all \(\scalo_i\) by some constant factor and scale down all \(\unpixlo{i}\) by the same scaling factor. This does not change \(\pixlo{i}\), the abundances, or shape and relative magnitude of endmember signatures. In fact from \eqref{LMM-scaling}, we get, \(\pixlo{i} = \frac{\scalo_i}{\gamma} (\gamma\matsig) \aai{i}\).

Therefore, by selecting \(\gamma\) to be the mean of all \(\scalo_i\) \(\left(\mean{\scalo_i}\right)\), we can make the mean of the new scaling factors to be 1.

Consequently, from here onwards, \(\mean{\scalo_i}\) will be taken to be 1 without loss of generality. i.e.

\begin{align}
    \label{eq:mean-1}
    \mean{\scalo_i} = 1
\end{align}


As shown in Appendix~\ref{appendix:hyperplane-proof}, any point of the form given in \eqref{eq:point_formula} lies on the hyperplane defined by \eqref{eq:hyperplane}.

\begin{align}
    \label{eq:point_formula}
    \mathbf{x} = \sum_{i=1}^{\npix} \lambda_i \yystar{i} \;\;\text{s.t.}\;\; \sum_{i=1}^{\npix} \lambda_i = 1
\end{align}


The DOF in \(\scalo\) corresponds to an offset ambiguity in the hyperplane.
We resolve this ambiguity by enforcing \(\mean{\scalo_i} = 1\),
which implies that the consistent choice for the point \(\ccstar\) on the hyperplane is the mean of the dimensionally reduced pixels, \(\mean{\yy{}}\).

\begin{align}
    \ccstar
     & = \frac{1}{N}\sum_{i=1}^{N} \yy{i} \label{eq:cstar} \\
     & = \frac{1}{N}\sum_{i=1}^{N} \scalo_i \yystar{i}
    = \sum_{i=1}^{N} \lambda_i \yystar{i};\; \text{where}\;
    \lambda_i \triangleq \frac{\scalo_i}{N} \nonumber
\end{align}
Moreover,
\begin{align*}
    \sum_{i=1}^{N}\lambda_i
    = \frac{1}{N}\sum_{i=1}^{N}\scalo_i
    = \mean{\scalo_i}
    = 1,
\end{align*}

Therefore, \(\ccstar\), defined by \eqref{eq:cstar},  satisfies \eqref{eq:point_formula} and lies on the desired hyperplane.

Note that the decision to \(\mean{\scalo_i}=1\) also makes physical sense, since this has the effect of setting the hyperplane corresponding to average illumination to be the desired hyperplane. This results in the average pixel being the point on the desired hyperplane, resulting in the above equation for \(\ccstar\)

Now all that is left to do is to estimate \(\nnstar\), which will allow us to estimate \(\scalo_i\).

%% file: Chapters/estimating_nstar.tex
\subsection{Estimating \texorpdfstring{\(\nnstar\)}{nstar}}
\label{sec:estimate-n}
As previously noted, the original hyperplane cannot be uniquely recovered without assumptions on the distribution of scaling factors. However, since these scaling factors arise from natural phenomena, their distributions often exhibit certain regularities, which can enable the estimation of the true hyperplane under appropriate constraints. This section demonstrates the derivation and justification of a methodology for estimating the normal to the hyperplane.

It can be observed from Equation \eqref{eq:mu} that, the magnitude of \(\nnstar\) has no consequence on \(\scalo_{i}\). Therefore, we will assume \(\nnstar\) and its estimate \(\nhat\) to be unit vectors. (i.e. \(\norm{\nnstar} = \norm{\nhat} = 1\))

Therefore, taking the estimate of \(\nnstar\) to be \(\nhat\), we have the estimated scaling factors \(\muhat{i}\) to be,

\begin{align}
    \label{eq:muhat}
    \muhat{i} = \frac{\yy{i}^T\nhat}{(\ccstar)^T\nhat}
\end{align}
Note that we need \(\muhat{i}\) to be positive for all pixels. Specific conditions under which this is guaranteed is discussed in Appendix~\ref{appendix:non-negativity-of-scaling-factors}. However in practice, this issue is not typically encountered, and the algorithms includes safeguards to ensure that the estimated scaling factors are positive.

In this section, we introduce several assumptions regarding the distribution of the scaling factors, which will enable the formulation of a method for estimating \(\nnstar\). Note that the hypothetical pixel vectors unaffected by scaling, denoted \(\yystar{i}\), depend solely on their corresponding abundance vectors \(\aai{i}\). Consequently, any variability observed among the actual pixel vectors \(\yy{i}\) with the same abundance must be attributed to the scaling factors \(\scalo_{i}\).

To capture this, we model each \(\scalo_{i}\) as being independently distributed. Each \(\scalo_i\), is distributed according to a distribution that is potentially conditioned on the corresponding abundance vector \(\aai{i}\). Thus, for pixels \(\yy{i}\) sharing the same abundance \(\aai{i} = \aai{}\), the associated scaling factors \(\scalo_{i}\) are identically distributed (but not identically valued). i.e., \(\scalo_{i} \sim f(\scalo_{i};\aai{i})\). The reason for allowing the distribution to vary with abundance is that the textural properties can vary depending on the constituents of the pixel. Therefore, pixels with a larger portion of certain endmembers might show larger variability in the scale of the pixel signatures \cite{SVBible}.

However, we will assume that the expected value of any scaling factor \(\scalo_{i}\), for any pixel, regardless of its corresponding abundance, is constant, i.e for some \(k>0\),

\begin{align}
    \label{eq:mu-expected}
    \forall \aai{i}, E\left[\scalo_{i}\right] = k
\end{align}

This assumption implies that the scaling distortions are unbiased with respect to material composition, although their variance may differ based on textural properties. Specifically, although individual pixels with the same abundance may exhibit different scaling factors due to variations in illumination, topography, or other effects, the expected scaling factor, averaged over all possible realizations of such variations, is constant across all abundance vectors. In other words, the average illumination effect does not favor any particular composition.

Similar to the case of choosing the mean of the scaling factors this can be chosen to be any positive value without loss of generality because of the degree of freedom. The expected value will also be chosen to be 1 so as to be consistent with the decision to fix \(\mean{\scalo_i} =1\) made in Section \ref{sec:dof-mu}, Equation \eqref{eq:mean-1}. i.e. we will choose \(E\left[\scalo_{i}\right] = 1\). However, this is only a design decision. It can be shown that the results will be identical (up to a constant scaling factor for all pixels) for any choice. For example, if \(\mean{\scalo_{i}}\) and \(E\left[\scalo_{i}\right]\) are both chosen to be 0.7 (and the equations and algorithm are modified accordingly) the corrected pixels signatures will be the same as those if 1 was chosen except for all being scaled by 0.7.

Furthermore, it should be noted that no further restrictions are made on the distribution of \(\scalo_{i}\). For instance, the variance of the distribution can vary based on abundance. (This can in fact be observed in real datasets - see Figure \ref{fig:signature-correction-samson} where the tree endmember show larger variance compared to others.)

Under these assumptions, Appendix~\ref{appendix:estimator-proof} shows that the minimizer of \(\ee{\Nsum{\norm{\yy{i}-\yyhat{i}}^2}}\) is a good estimate for \(\nnstar\). Where, \(\yyhat{i} = \frac{\yy{i}}{\muhat{i}}\).

In fact it is shown that the RMS value of the distance between the corrected pixel signatures \(\yyhat{i}\) and the hypothetical pixel signatures \(\yystar{i}\) is upper bounded by Equation \eqref{eq:Upper-bound-on-the-Relative-RMSE-Error}.

\begin{align}
    \label{eq:Upper-bound-on-the-Relative-RMSE-Error}
    \frac{\delta_{\text{rms}}}{\sqrt{\mean{\norm{\yystar{i}}^2}}}
    \leq
    \sqrt{\frac{\maxvar- \minvar \frac{\left(\mean{\norm{\yystar{i}}}\right)^2}{\mean{\norm{\yystar{i}}^2}}}{N}}
\end{align}

Therefore, the relative error is well bounded and, as the number of pixels increases, it approaches zero. Given this, we can claim that \(\nhat\) can be made as accurate as desired with enough pixels. Thus the current estimate \(\nhat\) is a reasonable estimate of \(\nnstar\)

Since in many practical situations the image is only available under one possible illumination, each pixel is not available under different illumination conditions in order to estimate this expected value. However, the image often contains many realizations of pixels with similar enough abundance values under varying illumination conditions. Therefore, the simple summation given by Equation \eqref{eq:Psi} will reasonably approximate \(\zeta(\nn)\)
\begin{align}
    \label{eq:Psi}
    \Psi(\nn) & = {\Nsum{\norm{\yy{i}-\yyhat{i}}^2}}
\end{align}

Therefore  in the algorithm \(\mathbf{\nhat}\) will be estimated with Equation \eqref{eq:nhat-Psi}
\begin{align}
    \label{eq:nhat-Psi}
    \nhat & = \arg\min_{\nn} \left\{\Psi(\nn)\right\}
\end{align}

Although the assumption in \eqref{eq:mu-expected} is required for theoretical justification, the algorithm performs remarkably well in many practical situations where this condition is not strictly satisfied. This can be seen in the notable performance gains on the Samson and Urban datasets discussed in the results section, as well as the strong outcomes on the synthetic dataset, despite the use of spatially correlated scaling factors that simulate natural shadow effects.

%% file: Chapters/algorithm.tex
\subsection{Preprocessing Algorithm for Scale Correction}
\label{sec:Algorithm}
\renewcommand{\algorithmicrequire}{\textbf{Input:}}
\renewcommand{\algorithmicensure}{\textbf{Output:}}

\begin{algorithm}
    \caption{Correcting for Scaling Factors}
    \begin{algorithmic}[1]
        \label{alg:Algorithm}
        \REQUIRE \(\img = \{\pixlo{1},\pixlo{2}\, \dots \pixlo{N} \}\) -- Set of observed pixel vectors
        \ENSURE \(\corpix = \{\corpixlo{1},\corpixlo{2}\, \dots \corpixlo{N} \}\) -- Set of corrected Pixels

        \STATE \(\{\yy{1},\yy{2},\dots \yy{N}\}\) \(\gets\) SVD\textunderscore Dimensionality\textunderscore Reduction(\(\img\))

        \STATE Compute point on hyperplane: \(\ccstar \gets \frac{1}{N} \sum_{i=1}^N \mathbf{y}_i\)
        \STATE  \(\mathcal{N} \gets\) Generate candidate normals list:

        \STATE Define objective function:
        \begin{align*}
            \Psi(\mathbf{n}) = \sum_{i=1}^N \| \mathbf{y}_i - \hat{\mathbf{y}}_i \|^2,\quad \\\text{where } \hat{\mathbf{y}}_i = \frac{\yy{i}}{\muhat{i}} ,\quad \muhat{i} = \frac{\mathbf{y}_i^\top \mathbf{n}}{(\mathbf{c}^*)^\top \mathbf{n}}
        \end{align*}

        \STATE \(\nn_{\text{pso}} \gets \) Particle Swarm Optimization (PSO) to minimize \(\Psi(\mathbf{n})\) using initial set \(\mathcal{N}\)

        \STATE \(\nn_{\text{best}} \gets\) Fine-tune \(\mathbf{n}_\text{pso}\) using Gradient Descent on \(\Psi(\mathbf{n})\)

        \FOR{\(i = 1\) to \(N\)}
        \STATE Compute scaling factor: \(\muhat{i} \gets \frac{\mathbf{y}_i^\top \mathbf{n}_{\text{best}}}{(\mathbf{c}^*)^\top \mathbf{n}_{\text{best}}}\)
        \vspace{1em}

        \IF{\(\muhat{i} \leq 0\)}
        \STATE Remove pixel \(i\) from the dataset marking it as an outlier and go to Step 2
        \ENDIF

        \vspace{1em}
        \STATE Compute corrected pixel: \(\corpixlo{i} \gets \frac{\pixlo{i}}{\muhat{i}}\)
        \ENDFOR

        \RETURN \(\{\corpixlo{1}, \corpixlo{2}, \dots \corpixlo{N}\}\)

    \end{algorithmic}
\end{algorithm}

\input{Figures_Latex/Loss_landscape}
Based on the results obtained in previous sections, the algorithm for correcting the scaling factors was developed. This section discusses the implementation of the algorithm and specifically addresses the optimization strategies for minimizing the objective function.

As discussed previously, the proposed algorithm aims to correct for the variations of scale due to illumination and other effects so as to enhance the performance of other HU Algorithms.

It does this by estimating the scaling factors, which are then used to correct the pixels as per \eqref{eq-Correction-Equation}. The algorithm estimates these scaling factors using \eqref{eq:muhat}. This equation has two parameters \(\ccstar\) and \(\nhat\). \(\ccstar\) is obtained using \eqref{eq:cstar}. \(\nhat\) is obtained as the minimizer of \(\Psi(\nn)\) (see \eqref{eq:nhat-Psi}) where \(\Psi(\nn)\) is given by \eqref{eq:Psi}.

A summary of the algorithm is given in Algorithm \ref{alg:Algorithm}. First the pixels are dimensionally reduced from \(\nband\) to \(\KK\) dimensions using SVD as described in Section \ref{sec:dimred}. This is crucial for the subsequent steps. These \(\yy{i}\) are then used to calculate \(\ccstar\).

Now, it is necessary to find \(\nhat\) by finding the minima of \(\Psi(\nn)\). However, it was discovered that due to the complexity of the loss landscape with many local minima, Gradient Descent alone was not reliable for finding the true minima and making a good estimate for \(\nhat\). This can be visualized in Figure \ref{fig:loss-landscape}. It can be seen that there are many local minima that gradient descent tends to get stuck at.

In order to alleviate this, initially \(\Psi(\nn)\) is optimized using Particle-Swarm-Optimization (PSO) with carefully generated initial points for the particles to ensure convergence. In particular, a set of candidate normals is generated that is likely to be in the neighborhood of the true normal. This is done by taking random \(\KK\)-sets of dimensionally reduced pixels (\(\yy{i}\)) which are reasonably far apart and calculating the normal of the hyperplane on which these pixels lie (this is well defined if the pixels are chosen to be linearly independent and ill-conditioned sets are rejected). If \(B \in \real^{\KK \times \KK}\) is a matrix that has the chosen \(\KK\) pixels as columns, the candidate normal corresponding to this random set, \(\nn\), is given by \eqref{eq:nn-candidate} (see Appendix~\ref{appendix:candidate-normal}).
\begin{align}
    \label{eq:nn-candidate}
    B^T \nn = \onemat \;\;\;\text{where}\; \onemat = (1,1, \dots 1)^T \in \real^K
\end{align}
Importantly, the pixels used to form \(B\) are \emph{not} assumed to be pure (they may be highly mixed).
This is done several times to generate a set of candidates \(\mathcal{N}\). It can be shown that if the candidate pixels are reasonably uniformly lit the candidate normal will be close to the true normal. Since many such candidates are generated, certain portion is likely to be in the neighborhood of the true normal.
This set \(\mathcal{N}\) is used as initial points in the PSO algorithm to minimize \(\Psi(\nn)\), thus obtaining an optimal normal \(\nn_{\text{PSO}}\). This is then further fine-tuned using Gradient Descent with \(\Psi(\nn)\) as a loss function to obtain the final estimate \(\mathbf{n}_{\text{best}}\).

Finally, the obtained estimate \(\mathbf{n}_{\text{best}}\) for the minimizer \(\nhat\) is used to calculate the estimate the predicted scaling factors \(\muhat{i}\) for each of the \(i^\tth\) pixel as per Equation \eqref{eq:muhat}. This is then used to estimate the corrected pixels \(\corpixlo{i}\) using Equation \(\eqref{eq-Correction-Equation}\).

While not encountered in practice, if any of the estimated scaling factors are negative or zero, the corresponding pixel is removed from the dataset and the algorithm is restarted. This is because a negative or zero scaling factor is not physically meaningful and likely caused by a pixel with large unmodelled effects or noise. This can be further restricted to a specific range if relavent domain knowledge is available.

%% file: Figures_Latex/Loss_landscape.tex
\begin{figure}[!t]
    \centering
    \includegraphics[width=0.5\linewidth]{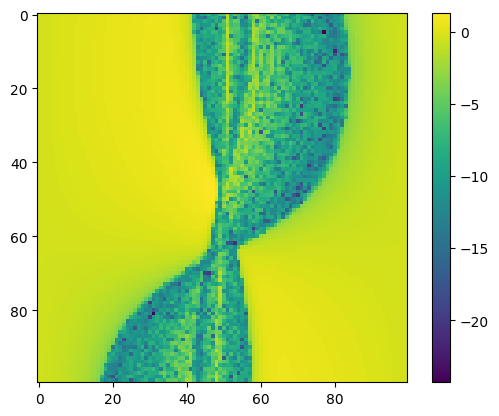}
    \caption{Heat map of $\log(\Psi(\nn))$ for different $\nn$ with $\nn$ parametrized in polar co-ordinates for the Samson dataset. Many local minima of $\Psi(\nn)$ can be observed}
    \label{fig:loss-landscape}
\end{figure}

%% file: Chapters/Experiments/experiments.tex
\input{Figures_Latex/rmse_vs_std}
\section{Experiments}
This section presents the experimental validation of the proposed preprocessing algorithm and its impact on hyperspectral unmixing performance. We evaluate its ability to accurately estimate per-pixel scaling factors and demonstrate how this correction improves abundance estimation of SOTA HU methods.

For the experimental validation of the algorithm, two synthetic datasets with synthetic scaling factors are used. Then the ability of the proposed algorithm to correctly predict these scaling factors is tested. This aims to demonstrate that the algorithm performs as theoretically predicted. Additionally, the robustness of the algorithm's prediction under different conditions is also tested.

Afterwards, the effect of the preprocessing algorithm on HU performance is investigated.
This is achieved by comparing the results of SOTA algorithms for datasets preprocessed with the proposed algorithm against the results for the raw datasets.


The approach is evaluated across three distinct categories of algorithms. First, traditional signal processing methods are represented by FCLSU~\cite{FCLSU} and SUnSAL~\cite{SUnSAL} with NFINDR~\cite{NFINDR} to analyze the impact of the preprocessing step. Second, to capture the evolution of deep learning-based techniques, CyCU~\cite{CYCU}, MiSiCNet~\cite{MiSiCNet}, DeepTrans~\cite{Deeptrans}, and UST-Net~\cite{USTNet} were selected. Third, algorithms explicitly designed to account for spectral variability (SV) were examined. This final category includes PGMSU~\cite{PGMSU} and IDNet~\cite{IDNet}, which rely on probabilistic frameworks~\cite{SVBible}; SSAF-Net~\cite{SSAFNet} and PPMNet~\cite{PPMNet}, which utilize the Perturbed Linear Mixture Model (PLMM)~\cite{SVPhysics}; and GLMM~\cite{GLMM}, which is based on the Generalized Linear Mixture Model.

By choosing several algorithm form different domains and different stages HU algorithm evolution, the universal nature of the performance improvements brought by the proposed preprocessing algorithm will be demonstrated.

This evaluation is conducted on two synthetic datasets and two real datasets. The real datasets used in this study are the Samson and Urban datasets.

Finally, an ablation study on the different components of the algorithms is performed to demonstrate the importance of each step.

\input{Chapters/Experiments/datasets}

\input{Figures_Latex/scaling_factor_compare}

\input{Figures_Latex/scaling_factor_compare_spheric}

\input{Chapters/Experiments/Algo_evaluation}



\input{Figures_Latex/samson_and_urban_shadow}

\input{Chapters/Experiments/SOTA_Performance_Improvement}
\input{Chapters/Experiments/signature_visualization.tex}
\input{Chapters/Experiments/preprocessing_comparison.tex}
\input{Chapters/Experiments/noise_robustness.tex}
\input{Chapters/Experiments/non_linear_evaluation.tex}
\input{Chapters/Experiments/low_purity.tex}

\input{Chapters/Experiments/ablation_study}
\input{Chapters/Experiments/pso_robustness.tex}

%% file: Figures_Latex/rmse_vs_std.tex
\begin{figure*}[!t]
\centering
\subfloat[]{\includegraphics[width=2.5in]{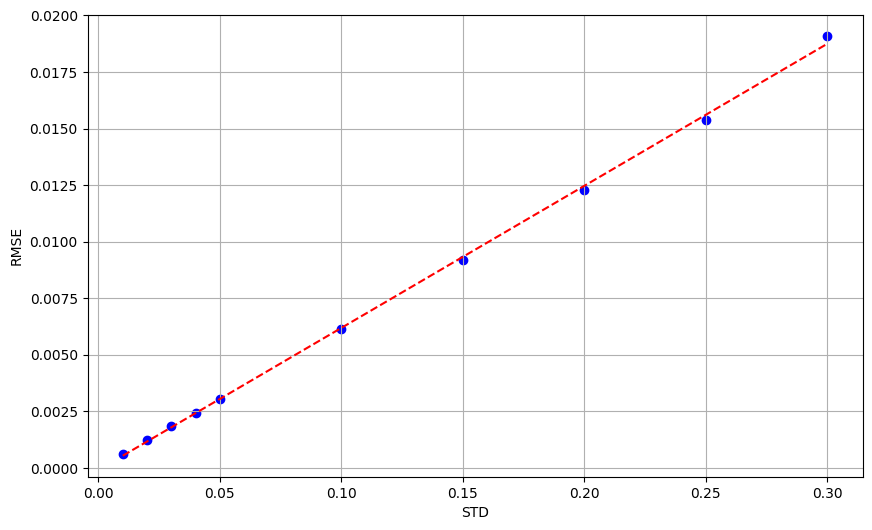}%
\label{fig:matern-std-rmse}}
\hfil
\subfloat[]{\includegraphics[width=2.5in]{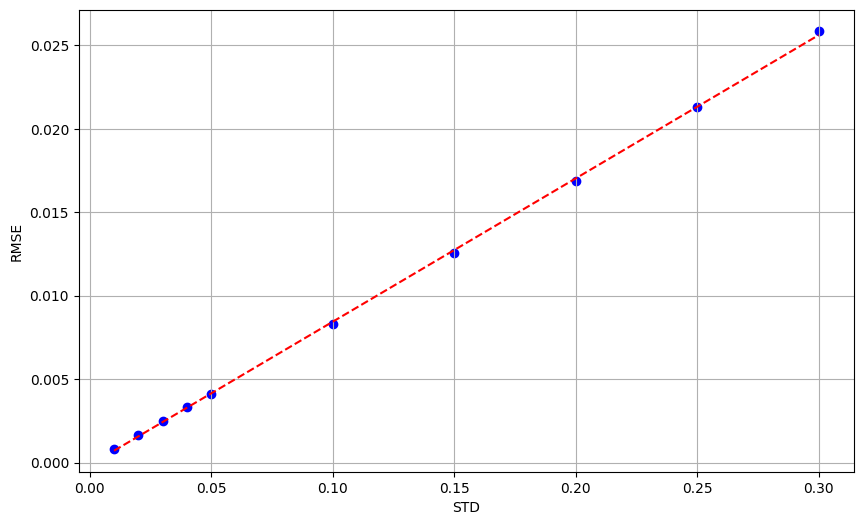}%
\label{fig:spheric-std-rmse}}
\caption{Variation of the RMSE of estimated scaling factors with the standard deviation of scaling factors. (a) RMSE vs. STD for Matern Dataset. (b) RMSE vs. STD for Spheric Dataset}
\label{fig:std-rmse}
\end{figure*}

%% file: Chapters/Experiments/datasets.tex
\subsection{Hyperspectral Dataset Description}

\subsubsection{Synthetic Datasets}

For the synthetic experiments, we use datasets from the IC Synthetic Hyperspectral Collection, generated using the MATLAB toolbox provided by the Grupo de Inteligencia Computacional at UPV/EHU \cite{SynthToolBox}. Each image consists of \(128 \times 128\) pixels and 431 spectral bands, synthesized from five endmembers selected from the USGS spectral library. Specifically, we use the \emph{Spheric Gaussian Field} and \emph{Matern Gaussian Field} mixtures, both of which are commonly used in unmixing benchmarks.

To simulate topographic and illumination-induced spectral variability, we apply a pixel-wise scaling factor to the original data, producing scaled versions of each dataset. An example of a scalar field used for this transformation is shown in Figure \ref{fig:True-mu}. This is generated with spatial correlations in the scalar field to mimic natural variations in topography and illumination.

Both the original and scaled versions are used to validate the proposed preprocessing algorithm and assess its impact on downstream HU performance.

\subsubsection{Samson Dataset}
The dataset consists of \(95\times 95\) pixels, with each pixel containing 156 bands. The image contains three endmembers, namely soil, tree, and water. This dataset is especially relevant for this research since, as can be observed from nearly pure pixels given in Figure \ref{fig:signature-correction-samson}, the dataset contains significant variations in scale.

\subsubsection{Urban Dataset}
This dataset consists of \(307 \times 307\) pixels, with each pixel having data from 210 spectral bands. After removing some of the noisy bands, the final Urban dataset contains 162 bands, which is commonly used in HU benchmarks. The dataset consists of four endmembers: asphalt, grass, tree, and roof. Similar to the Samson dataset, Spectral Variability in Scale can be observed in this dataset as well (See Figure \ref{fig:signature-correction-urban}).

%% file: Figures_Latex/scaling_factor_compare.tex
\begin{figure}[!t]
    \centering
    \subfloat[]{\includegraphics[width=1.5in]{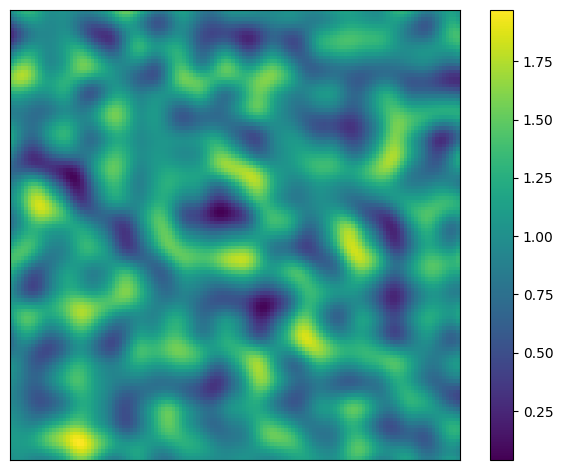}%
        \label{fig:True-mu}}
    \hfil
    \subfloat[]{\includegraphics[width=1.5in]{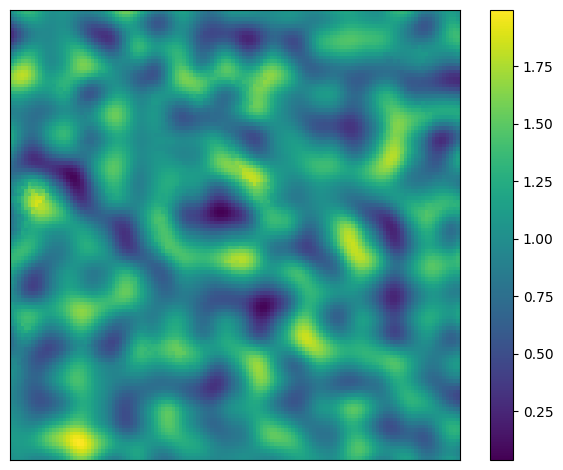}%
        \label{fig:Pred-mu}}
    \caption{Comparison of the Original and Predicted Scaling Factors for the Matern Dataset with scaling factors having a standard deviation of 0.3. (a) Original Scaling Factors. (b) Predicted Scaling Factors}
    \label{fig:mu-compare}
\end{figure}

%% file: Figures_Latex/scaling_factor_compare_spheric.tex
\begin{figure}[!t]
\centering
\subfloat[]{\includegraphics[width=1.5in]{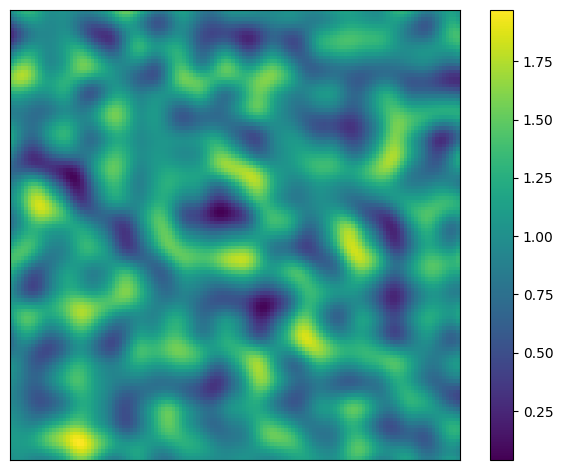}%
\label{fig:True-mu-spheric}}
\hfil
\subfloat[]{\includegraphics[width=1.5in]{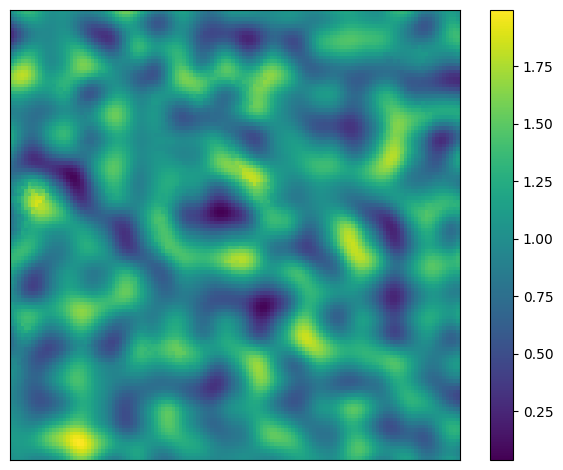}%
\label{fig:Pred-mu-spheric}}
\caption{Comparison of the Original and Predicted Scaling Factors for the Spheric Dataset with scaling factors having a standard deviation of 0.3. (a) Original Scaling Factors. (b) Predicted Scaling Factors}
\label{fig:mu-compare-spheric}
\end{figure}

%% file: Chapters/Experiments/Algo_evaluation.tex
\subsection{Evaluation of Accuracy with Synthetic Datasets}

To verify that the algorithm behaves as theoretically expected, it is applied to the synthetic datasets that were previously modified using known pixel-wise scaling factors. The predicted scaling factors are then compared with the ground truth to assess the accuracy of the preprocessing algorithm.

Accuracy is quantified using the RMSE between the estimated and true scaling factors as given in \eqref{eq:mu-rmse}. This evaluation is performed on both synthetic datasets, each modified using scaling fields with different standard deviations. The Root Mean Square Error (RMSE) values corresponding to each standard deviation, for each synthetic dataset, are reported in the graph given in Figure \ref{fig:std-rmse}.

\begin{align}
    \label{eq:mu-rmse}
    \text{RMSE}_{\mu} =\sqrt{
        \frac{1}{\npix}
        \sum_{i=1}^{\npix}{
                (\muhat{i} - \mu_i)^2}
    }
\end{align}

As can be observed, the algorithm performs well with very low estimation errors for both datasets. These results highlight the excellent accuracy of the algorithm.

Furthermore, notice that the estimation error increases linearly with the standard deviation of the scaling factors. This is in agreement with the prediction in Equation \eqref{eq:Upper-bound-on-the-Relative-RMSE-Error}. This property indicates that the algorithm performs as optimally as expected and does not break down at larger scale variations in the scaling factors.

Furthermore, a visual map of predicted scaling factors vs. ground truth scaling factors is shown in Figure \ref{fig:mu-compare} for a qualitative analysis. The original scaling factors are given in Figure \ref{fig:True-mu}, while the predicted scaling factors are given in Figure \ref{fig:Pred-mu}. Similarly, the scaling factor comparisons for the Spheric Dataset are given in Figure \ref{fig:Pred-mu-spheric}. Specifically, these visual comparisons are provided for the synthetic datasets with the scaling factors having a standard deviation of 0.3.

A detailed inspection of these heatmaps reveals that the proposed algorithm accurately captures not only the global distribution of the scaling factors but also the fine-grained, spatially correlated local variations that simulate complex topographical and illumination effects. For instance, the highly illuminated regions (areas with elevated scaling factors, shown in yellow) and the deep shadow regions (areas with low scaling factors, shown in dark blue) in the predicted maps perfectly align with the structural patterns of the ground truth. The preservation of these distinct spatial structures demonstrates that the geometric projection method does not merely fit a global average, but successfully reconstructs the exact per-pixel scaling field. As one can observe, the predicted scaling factors are nearly identical to the ground truth scaling factors, which strongly corroborates the low RMSE values and stands as a testament to the high fidelity and accuracy of the algorithm.

%% file: Figures_Latex/samson_and_urban_shadow.tex
\begin{figure}[!t]
    \centering
    \subfloat[]{%
        \includegraphics[width=0.48\linewidth]{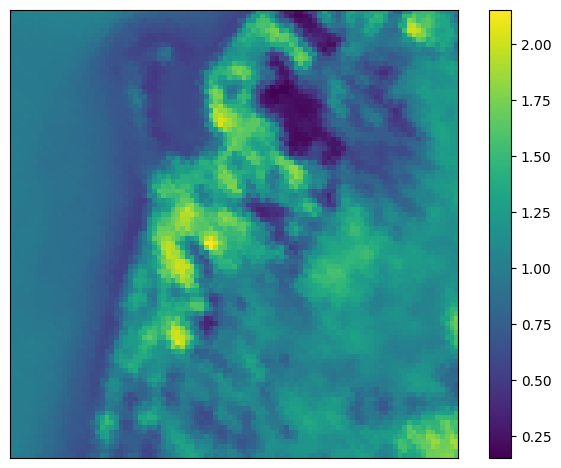}
        \label{fig:samson-pred-mu}
    }
    \hfill
    \subfloat[]{%
        \includegraphics[width=0.48\linewidth]{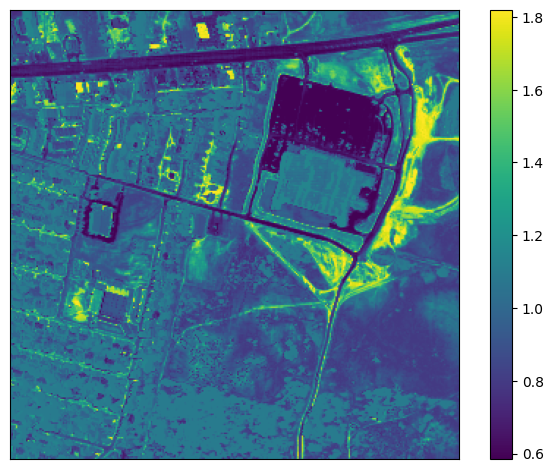}
        \label{fig:urban-pred-mu}
    }
    \caption{Predicted Scaling Factors $\muhat{i}$ for two real datasets. (a) Predicted Scaling Factors for the Samson Dataset. (b) Predicted Scaling Factors for the Urban Dataset.}
    \label{fig:pred-mu-combined}
\end{figure}

%% file: Chapters/Experiments/SOTA_Performance_Improvement.tex
\subsection{Effect of the Proposed Preprocessing on HU Algorithms}
This section aims to demonstrate that the scale corrections produced by the proposed algorithm significantly improve the accuracy of unmixing algorithms compared to the case where the raw datasets are used, particularly by reducing errors in abundance estimation.

To this effect, the RMSE between abundance estimates and ground truth abundance, for the SOTA algorithm, was calculated on both preprocessed datasets and raw datasets. This was done for the two synthetic datasets, Spheric and Matern, as well as for the two real datasets, Urban and Samson. The RMSE between ground truth abundances (\(\matabd\)) and the estimated abundances (\(\hat{\matabd}\)) is calculated using Equation \eqref{eq:abundance-rmse}. The Mean RMSE is calculated by taking the root mean square of the RMSE values for each endmember. Experiments are conducted over 10 independent runs, and average results are reported to ensure robustness against random fluctuations.

\begin{align}
    \label{eq:abundance-rmse}
    RMSE(A,\hat{A}) = \sqrt{
    \frac{1}{N} \sum_{i=1}^N{||\abdo_i - \hat{\abdo}_i||^2}
    }
\end{align}

The preprocessing algorithm was mainly designed with the objective of reducing abundance estimation errors, and improving endmember estimation is not our main focus. However, for the sake of completeness, the accuracy of the endmember estimation was evaluated and summarized. The mean SAD error between the ground truth (\(\matsig\)) and predicted endmember signatures (\(\widehat{\matsig}\)) given by Equation \eqref{eq:SAD-error}, was used to evaluate the accuracy of the endmember estimation.
\begin{align}
    \label{eq:SAD-error}
    {\text{SAD}}(\matsig,\widehat{\matsig}) & = \frac{1}{K} \sum_{i=1}^K{\arccos{\frac{\sigo{i}^T \widehat{\sigo{i}}}{\norm{\sigo{i}} \norm{\widehat{\sigo{i}}}} }}
\end{align}

These numerical results for these algorithms are tabulated, and the results without any preprocessing step are labeled ``Before`` and the results for the dataset with the preprocessing algorithm are labeled ``After''. The better results (lower error) of the two cases for the SAD and RMSE metrics are given in \color{blue} \textbf{bold text colored blue} \color{black} for easier comparison. Since the effect of the preprocessing algorithm on abundance estimation is our main focus, these results are given in greater detail.

Finally, for a qualitative analysis of the improvements for abundance estimation with visual comparison, the estimated abundance maps are also given as figures with the same labeling as the tables.

In each of the following sections, the abundance estimation errors will be analyzed along with a qualitative analysis of the predicted abundance maps. In the final section analysis of the effect on endmember estimation will be done for the sake of completeness.

\input{Tables_Latex/RMSE_Tables/Matern_Synthetic}
\input{Tables_Latex/RMSE_Tables/Spheric_Synthetic}
\input{Tables_Latex/RMSE_Tables/samson}
\input{Tables_Latex/RMSE_Tables/urban_4}

\subsubsection{Experiments with Synthetic Datasets}

\input{Figures_Latex/matern_sota_compare_abd}
\input{Figures_Latex/spheric_sota_compare_abd}
For this section, synthetic datasets scaled by synthetic scaling factors with a standard deviation of 0.3 were used. The accuracy of abundance estimation was compared between the scaled synthetic dataset and the dataset corrected using the proposed preprocessing algorithm.

First, consider the RMSE values for abundance estimation on the Matern and Spheric datasets, as presented in Table \ref{tab:matern-rmse} and Table \ref{tab:spheric-rmse}. For the Matern dataset, the smallest improvement in total RMSE from the scaled dataset to the corrected dataset is observed with the IDNet algorithm, showing approximately a 17\% reduction. In contrast, the largest improvement is achieved by the SUnSAL algorithm, with an RMSE reduction of nearly 97\%. Many algorithms exhibit improvements on the order of 50\%. A similar pattern is observed for the Spheric dataset, where improvements range from a minimum of approximately 23\% to nearly 99\%, with many algorithms again achieving improvements close to 50\%.

Examining the estimated abundance maps for both datasets, illustrated in Figure \ref{fig:matern-sota-abd} and Figure \ref{fig:spheric-sota-abd}, it is evident that the presence of scaling factors distorts the estimated abundance maps. These distortions are often misinterpreted by the algorithms as genuine abundance variations, as indicated by the visible patterns of the scaling factors in the abundance maps. Even when the scaling factors themselves are not directly reflected, their influence causes the algorithms to converge to incorrect endmembers, such as the first endmember estimated by the UST-Net algorithm.

Finally, considering the results without the proposed preprocessing step, the best-performing algorithm in terms of Mean RMSE for the Matern dataset is PGMSU with a value of 0.1410. After applying the preprocessing step, SUnSAL becomes the best performer with an overall error of 0.0048, closely followed by GLMM with an error of 0.0049. This corresponds to an error reduction of approximately 96\%, highlighting the significant contribution of the preprocessing step to improving the unmixing results. Similarly, for the Spheric dataset, the best performer changes from MiSiCNet to GLMM and SUnSAL, with the error reducing from 0.1789 to 0.0043, corresponding to an error reduction of approximately 97\%.
\subsubsection{Experiments with Samson Dataset}

\input{Figures_Latex/samson_sota_compare_abd.tex}

In this section, the performance improvement resulting from the preprocessing algorithm on the Samson dataset is analyzed. First, the scaling factors for the Samson dataset were estimated, and subsequently, the dataset was corrected using these factors. The estimated scaling factors are presented in Figure \ref{fig:samson-pred-mu}.

Next, error metrics for both the original and corrected datasets are evaluated across several SOTA algorithms.

From the numerical results for abundance estimation shown in Table \ref{tab:samson-rmse}, it is evident that the majority of algorithms exhibit significant improvements. The total RMSE reductions range between 20\% and 50\%, with the highest improvement of 55\% observed for the DeepTrans algorithm. The sole exception is the SUnSAL algorithm, which demonstrates a slight decline in overall performance. This reduction is of little consequence since SUnSAL hasn't performed well for this dataset in either case.

Visual inspection of the estimated abundance maps given in Figure \ref{fig:samson-sota-abd} reveals that the scale correction substantially enhances the quality of the abundance estimations.

For instance, consider the predicted abundance maps corresponding to the water endmember for the cases without the proposed preprocessing algorithm. Several algorithms erroneously predict elevated water abundance in specific regions compared to the ground truth. Examination of the estimated scaling factors (Figure \ref{fig:samson-pred-mu}) indicates that these regions have been substantially down-scaled, with values close to 0.25. This down-scaling attenuates the magnitude of the tree endmember spectra, causing them to resemble the water signature, which typically exhibits a lower magnitude. Consequently, this spectral distortion misleads the algorithms into misidentifying these regions as water.

Finally, the best performing algorithm in both cases is UST-Net, which shows an improvement with the proposed preprocessing applied. The error has dropped from 0.0832 to 0.0645, corresponding to an error reduction of 22\%. Also note that PGMSU, which had quite a high error value, has achieved an RMSE value of 0.0776, getting closer to the result of the best performer.
\subsubsection{Experiments with Urban Dataset}

\input{Figures_Latex/urban_sota_compare_abd}
Similar to the Samson dataset, the estimated scaling factors for the Urban dataset were used to generate a corrected version of the dataset. These predicted scaling factors are shown in Figure \ref{fig:urban-pred-mu}. Subsequently, HU performance was compared between the original and corrected datasets.

Consider the RMSE errors given in Table \ref{tab:urban-rmse}. The most RMSE improvements range from 25\% to 73\% with the exception being IDNet where only around 5\% improvement was observed, with the maximum improvement of 73\% observed for SUnSAL and the minimum improvement of 25\% shown by CYCU.

Qualitative improvements are also evident in the estimated abundance maps given in Figure \ref{fig:urban-sota-abd}. For example, several algorithms, such as PGMSU, MiSiCNet, and SSAF-Net, converged toward incorrect abundance maps for the Grass endmember in the raw dataset. The same region, which is prominent in these incorrect abundance estimates, corresponds to large estimated scaling factors close to 1.8 (see Figure \ref{fig:urban-pred-mu}), likely misleading these algorithms. The correction using the proposed method has rectified this issue, resulting in improved abundance estimation.

Finally, considering the overall improvement due to the proposed algorithm, one can observe that the best performer without any preprocessing algorithm is IDNet, with an RMSE of 0.0953. After the preprocessing algorithm is applied, MiSiCNet shows the best performance with an RMSE of 0.0799. This shows an overall contribution of 18\% reduction in error by the preprocessing algorithm for the abundance estimation problem.

\input{Tables_Latex/SAD_Tables/sad_summary}

\subsubsection{Analysis of Endmember Signature Estimation}
Although the primary objective of this study is to assess the impact of the proposed algorithm on abundance estimation, the performance with respect to endmember signature estimation was also examined for completeness. Given that the algorithm primarily addresses the scale of pixel signatures, substantial improvements in endmember estimates were not anticipated. This expectation was largely confirmed by the results, with the exception of a few instances where notable improvements were observed.

Accordingly, a detailed breakdown of endmember estimation errors based on the SAD metric is omitted for brevity. Instead, a concise overview of the most relevant findings is provided in the following discussion and summarized in Table \ref{tab:sad-summary}. Since FCLS and SUnSAL both use NFINDR for endmember estimation, the results for NFINDR are presented.

For the Synthetic Matern dataset, preprocessing yields substantial improvements across most methods. Particularly noteworthy cases are NFINDR and GLMM, which achieves near-perfect estimation of the endmember signatures, with a SAD error approaching zero (in the order \(1\times 10^{-6}\)) once the scaling factors are corrected. Similarly, for the Synthetic Spheric dataset, most algorithms benefit significantly from preprocessing. NFINDR again demonstrates near-perfect endmember estimation under corrected scaling, whereas CYCU shows a slight increase in estimation error compared to the unprocessed case.

For the Samson dataset, the overall effect of preprocessing is more moderate. While some algorithms show clear improvements, in several cases the changes are marginal, and in a few instances performance remains nearly unchanged or slightly reduced. In contrast, the Urban dataset exhibits more consistent gains, with substantial improvements observed in most methods and smaller positive and negative changes in a few cases.

As evident from the results, while significant improvements were achieved in certain cases, in several others the gains were marginal or negligible. The improvements in endmember signature estimates are likely a secondary effect of enhanced abundance estimation, which provides more reliable guidance for the endmember estimation process. An additional contributing factor may be the more consistent geometric structure introduced by the preprocessing algorithm.

\subsubsection{Overall Analysis of Results}
Overall, the proposed preprocessing algorithm delivers consistently strong gains in abundance estimation accuracy across a wide range of hyperspectral unmixing methods. In most cases, the improvement is substantial, often in the order of 50\% reduction in RMSE. This indicates that the algorithm effectively mitigates the detrimental influence of scale-induced spectral variability, allowing diverse unmixing approaches to operate closer to their full potential.

For endmember signature estimation, the improvements are more algorithm dependent. While several methods exhibit significant reductions in SAD error, sometimes exceeding 60\%, others show smaller or negligible changes. This suggests that the proposed preprocessing algorithm most directly benefits abundance estimation, but can also lead to notable gains in endmember accuracy when the downstream algorithm is sensitive to scale distortions. Taken together, the results demonstrate that the proposed algorithm is a powerful and broadly applicable preprocessing tool, particularly valuable when reliable abundance maps are crucial.

%% file: Tables_Latex/RMSE_Tables/Matern_Synthetic.tex
\begin{table*}[tbp]
	\caption{Abundance RMSE for the Matern Synthetic Dataset without Preprocessing (Before) and with Preprocessing (After)}
	\label{tab:matern-rmse}
	\begin{adjustbox}{width=\textwidth}
		\begin{tabular}{|l|ll|ll|ll|ll|ll|ll|ll|ll|ll|rr|rr|}
			\hline
			\multicolumn{1}{|c|}{}                                      & \multicolumn{2}{c|}{\textbf{CYCU}}  & \multicolumn{2}{c|}{\textbf{DeepTrans}} & \multicolumn{2}{c|}{\textbf{FCLS}}  & \multicolumn{2}{c|}{\textbf{MiSiCNet}} & \multicolumn{2}{c|}{\textbf{PGMSU}} & \multicolumn{2}{c|}{\textbf{PPM Net}}  & \multicolumn{2}{c|}{\textbf{SSAF-net}} & \multicolumn{2}{c|}{\textbf{SUnSAL}}   & \multicolumn{2}{c|}{\textbf{UST Net}} & \multicolumn{2}{c|}{\textbf{GLMM}}     & \multicolumn{2}{c|}{\textbf{IDNet}}                                                                                                                                                                                                                                                                                                                                                                                                                                                              \\ \cline{2-23}
			\multicolumn{1}{|c|}{}                                      & \multicolumn{1}{c}{\textbf{Before}} & \multicolumn{1}{c|}{\textbf{After}}     & \multicolumn{1}{c}{\textbf{Before}} & \multicolumn{1}{c|}{\textbf{After}}    & \multicolumn{1}{c}{\textbf{Before}} & \multicolumn{1}{c|}{\textbf{After}}    & \multicolumn{1}{c}{\textbf{Before}}    & \multicolumn{1}{c|}{\textbf{After}}    & \multicolumn{1}{c}{\textbf{Before}}   & \multicolumn{1}{c|}{\textbf{After}}    & \multicolumn{1}{c}{\textbf{Before}} & \multicolumn{1}{c|}{\textbf{After}}    & \multicolumn{1}{c}{\textbf{Before}} & \multicolumn{1}{c|}{\textbf{After}}    & \multicolumn{1}{c}{\textbf{Before}} & \multicolumn{1}{c|}{\textbf{After}}          & \multicolumn{1}{c}{\textbf{Before}} & \multicolumn{1}{c|}{\textbf{After}}    & \multicolumn{1}{l}{\textbf{Before}} & \multicolumn{1}{l|}{\textbf{After}}    & \multicolumn{1}{l}{\textbf{Before}}    & \multicolumn{1}{l|}{\textbf{After}}    \\ \cline{2-23}
			\multicolumn{1}{|c|}{\multirow{-2}{*}{\textbf{Endmembers}}} & 0.1366                              & {\color[HTML]{3531FF} \textbf{0.1344}}  & 0.1887                              & {\color[HTML]{3531FF} \textbf{0.0626}} & 0.0785                              & {\color[HTML]{3531FF} \textbf{0.0066}} & 0.1131                                 & {\color[HTML]{3531FF} \textbf{0.0155}} & 0.1374                                & {\color[HTML]{3531FF} \textbf{0.0762}} & 0.1274                              & {\color[HTML]{3531FF} \textbf{0.0448}} & 0.1712                              & {\color[HTML]{3531FF} \textbf{0.0684}} & 0.0753                              & {\color[HTML]{3531FF} \textbf{0.0037}}       & 0.2189                              & {\color[HTML]{3531FF} \textbf{0.1747}} & 0.1069                              & {\color[HTML]{3531FF} \textbf{0.0038}} & {\color[HTML]{3531FF} \textbf{0.0895}} & 0.0934                                 \\ \hline
			Endmember 2                                                 & 0.2325                              & {\color[HTML]{3531FF} \textbf{0.0876}}  & 0.2287                              & {\color[HTML]{3531FF} \textbf{0.0505}} & 0.3777                              & {\color[HTML]{3531FF} \textbf{0.0055}} & 0.2619                                 & {\color[HTML]{3531FF} \textbf{0.0515}} & 0.1789                                & {\color[HTML]{3531FF} \textbf{0.0605}} & 0.2178                              & {\color[HTML]{3531FF} \textbf{0.0537}} & 0.2891                              & {\color[HTML]{3531FF} \textbf{0.1522}} & 0.2722                              & {\color[HTML]{3531FF} \textbf{0.0038}}       & 0.2791                              & {\color[HTML]{3531FF} \textbf{0.0765}} & 0.1972                              & {\color[HTML]{3531FF} \textbf{0.0038}} & 0.2787                                 & {\color[HTML]{3531FF} \textbf{0.2317}} \\ \hline
			Endmember 3                                                 & 0.264                               & {\color[HTML]{3531FF} \textbf{0.2484}}  & 0.1664                              & {\color[HTML]{3531FF} \textbf{0.0402}} & 0.1576                              & {\color[HTML]{3531FF} \textbf{0.0042}} & 0.2427                                 & {\color[HTML]{3531FF} \textbf{0.0145}} & 0.1156                                & {\color[HTML]{3531FF} \textbf{0.0398}} & 0.1155                              & {\color[HTML]{3531FF} \textbf{0.096}}  & 0.2159                              & {\color[HTML]{3531FF} \textbf{0.082}}  & 0.1463                              & {\color[HTML]{3531FF} \textbf{0.0042}}       & 0.1474                              & {\color[HTML]{3531FF} \textbf{0.0872}} & 0.1184                              & {\color[HTML]{3531FF} \textbf{0.0042}} & {\color[HTML]{3531FF} \textbf{0.1236}} & 0.1591                                 \\ \hline
			Endmember 4                                                 & 0.1604                              & {\color[HTML]{3531FF} \textbf{0.1499}}  & 0.0857                              & {\color[HTML]{3531FF} \textbf{0.0602}} & 0.0975                              & {\color[HTML]{3531FF} \textbf{0.0073}} & 0.0896                                 & {\color[HTML]{3531FF} \textbf{0.0305}} & 0.0927                                & {\color[HTML]{3531FF} \textbf{0.0747}} & 0.0975                              & {\color[HTML]{3531FF} \textbf{0.0373}} & 0.1072                              & {\color[HTML]{3531FF} \textbf{0.0799}} & 0.0933                              & {\color[HTML]{3531FF} \textbf{0.0071}}       & 0.1522                              & {\color[HTML]{3531FF} \textbf{0.113}}  & 0.0633                              & {\color[HTML]{3531FF} \textbf{0.0071}} & {\color[HTML]{3531FF} \textbf{0.0928}} & 0.1257                                 \\ \hline
			Endmember 5                                                 & 0.2832                              & {\color[HTML]{3531FF} \textbf{0.1978}}  & 0.1716                              & {\color[HTML]{3531FF} \textbf{0.059}}  & 0.1785                              & {\color[HTML]{3531FF} \textbf{0.0095}} & 0.1681                                 & {\color[HTML]{3531FF} \textbf{0.0496}} & 0.1629                                & {\color[HTML]{3531FF} \textbf{0.1224}} & 0.195                               & {\color[HTML]{3531FF} \textbf{0.0619}} & 0.2266                              & {\color[HTML]{3531FF} \textbf{0.0788}} & 0.1795                              & {\color[HTML]{3531FF} \textbf{0.0045}}       & 0.1558                              & {\color[HTML]{3531FF} \textbf{0.1329}} & 0.0892                              & {\color[HTML]{3531FF} \textbf{0.0046}} & 0.1881                                 & {\color[HTML]{3531FF} \textbf{0.1756}} \\ \hline
			\textbf{Mean RMSE}                                          & 0.2229                              & {\color[HTML]{3531FF} \textbf{0.1727}}  & 0.1746                              & {\color[HTML]{3531FF} \textbf{0.0551}} & 0.2074                              & {\color[HTML]{3531FF} \textbf{0.0068}} & 0.1879                                 & {\color[HTML]{3531FF} \textbf{0.036}}  & {\ul 0.1410}                          & {\color[HTML]{3531FF} \textbf{0.0795}} & 0.1578                              & {\color[HTML]{3531FF} \textbf{0.0622}} & 0.2109                              & {\color[HTML]{3531FF} \textbf{0.0971}} & 0.1686                              & {\color[HTML]{3531FF} {\ul \textbf{0.0048}}} & 0.1975                              & {\color[HTML]{3531FF} \textbf{0.122}}  & 0.1235                              & {\color[HTML]{3531FF} \textbf{0.0049}} & 0.1703                                 & {\color[HTML]{3531FF} \textbf{0.1639}} \\ \hline
		\end{tabular}
	\end{adjustbox}
\end{table*}

%% file: Tables_Latex/RMSE_Tables/Spheric_Synthetic.tex
\begin{table*}[tbp]
	\caption{Abundance RMSE for the Spheric Synthetic Dataset without Preprocessing (Before) and with Preprocessing (After)}
	\label{tab:spheric-rmse}
	\begin{adjustbox}{width=\textwidth}
		\begin{tabular}{|l|ll|ll|ll|ll|ll|ll|ll|ll|ll|rr|rr|}
			\hline
			\multicolumn{1}{|c|}{}                            & \multicolumn{2}{c|}{\textbf{CYCU}}   & \multicolumn{2}{c|}{\textbf{DeepTrans}} & \multicolumn{2}{c|}{\textbf{FCLS}}  & \multicolumn{2}{c|}{\textbf{MiSiCNet}} & \multicolumn{2}{c|}{\textbf{PGMSU}} & \multicolumn{2}{c|}{\textbf{PPM Net}} & \multicolumn{2}{c|}{\textbf{SSAF-net}} & \multicolumn{2}{c|}{\textbf{SUnSAL}} & \multicolumn{2}{c|}{\textbf{UST Net}} & \multicolumn{2}{c|}{\textbf{GLMM}}   & \multicolumn{2}{c|}{\textbf{IDNet}}                                                                                                                                                                                                                                                                                                                                                                                                                                                         \\ \cline{2-23}
			\multicolumn{1}{|c|}{}                            & \multicolumn{1}{c}{\textbf{Before}}  & \multicolumn{1}{c|}{\textbf{After}}     & \multicolumn{1}{c}{\textbf{Before}} & \multicolumn{1}{c|}{\textbf{After}}    & \multicolumn{1}{c}{\textbf{Before}} & \multicolumn{1}{c|}{\textbf{After}}   & \multicolumn{1}{c}{\textbf{Before}}    & \multicolumn{1}{c|}{\textbf{After}}  & \multicolumn{1}{c}{\textbf{Before}}   & \multicolumn{1}{c|}{\textbf{After}}  & \multicolumn{1}{c}{\textbf{Before}} & \multicolumn{1}{c|}{\textbf{After}}  & \multicolumn{1}{c}{\textbf{Before}} & \multicolumn{1}{c|}{\textbf{After}}  & \multicolumn{1}{c}{\textbf{Before}} & \multicolumn{1}{c|}{\textbf{After}}        & \multicolumn{1}{c}{\textbf{Before}}  & \multicolumn{1}{c|}{\textbf{After}}  & \multicolumn{1}{l}{\textbf{Before}} & \multicolumn{1}{l|}{\textbf{After}}         & \multicolumn{1}{l}{\textbf{Before}} & \multicolumn{1}{l|}{\textbf{After}}    \\ \cline{2-23}
			\multicolumn{1}{|c|}{\multirow{-2}{*}{Endmember}} & 0.2123                               & \textbf{\color[HTML]{3531FF} 0.1116}    & 0.1923                              & \textbf{\color[HTML]{3531FF} 0.0685}   & 0.3654                              & \textbf{\color[HTML]{3531FF} 0.0076}  & 0.2211                                 & \textbf{\color[HTML]{3531FF} 0.0342} & 0.2461                                & \textbf{\color[HTML]{3531FF} 0.0792} & 0.2679                              & \textbf{\color[HTML]{3531FF} 0.1174} & 0.2751                              & \textbf{\color[HTML]{3531FF} 0.1045} & 6.6446                              & \textbf{\color[HTML]{3531FF} 0.0058}       & 0.1906                               & \textbf{\color[HTML]{3531FF} 0.0795} & 0.2025                              & {\color[HTML]{3531FF} \textbf{0.0059}}      & 0.2890                              & {\color[HTML]{3531FF} \textbf{0.1167}} \\ \hline
			Endmember 2                                       & 0.1530                               & \textbf{\color[HTML]{3531FF} 0.143}     & 0.0995                              & \textbf{\color[HTML]{3531FF} 0.0814}   & 0.1247                              & \textbf{\color[HTML]{3531FF} 0.0028}  & 0.1001                                 & \textbf{\color[HTML]{3531FF} 0.0184} & 0.0934                                & \textbf{\color[HTML]{3531FF} 0.0375} & 0.1237                              & \textbf{\color[HTML]{3531FF} 0.0488} & 0.1931                              & \textbf{\color[HTML]{3531FF} 0.0887} & 0.1318                              & \textbf{\color[HTML]{3531FF} 0.0020}       & \textbf{\color[HTML]{3531FF} 0.1027} & 0.1098                               & 0.0759                              & {\color[HTML]{3531FF} \textbf{0.0021}}      & 0.1047                              & {\color[HTML]{3531FF} \textbf{0.0999}} \\ \hline
			Endmember 3                                       & 0.2553                               & \textbf{\color[HTML]{3531FF} 0.1961}    & 0.1894                              & \textbf{\color[HTML]{3531FF} 0.0855}   & 0.1133                              & \textbf{\color[HTML]{3531FF} 0.0102}  & 0.1593                                 & \textbf{\color[HTML]{3531FF} 0.0182} & 0.1865                                & \textbf{\color[HTML]{3531FF} 0.0873} & 0.1978                              & \textbf{\color[HTML]{3531FF} 0.1235} & 0.2124                              & \textbf{\color[HTML]{3531FF} 0.0649} & 0.1029                              & \textbf{\color[HTML]{3531FF} 0.0020}       & 0.3148                               & \textbf{\color[HTML]{3531FF} 0.1043} & 0.0721                              & {\color[HTML]{3531FF} \textbf{0.0022}}      & 0.3337                              & {\color[HTML]{3531FF} \textbf{0.1553}} \\ \hline
			Endmember 4                                       & \textbf{\color[HTML]{3531FF} 0.1730} & 0.1824                                  & 0.1653                              & \textbf{\color[HTML]{3531FF} 0.0446}   & 0.1444                              & \textbf{\color[HTML]{3531FF} 0.0062}  & 0.1898                                 & \textbf{\color[HTML]{3531FF} 0.0115} & 0.1356                                & \textbf{\color[HTML]{3531FF} 0.0399} & 0.1852                              & \textbf{\color[HTML]{3531FF} 0.0360} & 0.2096                              & \textbf{\color[HTML]{3531FF} 0.0709} & 0.1504                              & \textbf{\color[HTML]{3531FF} 0.0062}       & 0.0922                               & \textbf{\color[HTML]{3531FF} 0.0852} & 0.0905                              & {\color[HTML]{3531FF} \textbf{0.0062}}      & 0.1484                              & {\color[HTML]{3531FF} \textbf{0.0848}} \\ \hline
			Endmember 5                                       & 0.2178                               & \textbf{\color[HTML]{3531FF} 0.1412}    & 0.2345                              & \textbf{\color[HTML]{3531FF} 0.1017}   & 0.1438                              & \textbf{\color[HTML]{3531FF} 0.0075}  & 0.1994                                 & \textbf{\color[HTML]{3531FF} 0.0065} & 0.2268                                & \textbf{\color[HTML]{3531FF} 0.0611} & 0.2410                              & \textbf{\color[HTML]{3531FF} 0.0599} & 0.2556                              & \textbf{\color[HTML]{3531FF} 0.0639} & 0.1399                              & \textbf{\color[HTML]{3531FF} 0.0033}       & 0.3024                               & \textbf{\color[HTML]{3531FF} 0.1685} & 0.1248                              & {\color[HTML]{3531FF} \textbf{0.0033}}      & 0.1370                              & {\color[HTML]{3531FF} \textbf{0.0865}} \\ \hline
			\textbf{Mean RMSE}                                & 0.2055                               & \textbf{\color[HTML]{3531FF} 0.1579}    & 0.1817                              & \textbf{\color[HTML]{3531FF} 0.0787}   & 0.2017                              & \textbf{\color[HTML]{3531FF} 0.0073}  & {\ul 0.0197}                           & \textbf{\color[HTML]{3531FF} 0.0201} & 0.1865                                & \textbf{\color[HTML]{3531FF} 0.0642} & 0.2091                              & \textbf{\color[HTML]{3531FF} 0.0852} & 0.2313                              & \textbf{\color[HTML]{3531FF} 0.0801} & 2.9739                              & {\ul \textbf{\color[HTML]{3531FF} 0.0043}} & 0.2218                               & \textbf{\color[HTML]{3531FF} 0.1140} & 0.1231                              & {\ul{\color[HTML]{3531FF} \textbf{0.0043}}} & 0.2221                              & {\color[HTML]{3531FF} \textbf{0.1117}} \\ \hline
		\end{tabular}
	\end{adjustbox}
\end{table*}

%% file: Tables_Latex/RMSE_Tables/samson.tex
\begin{table*}[tbp]
	\caption{Abundance RMSE for the Samson Dataset without Preprocessing (Before) and with Preprocessing (After)}
	\label{tab:samson-rmse}
	\begin{adjustbox}{width=\textwidth}
		\begin{tabular}{|l|ll|ll|ll|ll|ll|ll|ll|ll|ll|rr|rr|}
			\hline
			\multicolumn{1}{|c|}{}                                      & \multicolumn{2}{c|}{\textbf{CYCU}} & \multicolumn{2}{c|}{\textbf{DeepTrans}} & \multicolumn{2}{c|}{\textbf{FCLS}} & \multicolumn{2}{c|}{\textbf{MiSiCNet}} & \multicolumn{2}{c|}{\textbf{PGMSU}} & \multicolumn{2}{c|}{\textbf{PPM Net}} & \multicolumn{2}{c|}{\textbf{SSAF-net}} & \multicolumn{2}{c|}{\textbf{SunSal}} & \multicolumn{2}{c|}{\textbf{UST Net}} & \multicolumn{2}{c|}{\textbf{GLMM}}   & \multicolumn{2}{c|}{\textbf{IDNet}}                                                                                                                                                                                                                                                                                                                                                                                                            \\ \cline{2-23}
			\multicolumn{1}{|c|}{}                                      & \textbf{Before}                    & \textbf{After}                          & \textbf{Before}                    & \textbf{After}                         & \textbf{Before}                     & \textbf{After}                        & \textbf{Before}                        & \textbf{After}                       & \textbf{Before}                       & \textbf{After}                       & \textbf{Before}                     & \textbf{After}                       & \textbf{Before} & \textbf{After}                       & \textbf{Before}                      & \textbf{After}                       & \textbf{Before} & \textbf{After}                             & \multicolumn{1}{l}{\textbf{Before}} & \multicolumn{1}{l|}{\textbf{After}}    & \multicolumn{1}{l}{\textbf{Before}} & \multicolumn{1}{l|}{\textbf{After}}    \\ \cline{2-23}
			\multicolumn{1}{|c|}{\multirow{-2}{*}{\textbf{Endmembers}}} & 0.2392                             & \textbf{\color[HTML]{3531FF} 0.1542}    & 0.168                              & \textbf{\color[HTML]{3531FF} 0.1018}   & 0.2658                              & \textbf{\color[HTML]{3531FF} 0.232}   & 0.1819                                 & \textbf{\color[HTML]{3531FF} 0.1355} & 0.1675                                & \textbf{\color[HTML]{3531FF} 0.0889} & 0.2212                              & \textbf{\color[HTML]{3531FF} 0.1684} & 0.1684          & \textbf{\color[HTML]{3531FF} 0.142}  & \textbf{\color[HTML]{3531FF} 0.2214} & 0.2319                               & 0.0912          & \textbf{\color[HTML]{3531FF} 0.0689}       & 0.1635                              & {\color[HTML]{3531FF} \textbf{0.1063}} & 0.2067                              & {\color[HTML]{3531FF} \textbf{0.1287}} \\ \hline
			Tree                                                        & 0.27                               & \textbf{\color[HTML]{3531FF} 0.1133}    & 0.186                              & \textbf{\color[HTML]{3531FF} 0.0582}   & 0.2519                              & \textbf{\color[HTML]{3531FF} 0.1643}  & 0.179                                  & \textbf{\color[HTML]{3531FF} 0.0913} & 0.0882                                & \textbf{\color[HTML]{3531FF} 0.043}  & 0.1441                              & \textbf{\color[HTML]{3531FF} 0.0485} & 0.1576          & \textbf{\color[HTML]{3531FF} 0.065}  & 0.2768                               & \textbf{\color[HTML]{3531FF} 0.1472} & 0.0932          & \textbf{\color[HTML]{3531FF} 0.0639}       & 0.1853                              & {\color[HTML]{3531FF} \textbf{0.0679}} & 0.2186                              & {\color[HTML]{3531FF} \textbf{0.1238}} \\ \hline
			Water                                                       & 0.2054                             & \textbf{\color[HTML]{3531FF} 0.1559}    & 0.2938                             & \textbf{\color[HTML]{3531FF} 0.1279}   & 0.4237                              & \textbf{\color[HTML]{3531FF} 0.3338}  & 0.3131                                 & \textbf{\color[HTML]{3531FF} 0.1863} & 0.2014                                & \textbf{\color[HTML]{3531FF} 0.091}  & 0.2594                              & \textbf{\color[HTML]{3531FF} 0.169}  & 0.272           & \textbf{\color[HTML]{3531FF} 0.1678} & \textbf{\color[HTML]{3531FF} 0.1262} & 0.3226                               & 0.0614          & \textbf{\color[HTML]{3531FF} 0.0604}       & 0.2831                              & {\color[HTML]{3531FF} \textbf{0.1212}} & 0.1380                              & {\color[HTML]{3531FF} \textbf{0.1249}} \\ \hline
			\textbf{Mean RMSE}                                          & 0.2396                             & \textbf{\color[HTML]{3531FF} 0.1425}    & 0.223                              & \textbf{\color[HTML]{3531FF} 0.1002}   & 0.3233                              & \textbf{\color[HTML]{3531FF} 0.2531}  & 0.2332                                 & \textbf{\color[HTML]{3531FF} 0.1431} & 0.1596                                & \textbf{\color[HTML]{3531FF} 0.0776} & 0.2137                              & \textbf{\color[HTML]{3531FF} 0.1406} & 0.2059          & \textbf{\color[HTML]{3531FF} 0.1324} & \textbf{\color[HTML]{3531FF} 0.2172} & 0.2446                               & {\ul 0.0832}    & {\ul \textbf{\color[HTML]{3531FF} 0.0645}} & 0.2170                              & {\color[HTML]{3531FF} \textbf{0.1010}} & 0.1911                              & {\color[HTML]{3531FF} \textbf{0.1258}} \\ \hline
		\end{tabular}
	\end{adjustbox}
\end{table*}

%% file: Tables_Latex/RMSE_Tables/urban_4.tex
\begin{table*}[tbp]
	\caption{Abundance RMSE for the Urban Dataset without Preprocessing (Before) and with Preprocessing (After)}
	\label{tab:urban-rmse}
	\begin{adjustbox}{width=\textwidth}
		\begin{tabular}{|l|ll|ll|ll|ll|ll|ll|ll|ll|ll|rr|rr|}
			\hline
			\multicolumn{1}{|c|}{}                                     & \multicolumn{2}{c|}{\textbf{CYCU}}   & \multicolumn{2}{c|}{\textbf{DeepTrans}} & \multicolumn{2}{c|}{\textbf{FCLS}}  & \multicolumn{2}{c|}{\textbf{MiSiCNet}} & \multicolumn{2}{c|}{\textbf{PGMSU}} & \multicolumn{2}{c|}{\textbf{PPM Net}} & \multicolumn{2}{c|}{\textbf{SSAF-net}} & \multicolumn{2}{c|}{\textbf{SUnSAL}}       & \multicolumn{2}{c|}{\textbf{UST Net}} & \multicolumn{2}{c|}{\textbf{GLMM}}   & \multicolumn{2}{c|}{\textbf{IDNet}}                                                                                                                                                                                                                                                                                                                                                                                                                                                  \\ \cline{2-23}
			\multicolumn{1}{|c|}{}                                     & \multicolumn{1}{c}{\textbf{Before}}  & \multicolumn{1}{c|}{\textbf{After}}     & \multicolumn{1}{c}{\textbf{Before}} & \multicolumn{1}{c|}{\textbf{After}}    & \multicolumn{1}{c}{\textbf{Before}} & \multicolumn{1}{c|}{\textbf{After}}   & \multicolumn{1}{c}{\textbf{Before}}    & \multicolumn{1}{c|}{\textbf{After}}        & \multicolumn{1}{c}{\textbf{Before}}   & \multicolumn{1}{c|}{\textbf{After}}  & \multicolumn{1}{c}{\textbf{Before}}  & \multicolumn{1}{c|}{\textbf{After}}  & \multicolumn{1}{c}{\textbf{Before}}  & \multicolumn{1}{c|}{\textbf{After}}  & \multicolumn{1}{c}{\textbf{Before}} & \multicolumn{1}{c|}{\textbf{After}}  & \multicolumn{1}{c}{\textbf{Before}} & \multicolumn{1}{c|}{\textbf{After}}  & \multicolumn{1}{c}{\textbf{Before}} & \multicolumn{1}{c|}{\textbf{After}}    & \multicolumn{1}{c}{\textbf{Before}}    & \multicolumn{1}{c|}{\textbf{After}}    \\ \cline{2-23}
			\multicolumn{1}{|c|}{\multirow{-2}{*}{\textbf{Endmember}}} & \textbf{\color[HTML]{3531FF} 0.1596} & 0.1814                                  & 0.2597                              & \textbf{\color[HTML]{3531FF} 0.1107}   & 0.1941                              & \textbf{\color[HTML]{3531FF} 0.1028}  & 0.2102                                 & \textbf{\color[HTML]{3531FF} 0.0934}       & 0.1995                                & \textbf{\color[HTML]{3531FF} 0.1145} & 0.1112                               & \textbf{\color[HTML]{3531FF} 0.1055} & 0.2179                               & \textbf{\color[HTML]{3531FF} 0.1202} & 0.549                               & \textbf{\color[HTML]{3531FF} 0.0919} & 0.2671                              & \textbf{\color[HTML]{3531FF} 0.109}  & 0.1654                              & {\color[HTML]{3531FF} \textbf{0.0996}} & 0.1028                                 & {\color[HTML]{3531FF} \textbf{0.0979}} \\ \hline
			Grass                                                      & 0.2372                               & \textbf{\color[HTML]{3531FF} 0.155}     & 0.1642                              & \textbf{\color[HTML]{3531FF} 0.1177}   & 0.3713                              & \textbf{\color[HTML]{3531FF} 0.0955}  & 0.3665                                 & \textbf{\color[HTML]{3531FF} 0.1021}       & 0.2425                                & \textbf{\color[HTML]{3531FF} 0.1084} & 0.1809                               & \textbf{\color[HTML]{3531FF} 0.1302} & 0.3835                               & \textbf{\color[HTML]{3531FF} 0.108}  & 0.2115                              & \textbf{\color[HTML]{3531FF} 0.0815} & 0.2478                              & \textbf{\color[HTML]{3531FF} 0.111}  & 0.1609                              & {\color[HTML]{3531FF} \textbf{0.0938}} & 0.1225                                 & {\color[HTML]{3531FF} \textbf{0.0942}} \\ \hline
			Tree                                                       & 0.1875                               & \textbf{\color[HTML]{3531FF} 0.0764}    & 0.1052                              & \textbf{\color[HTML]{3531FF} 0.0658}   & 0.2681                              & \textbf{\color[HTML]{3531FF} 0.0777}  & 0.2476                                 & \textbf{\color[HTML]{3531FF} 0.0547}       & 0.233                                 & \textbf{\color[HTML]{3531FF} 0.047}  & 0.143                                & \textbf{\color[HTML]{3531FF} 0.0828} & 0.2853                               & \textbf{\color[HTML]{3531FF} 0.057}  & 0.211                               & \textbf{\color[HTML]{3531FF} 0.1079} & 0.0962                              & \textbf{\color[HTML]{3531FF} 0.0747} & 0.0944                              & {\color[HTML]{3531FF} \textbf{0.0782}} & {\color[HTML]{3531FF} \textbf{0.0943}} & {\color[HTML]{3531FF} \textbf{0.0943}} \\ \hline
			Roof                                                       & 0.164                                & \textbf{\color[HTML]{3531FF} 0.1175}    & 0.3231                              & \textbf{\color[HTML]{3531FF} 0.0597}   & 0.0492                              & \textbf{\color[HTML]{3531FF} 0.0445}  & 0.0546                                 & \textbf{\color[HTML]{3531FF} 0.0464}       & \textbf{\color[HTML]{3531FF} 0.049}   & 0.0638                               & \textbf{\color[HTML]{3531FF} 0.0665} & \textbf{\color[HTML]{3531FF} 0.0665} & \textbf{\color[HTML]{3531FF} 0.0545} & 0.0639                               & 0.0528                              & \textbf{\color[HTML]{3531FF} 0.0423} & 0.2915                              & \textbf{\color[HTML]{3531FF} 0.071}  & 0.0555                              & {\color[HTML]{3531FF} \textbf{0.0435}} & {\color[HTML]{3531FF} \textbf{0.0428}} & 0.0692                                 \\ \hline
			\textbf{Mean RMSE}                                         & 0.1896                               & \textbf{\color[HTML]{3531FF} 0.1384}    & 0.2291                              & \textbf{\color[HTML]{3531FF} 0.0922}   & 0.2499                              & \textbf{\color[HTML]{3531FF} 0.0832}  & 0.2464                                 & {\ul \textbf{\color[HTML]{3531FF} 0.0779}} & 0.1971                                & \textbf{\color[HTML]{3531FF} 0.0882} & 0.1323                               & \textbf{\color[HTML]{3531FF} 0.0992} & 0.2641                               & \textbf{\color[HTML]{3531FF} 0.0914} & 0.3136                              & \textbf{\color[HTML]{3531FF} 0.0844} & 0.2382                              & \textbf{\color[HTML]{3531FF} 0.0933} & 0.1277                              & {\color[HTML]{3531FF} \textbf{0.0817}} & {\ul 0.0953}                           & {\color[HTML]{3531FF} \textbf{0.0896}} \\ \hline
		\end{tabular}
	\end{adjustbox}
\end{table*}

%% file: Figures_Latex/matern_sota_compare_abd.tex
\begin{figure*}[!t]
	\centering
	\includegraphics[width=\linewidth]{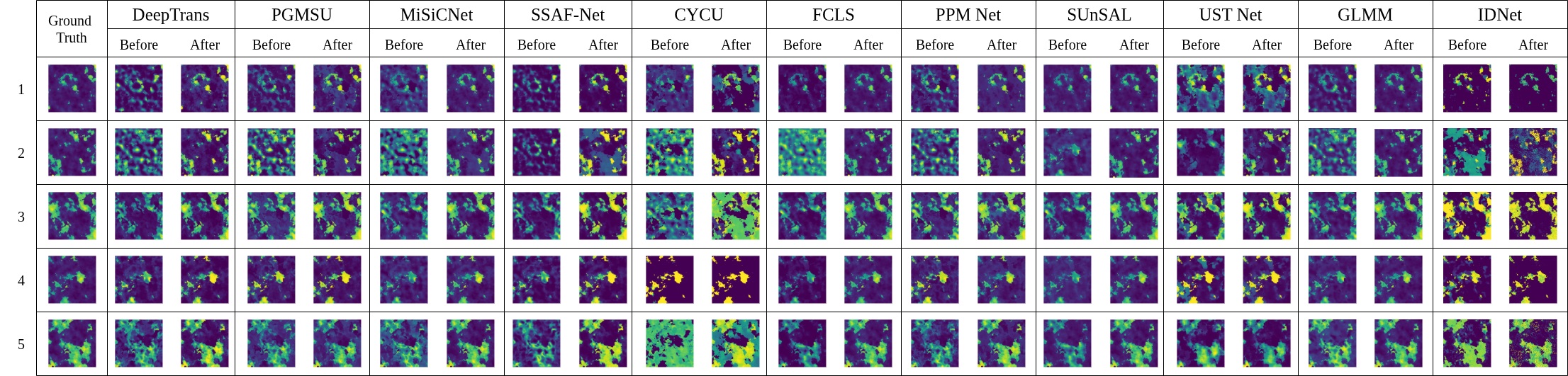}
	\caption{Comparison of Estimated Abundance Maps by SOTA algorithms for the datasets with the synthetic scaling factors (Before), and with the scale corrected HSI data using the preprocessing algorithm (After) for the Matern dataset }
	\label{fig:matern-sota-abd}
\end{figure*}

%% file: Figures_Latex/spheric_sota_compare_abd.tex
\begin{figure*}[!t]
	\centering
	\includegraphics[width=\linewidth]{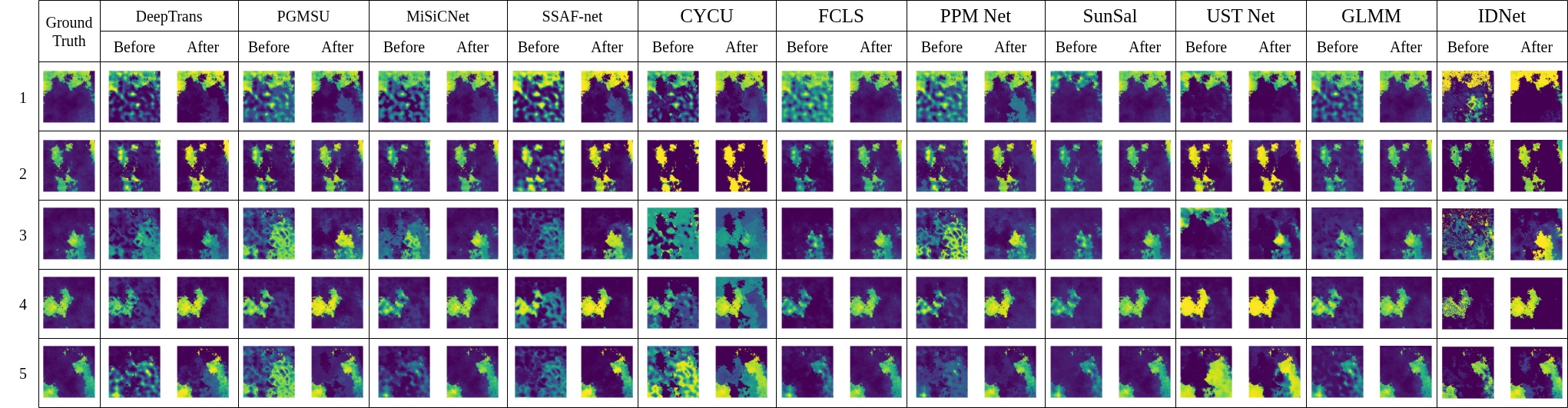}
	\caption{Comparison of Estimated Abundance Maps by SOTA algorithms for the datasets with the synthetic scaling factors (Before), and with the scale corrected HSI data using the preprocessing algorithm (After) for the Spheric dataset }
	\label{fig:spheric-sota-abd}
\end{figure*}

%% file: Figures_Latex/samson_sota_compare_abd.tex
\begin{figure*}[!t]
	\centering
	\includegraphics[width=\linewidth]{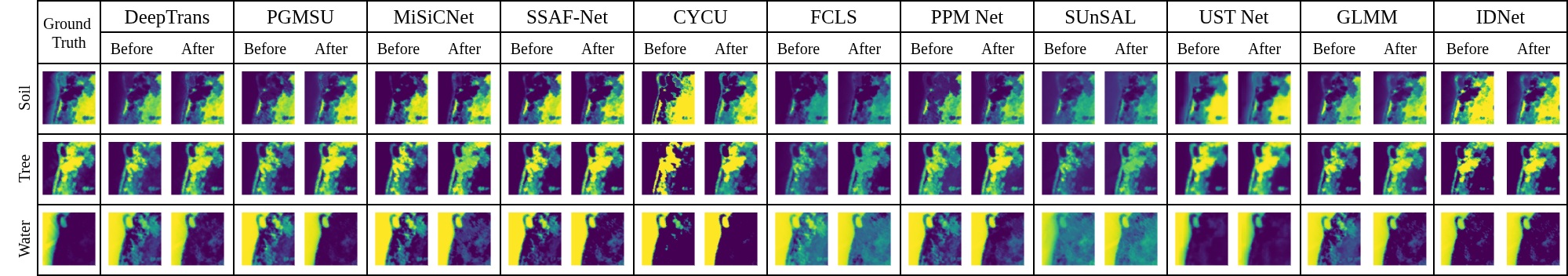}
	\caption{Comparison of Estimated Abundance Maps by SOTA algorithms for raw HSI data (Before) and with the preprocessed HSI data (After) for the Samson dataset }
	\label{fig:samson-sota-abd}
\end{figure*}

%% file: Figures_Latex/urban_sota_compare_abd.tex
\begin{figure*}[!t]
	\centering
	\includegraphics[width=\linewidth]{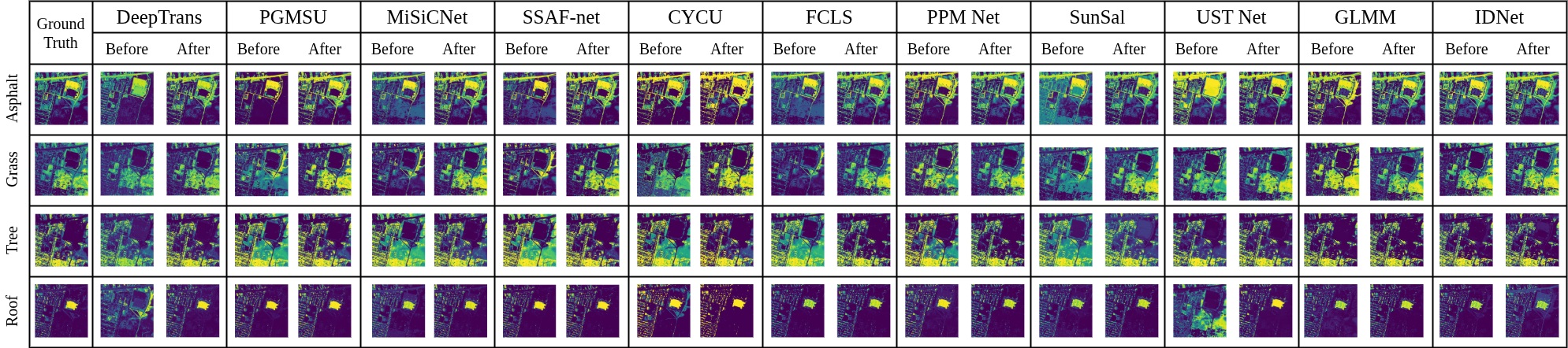}
	\caption{Comparison of Estimated Abundance Maps by SOTA algorithms for raw HSI data (Before) and with the preprocessed HSI data (After) for the Urban dataset }
	\label{fig:urban-sota-abd}
\end{figure*}

%% file: Tables_Latex/SAD_Tables/sad_summary.tex
\begin{table*}[tbp]
	\caption{Mean SAD for Endmember Estimation for Each Dataset without Preprocessing  (Before) and with Preprocessing (After)}
	\label{tab:sad-summary}
	\begin{adjustbox}{width=\textwidth}
		\begin{tabular}{|l|ll|ll|ll|ll|ll|ll|ll|ll|ll|ll|}
			\hline
			                           & \multicolumn{2}{c|}{\textbf{CYCU}}   & \multicolumn{2}{c|}{\textbf{DeepTrans}} & \multicolumn{2}{c|}{\textbf{NFINDR}} & \multicolumn{2}{c|}{\textbf{MiSiCNet}} & \multicolumn{2}{c|}{\textbf{PGMSU}}  & \multicolumn{2}{c|}{\textbf{PPM Net}} & \multicolumn{2}{c|}{\textbf{SSAF-Net}} & \multicolumn{2}{c|}{\textbf{UST Net}} & \multicolumn{2}{c|}{\textbf{GLMM}} & \multicolumn{2}{c|}{\textbf{IDNet}}                                                                                                                                                                                                                                                                                                                                                                      \\ \cline{2-21}
			\multirow{-2}{*}{Mean SAD} & \textbf{Before}                      & \textbf{After}                          & \textbf{Before}                      & \textbf{After}                         & \textbf{Before}                      & \textbf{After}                        & \textbf{Before}                        & \textbf{After}                        & \textbf{Before}                    & \textbf{After}                       & \textbf{Before} & \textbf{After}                       & \textbf{Before} & \textbf{After}                       & \textbf{Before}                      & \textbf{After}                       & \textbf{Before}                        & \textbf{After}                         & \textbf{Before}                        & \textbf{After}                         \\ \hline
			Synthetic Matern           & 0.4088                               & \textbf{\color[HTML]{3531FF} 0.1524}    & 0.0964                               & \textbf{\color[HTML]{3531FF} 0.0462}   & 0.0553                               & \textbf{\color[HTML]{3531FF} 0.0000}  & 0.1774                                 & \textbf{\color[HTML]{3531FF} 0.0209}  & 0.0517                             & \textbf{\color[HTML]{3531FF} 0.0481} & 0.0487          & \textbf{\color[HTML]{3531FF} 0.0343} & 0.0991          & \textbf{\color[HTML]{3531FF} 0.0566} & 0.1155                               & \textbf{\color[HTML]{3531FF} 0.0572} & 0.0033                                 & {\color[HTML]{3531FF} \textbf{0.0000}} & 0.0978                                 & {\color[HTML]{3531FF} \textbf{0.0964}} \\ \hline
			\textbf{Synthetic Spheric} & \textbf{\color[HTML]{3531FF} 0.1188} & 0.1684                                  & 0.1127                               & \textbf{\color[HTML]{3531FF} 0.0351}   & 0.0513                               & \textbf{\color[HTML]{3531FF} 0.0000}  & 0.2133                                 & \textbf{\color[HTML]{3531FF} 0.0149}  & 0.0607                             & \textbf{\color[HTML]{3531FF} 0.0325} & 0.3290          & \textbf{\color[HTML]{3531FF} 0.0260} & 0.1133          & \textbf{\color[HTML]{3531FF} 0.0584} & 0.0997                               & \textbf{\color[HTML]{3531FF} 0.0558} & 0.0037                                 & {\color[HTML]{3531FF} \textbf{0.0000}} & 0.1050                                 & {\color[HTML]{3531FF} \textbf{0.0284}} \\ \hline
			\textbf{Samson}            & 0.1022                               & \textbf{\color[HTML]{3531FF} 0.0898}    & 0.0614                               & \textbf{\color[HTML]{3531FF} 0.0300}   & \textbf{\color[HTML]{3531FF} 0.0702} & 0.0828                                & 0.1488                                 & \textbf{\color[HTML]{3531FF} 0.1297}  & 0.1568                             & \textbf{\color[HTML]{3531FF} 0.0949} & 0.1924          & \textbf{\color[HTML]{3531FF} 0.1752} & 0.1522          & \textbf{\color[HTML]{3531FF} 0.1343} & \textbf{\color[HTML]{3531FF} 0.0232} & 0.0250                               & {\color[HTML]{3531FF} \textbf{0.0537}} & 0.0647                                 & 0.0536                                 & {\color[HTML]{3531FF} \textbf{0.0492}} \\ \hline
			\textbf{Urban}             & 0.1865                               & \textbf{\color[HTML]{3531FF} 0.1333}    & 0.1651                               & \textbf{\color[HTML]{3531FF} 0.0689}   & 0.1414                               & \textbf{\color[HTML]{3531FF} 0.0774}  & 0.1562                                 & \textbf{\color[HTML]{3531FF} 0.0637}  & 0.1493                             & \textbf{\color[HTML]{3531FF} 0.0774} & 0.0894          & \textbf{\color[HTML]{3531FF} 0.0772} & 0.1607          & \textbf{\color[HTML]{3531FF} 0.0941} & 0.2106                               & \textbf{\color[HTML]{3531FF} 0.0738} & {\color[HTML]{3531FF} \textbf{0.0797}} & 0.0804                                 & {\color[HTML]{3531FF} \textbf{0.0839}} & 0.0895                                 \\ \hline
		\end{tabular}
	\end{adjustbox}
\end{table*}

%% file: Chapters/Experiments/signature_visualization.tex
\input{Figures_Latex/Signature_Visualization.tex}
\subsection{Qualitative Visualization of Signature Correction}

To provide a direct qualitative illustration of the effect of the proposed
preprocessing algorithm, Fig.~\ref{fig:signature-correction} shows
representative nearly pure pixel signatures before and after correction for
the considered datasets. For each endmember, a set of nearly pure pixels was
selected and their spectral signatures were plotted for both the raw data and
the scale-corrected data.

As can be observed, in the raw data, signatures corresponding to the same
material often exhibit a noticeable spread in magnitude while preserving a
similar overall shape. This is consistent with dominant wavelength-independent
scale variability caused by illumination, topography, and shadowing effects.
After applying the proposed preprocessing algorithm, the signatures become
substantially more compact within each material class, indicating that the
dominant scale variability has been significantly reduced.

Importantly, the characteristic spectral shape of each material, including
major peaks, valleys, and absorption features, is preserved after correction.
Moreover, the corrected signatures are not collapsed to a single identical
curve; small residual differences across wavelength can still be observed.
This indicates that the proposed method does not remove wavelength-dependent
variations, but instead primarily compensates for the dominant multiplicative
scale distortion. It can also be observed that the amount of residual spread
after correction is not the same for all endmembers, suggesting that the
strength of non-scaling variability differs across materials.

It should also be noted that this visualization is constructed using nearly
pure pixels, so it is intended to illustrate the effect of the proposed
correction on representative material signatures rather than to directly
characterize behaviour on mixed pixels. Nevertheless, these qualitative
observations are consistent with the quantitative results in Section IV-B and
help explain the downstream improvements in abundance estimation reported in
the following subsections.

For brevity, only one representative spectral visualization is shown for the
synthetic datasets. The Matern and Spheric datasets are generated from the
same set of endmembers, and the qualitative effect of the proposed correction
on nearly pure signatures is visually similar in both cases.

%% file: Figures_Latex/Signature_Visualization.tex
\begin{figure*}[t]
    \centering
    \subfloat[]{
        \includegraphics[width=0.95\textwidth]{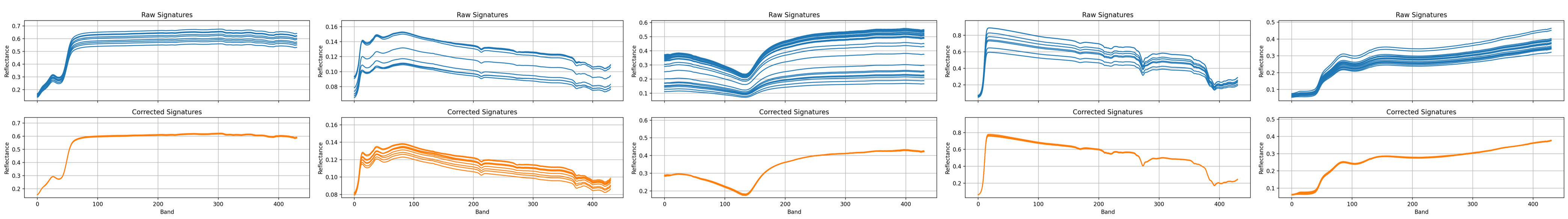}
        \label{fig:signature-correction-synth}
    }

    \vspace{0.5em}

    \subfloat[]{
        \includegraphics[width=0.6\textwidth]{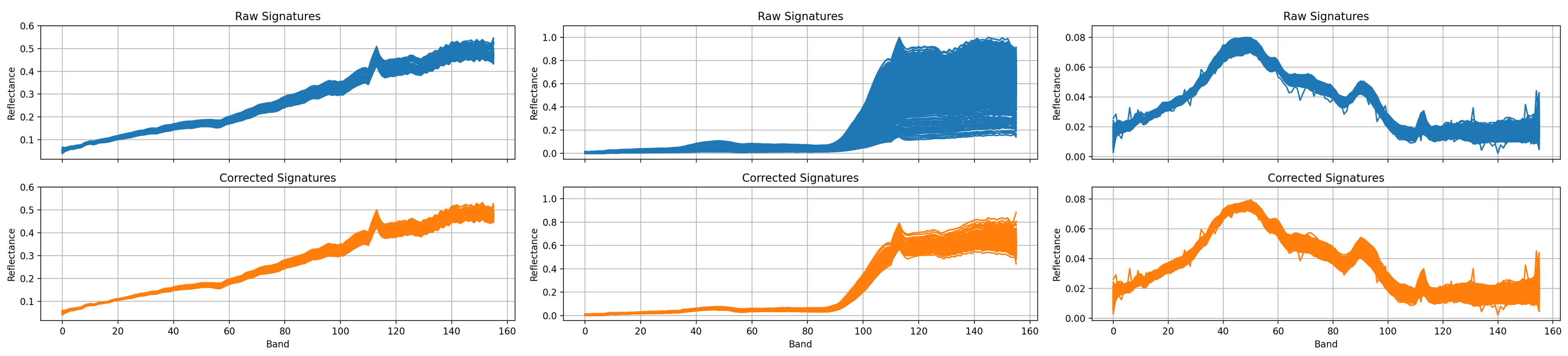}
        \label{fig:signature-correction-samson}
    }

    \vspace{0.5em}

    \subfloat[]{
        \includegraphics[width=0.85\textwidth]{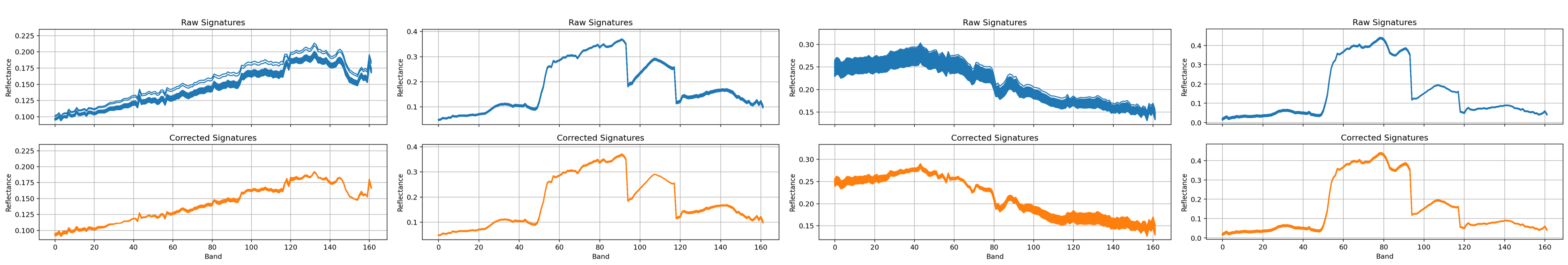}
        \label{fig:signature-correction-urban}
    }

    \caption{Representative nearly pure pixel signatures before and after the proposed scale correction. For each dataset, the top row shows the raw signatures and the bottom row shows the corrected signatures. For brevity, only one representative synthetic example is shown, since the Matern and Spheric datasets share the same endmember set and exhibit similar qualitative behavior after correction. (a) Matern and Spheric datasets. (b) Samson dataset. (c) Urban dataset.}
    \label{fig:signature-correction}
\end{figure*}

%% file: Chapters/Experiments/preprocessing_comparison.tex
\subsection{Comparison with Other Preprocessing Methods}
\input{Tables_Latex/preprocessing_comparison_tab.tex}
In addition to evaluating the effect of the proposed scale-correction
algorithm against raw data, we further compare it with commonly used
spectral normalization and perspective-projection-based preprocessing
strategies. Specifically, the following baseline preprocessing methods
are considered:

\begin{itemize}
    \item {Standard Normal Variate (SNV):} Per-pixel mean removal and variance normalization across spectral bands.
    \item {\(\ell_2\) Normalization:} Per-pixel normalization to unit Euclidean norm.
    \item {VCA-Style Perspective Transform:} A perspective projection using a fixed normal derived from the mean vector, similar in spirit to approaches used in endmember extraction methods such as VCA.
\end{itemize}

For the sake of brevity, this is compared against three of the hyperspectral unmixing algorithms
considered in the previous sections namely, PGMSU, PPM-Net and UST-Net as representatives of the
SOTA unmixing algorithms. For this comparison, abundance RMSE is used as the primary evaluation metric,
since the objective of the proposed preprocessing step is to improve abundance
estimation accuracy.

Table~\ref{tab:preprocessing_comparison} reports the mean abundance RMSE for the Samson, Urban, Matern, and Spheric datasets.

Across all datasets and all three unmixing algorithms,
the proposed method consistently achieves the lowest RMSE.
Notably, conventional normalization techniques such as SNV
and \(\ell_2\) normalization do not consistently improve performance
over the raw data. In fact, in most cases SNV significantly degrades
accuracy, indicating that removing per-pixel mean and variance
does not properly address illumination-induced scaling effects
and may distort the simplex geometry assumed by linear
mixture models.

Similarly, the VCA-style perspective transform shows unstable
behavior across datasets. While it provides moderate improvement
in a few settings, it often performs worse than
\(\ell_2\) normalization and, in some cases,
even worse than the raw input.
This suggests that selecting a fixed projection normal without
optimization is insufficient for robust scale correction.

These results demonstrate that the improvements observed in
Section IV-C are not merely due to generic normalization effects,
but arise specifically from the geometrically consistent
scale correction provided by the proposed hyperplane-based
formulation.


%% file: Tables_Latex/preprocessing_comparison_tab.tex
\begin{table*}[tbp]
    \caption{Comparison of Abundance RMSE for Different Preprocessing Methods}
    \label{tab:preprocessing_comparison}
    \begin{adjustbox}{width=\textwidth}
        \begin{tabular}{|l|l|l|l|l|l|}
            \hline
            \textbf{Hyperspectral Unmixing Algorithms} & \textbf{Preprocessing Algorithms} & \textbf{Samson Dataset}                & \textbf{Urban Dataset}                 & \textbf{Matern}                        & \textbf{Spheric}                       \\ \hline
                                                       & SNV                               & 0.3731                                 & 0.3204                                 & 0.2467                                 & 0.3087                                 \\ \cline{2-6}
                                                       & L2 Norm                           & 0.1876                                 & 0.165                                  & 0.1609                                 & 0.16                                   \\ \cline{2-6}
                                                       & VCA style Perspective Transform   & 0.2855                                 & 0.2363                                 & 0.266                                  & 0.2834                                 \\ \cline{2-6}
                                                       & Raw                               & 0.1596                                 & 0.1971                                 & 0.141                                  & 0.1865                                 \\ \cline{2-6}
            \multirow{-5}{*}{PGMSU}                    & Proposed Method                   & {\color[HTML]{3531FF} \textbf{0.0776}} & {\color[HTML]{3531FF} \textbf{0.0822}} & {\color[HTML]{3531FF} \textbf{0.0795}} & {\color[HTML]{3531FF} \textbf{0.0642}} \\ \hline
                                                       & SNV                               & 0.305                                  & 0.3449                                 & 0.234                                  & 0.247                                  \\ \cline{2-6}
                                                       & L2 Norm                           & 0.1631                                 & 0.2168                                 & 0.13                                   & 0.1166                                 \\ \cline{2-6}
                                                       & VCA style Perspective Transform   & 0.3546                                 & 0.2506                                 & 0.1237                                 & 0.1053                                 \\ \cline{2-6}
                                                       & Raw                               & 0.2137                                 & 0.1323                                 & 0.1578                                 & 0.2091                                 \\ \cline{2-6}
            \multirow{-5}{*}{PPM-Net}                  & Proposed Method                   & {\color[HTML]{3531FF} \textbf{0.1406}} & {\color[HTML]{3531FF} \textbf{0.0992}} & {\color[HTML]{3531FF} \textbf{0.0622}} & {\color[HTML]{3531FF} \textbf{0.0852}} \\ \hline
                                                       & SNV                               & 0.305                                  & 0.1728                                 & 0.313                                  & 0.3099                                 \\ \cline{2-6}
                                                       & L2 Norm                           & 0.1631                                 & 0.159                                  & 0.2157                                 & 0.1806                                 \\ \cline{2-6}
                                                       & VCA style Perspective Transform   & 0.3546                                 & 0.2081                                 & 0.1946                                 & 0.1794                                 \\ \cline{2-6}
                                                       & Raw                               & 0.0832                                 & 0.2382                                 & 0.1975                                 & 0.2218                                 \\ \cline{2-6}
            \multirow{-5}{*}{UST-Net}                  & Proposed Method                   & {\color[HTML]{3531FF} \textbf{0.0645}} & {\color[HTML]{3531FF} \textbf{0.0933}} & {\color[HTML]{3531FF} \textbf{0.122}}  & {\color[HTML]{3531FF} \textbf{0.114}}  \\ \hline
        \end{tabular}
    \end{adjustbox}
\end{table*}

%% file: Chapters/Experiments/noise_robustness.tex
\subsection{Noise Robustness Analysis}
In this section, the robustness of the proposed method to noise in the input data is evaluated. Specifically, the performance of the method is assessed under varying levels of additive noise applied to the Synthetic Matern Dataset by examining the accuracy of the predicted scaling factors used for correction.

As illustrated in \ref{fig:noise-robust}, the RMSE decreases monotonically from approximately 0.14 at 20 dB to 0.026 at 40 dB, corresponding to a reduction in estimation error exceeding 80\%. Notably, even at 20 dB, which represents a comparatively high noise regime, the estimation error remains bounded and moderate. This indicates that the proposed method does not exhibit instability or severe performance degradation under significant noise levels.

Furthermore, as the SNR increases beyond 30 dB, the rate of improvement diminishes progressively. This behaviour suggests that once the influence of additive noise is sufficiently reduced, the residual estimation error is primarily governed by modelling approximations and optimisation effects rather than by noise perturbations.
\input{Figures_Latex/noise_robustness.tex}

%% file: Figures_Latex/noise_robustness.tex
\begin{figure}[!t]
    \centering
    \includegraphics[width=\linewidth]{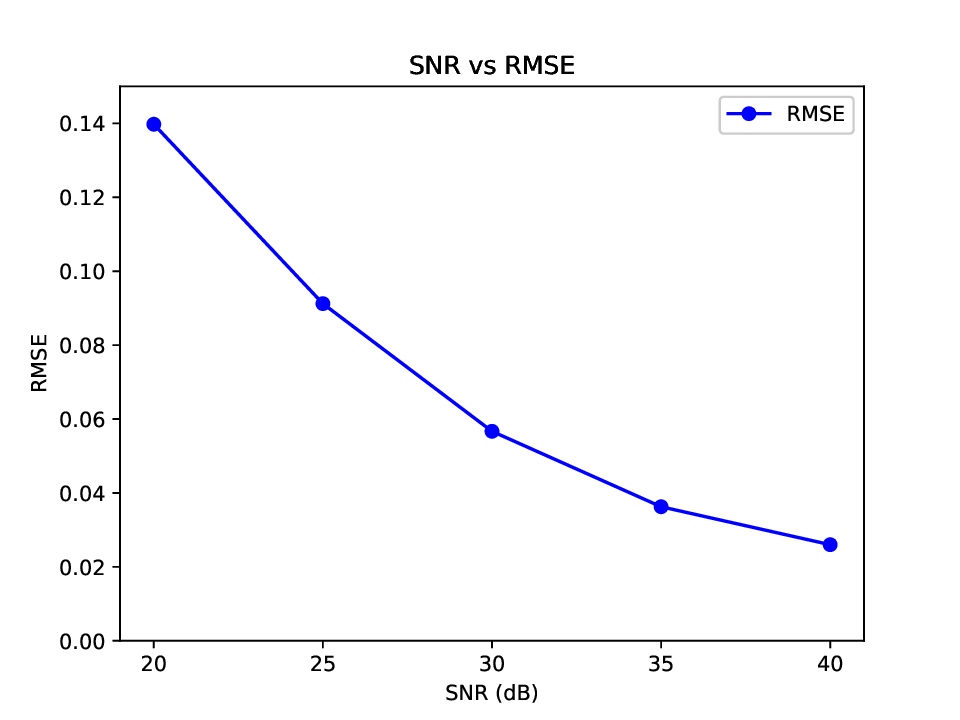}
    \caption{Noise robustness analysis of the proposed method evaluted for the matern dataset}
    \label{fig:noise-robust}
\end{figure}

%% file: Chapters/Experiments/non_linear_evaluation.tex
\input{Figures_Latex/non_linear_compare.tex}

\input{Figures_Latex/non_linear_rmse_variation.tex}
\subsection{Robustness to Nonlinear Mixing}
This section evaluates the robustness of the proposed method to nonlinear mixing effects. Motivated by bilinear mixing models (e.g., GBM/Fan-type formulations) \cite{NonLinearMM}, we generate synthetic observations with controlled nonlinearity using the abundance maps, endmember signatures, and per-pixel scaling factors of the Synthetic Matern dataset. Specifically, the data are synthesized according to
\begin{multline*}
    \pixlo{i} = \scalo_i \Bigg(
    \sum_{j=1}^\nend (\abdo_i)_j \sigo{j}
    +\\
    \gamma \sum_{j=1}^\nend \sum_{k=j+1}^\nend (\abdo_i)_j (\abdo_i)_k
    \big(\sigo{j}\odot \sigo{k}\big)
    \Bigg)
\end{multline*}
where \(\gamma\) controls the degree of nonlinearity: \(\gamma=0\) reduces to the LMM, while \(\gamma=1\) corresponds to the Fan-style bilinear model~\cite{FanModel}. The proposed method is applied to these data, and the estimated scaling factors are compared with the ground-truth scaling factors for different values of \(\gamma\) in Fig.~\ref{fig:Non-linear-mu-compare}. The corresponding RMSE values are summarized in Fig.~\ref{fig:non-linear-rmse-variation}.

From Fig.~\ref{fig:non-linear-rmse-variation}, the RMSE remains essentially unchanged for \(\gamma \leq 0.4\), indicating that mild nonlinearities have negligible impact on scale estimation accuracy. For \(\gamma > 0.4\), the RMSE increases gradually, yet remains bounded even at \(\gamma=1.0\). This suggests that the proposed method is robust to moderate levels of nonlinearity and continues to provide reliable scaling factor estimates under increasingly nonlinear mixing. The qualitative comparison in Fig.~\ref{fig:Non-linear-mu-compare} is consistent with this trend: the predicted scaling maps preserve the main spatial structure of the true scaling factors even as \(\gamma\) increases.

To complement the synthetic bilinear experiment, the proposed method was evaluated on two physically simulated orchard scenes \cite{orchard1,orchard2,orchard3} from the KU Leuven nonlinear-mixing benchmark, which exhibit realistic multiple-scattering effects. Since these scenes do not contain pronounced pixel-wise scale variability, multiplicative scaling factors with standard deviation \(0.3\) were additionally imposed.

The proposed algorithm was then used to estimate these scaling factors in the presence of the underlying nonlinear mixing effects, yielding RMSE values of \(0.0427\) and \(0.0488\) for the two and three endmember cases respectively. These results indicate that the method remains effective at estimating dominant scaling effects even under physically simulated nonlinear mixing. This experiment should be interpreted as a robustness study, since the scale variations were introduced rather than native to the dataset.

%% file: Figures_Latex/non_linear_compare.tex
\begin{figure}[!t]
    \centering
    \subfloat[]{\includegraphics[width=0.8in]{Figures/shadow_map.png}%
        \label{fig:Non-linear-True-mu}}
    \hfil
    \subfloat[]{\includegraphics[width=0.8in]{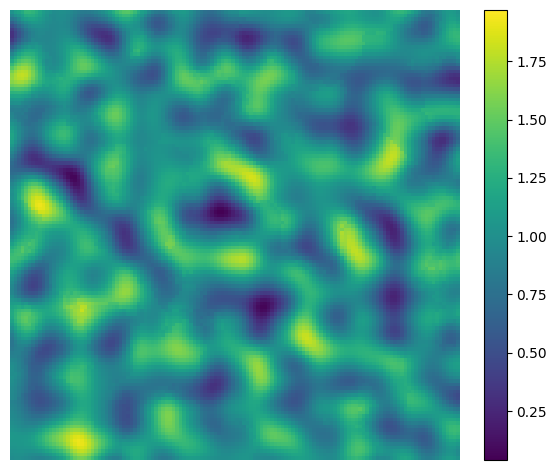}%
        \label{fig:Non-linear-gamma-0-0-mu}}
    \hfil
    \subfloat[]{\includegraphics[width=0.8in]{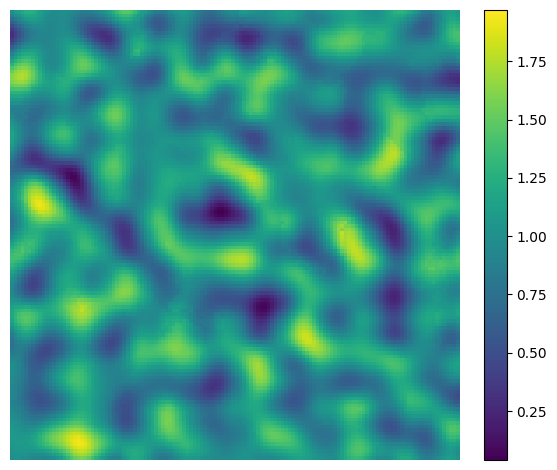}%
        \label{fig:Non-linear-gamma-0-5-mu}}
    \hfil
    \subfloat[]{\includegraphics[width=0.8in]{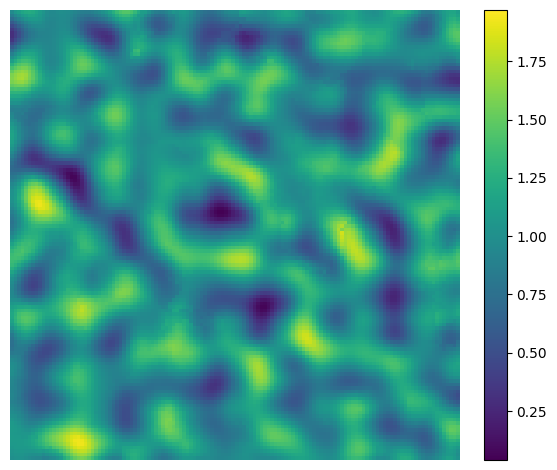}%
        \label{fig:Non-linear-gamma-1-0-mu}}
    \caption{Comparison of predicted and true scaling factors under different degrees of nonlinearity. (a) True Scaling Factors. (b) \(\gamma=0.0\) (c) \(\gamma=0.5\) (d) \(\gamma=1.0\).}
    \label{fig:Non-linear-mu-compare}
\end{figure}

%% file: Figures_Latex/non_linear_rmse_variation.tex
\begin{figure}[!t]
    \centering
    \includegraphics[width=\linewidth]{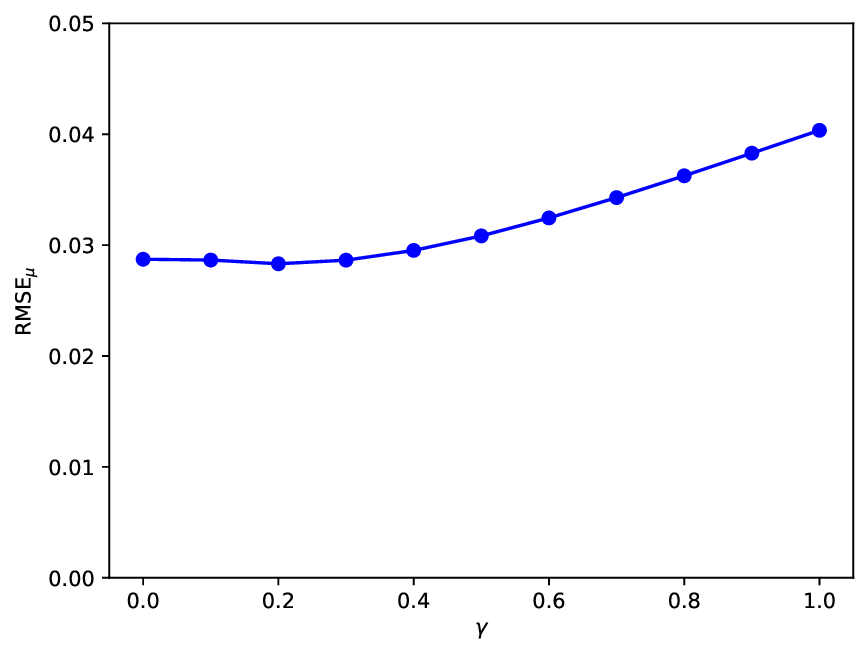}
    \caption{Variations in \(\text{RMSE}_\mu\) across different degrees of nonlinearity (\(\gamma\)).}
    \label{fig:non-linear-rmse-variation}
\end{figure}

%% file: Chapters/Experiments/low_purity.tex
\subsection{Robustness to Low-Purity (Highly Mixed) Pixels}
\input{Tables_Latex/low_purity_table.tex}
To assess whether the proposed initialization using candidate normals implicitly relies on the presence of pure pixels, we performed an additional experiment on the Synthetic Matern dataset where pixel purity was systematically reduced. Specifically, we enforced increasing mixture by capping the maximum abundance per pixel. Table~\ref{tab:low_purity_matern} shows that the RMSE of the estimated scaling factors remains essentially unchanged (mean \(\approx 0.0296\) with worst-case variation \(<1\%\)), indicating that the method does not depend on pure pixels and remains stable under low-purity conditions.

%% file: Tables_Latex/low_purity_table.tex
\begin{table}[tbp]
    \caption{Comparison of RMSE of Predicted Scaling Factors under Varying Pixel Purity}
    \label{tab:low_purity_matern}
    \centering
    \begin{tabular}{|l|l|}
        \hline
        Maximum Abundance & \(\text{RMSE}_\scalo\) \\ \hline
        1.0               & 0.0297                 \\ \hline
        0.9               & 0.0295                 \\ \hline
        0.8               & 0.0298                 \\ \hline
        0.7               & 0.0294                 \\ \hline
        0.6               & 0.0298                 \\ \hline
    \end{tabular}
\end{table}

%% file: Chapters/Experiments/ablation_study.tex
\input{Tables_Latex/Ablation_Study}
\subsection{Ablation Study}

In this section, the importance of each step in the optimization algorithm used for estimating $\nhat$ is investigated by exploring the effect on the results in the absence of certain steps in the optimization step. 

Specifically, the importance of the following steps in the algorithm will be evaluated.

\begin{enumerate}
    \item Importance of initial candidate normal generation using Equation \eqref{eq:nn-candidate}
    \item Importance of the Particle Swarm Optimization (PSO) algorithm for minimizing $\Psi(\nn)$
    \item Importance of the Gradient Descent (GD) fine-tuning step.
\end{enumerate}

This study was conducted using the Matern Synthetic Dataset with scaling factors having a standard deviation of 0.3. The results are summarized in Table \ref{tab:ablation_study}.

As can be observed from the table, the worst results are obtained when Gradient Descent is used in isolation, which supports the claim that Gradient Descent alone is insufficient.

Furthermore, it can be observed that without proper initialization, PSO also performs poorly. It can be observed that the initialization scheme proposed in Section \ref{sec:Algorithm} gives significant performance improvements of nearly tenfold, compared to randomly selected initial points.

Finally, fine-tuning the normal using gradient descent offered moderate improvement in performance compared to PSO in isolation, and also increased the consistency of the algorithm.

%% file: Tables_Latex/Ablation_Study.tex
\begin{table*}[!t]
    \caption{Ablation Study}
    \label{tab:ablation_study}
    \centering
    \begin{tabular}{|c|c|c|c|c|}
    \hline
        Experimental Setup &Candidate Generation & PSO  & GD Fine Tuning  & RMSE \\
        \hline
         Optimization with only Gradient Descent& $\times$ &$\times$  & \checkmark & 0.2607 \\
         \hline
        PSO with Random Initial Points and GD-Fine Tuning& $\times$ &\checkmark  &\checkmark  &0.1348 \\
         \hline
         PSO with Candidate Normals as Initial Points & \checkmark & \checkmark & $\times$  & 0.0214\\
         \hline
         Proposed Optimization Setup& \checkmark &\checkmark  &\checkmark  &0.0191 \\
         \hline
    \end{tabular}
\end{table*}

%% file: Chapters/Experiments/pso_robustness.tex
\input{Figures_Latex/pso_robustness_figure.tex}
\subsection{Parameter Robustness Analysis of Hyperplane Estimation}

Given the results of the previous section, it is evident that PSO plays a crucial role in the accurate estimation of the hyperplane parameters. Therefore, in this section, the sensitivity to initial conditions is analyzed. The estimation is performed over ten independent runs with different random seeds, and the RMSE of the estimated scaling factors is computed for each run using the Synthetic Matern dataset.

This analysis is conducted for different numbers of particles (5, 10, 100, 1000) while keeping the number of iterations fixed at 10, and for different numbers of iterations (5, 10, 100, 1000) while keeping the number of particles fixed at 100. The results are illustrated in Figure \ref{fig:pso-robust}.

Figure \ref{fig:pso-robust-particles} shows the effect of varying the number of particles on the estimation accuracy. With only 5 particles, the RMSE is relatively high and exhibits large variability across runs. Increasing the swarm size to 10 significantly reduces both the mean RMSE and its standard deviation. Beyond 100 particles, the improvement becomes marginal and the performance stabilizes, indicating that a moderate swarm size is sufficient for reliable estimation.

Figure \ref{fig:pso-robust-iterations} presents the sensitivity to the number of iterations while keeping the particle count fixed at 100. A small number of iterations leads to higher error and variability, whereas increasing the iterations to 10 substantially improves stability. Beyond 100 iterations, the variability becomes negligible, suggesting that the optimization converges reliably within a reasonable number of iterations.

In light of these observations, the proposed method uses 100 particles and 100 iterations to reduce estimation error and dependence on initial conditions.

%% file: Figures_Latex/pso_robustness_figure.tex
\begin{figure*}[!t]
    \centering
    \subfloat[]{\includegraphics[width=2.5in]{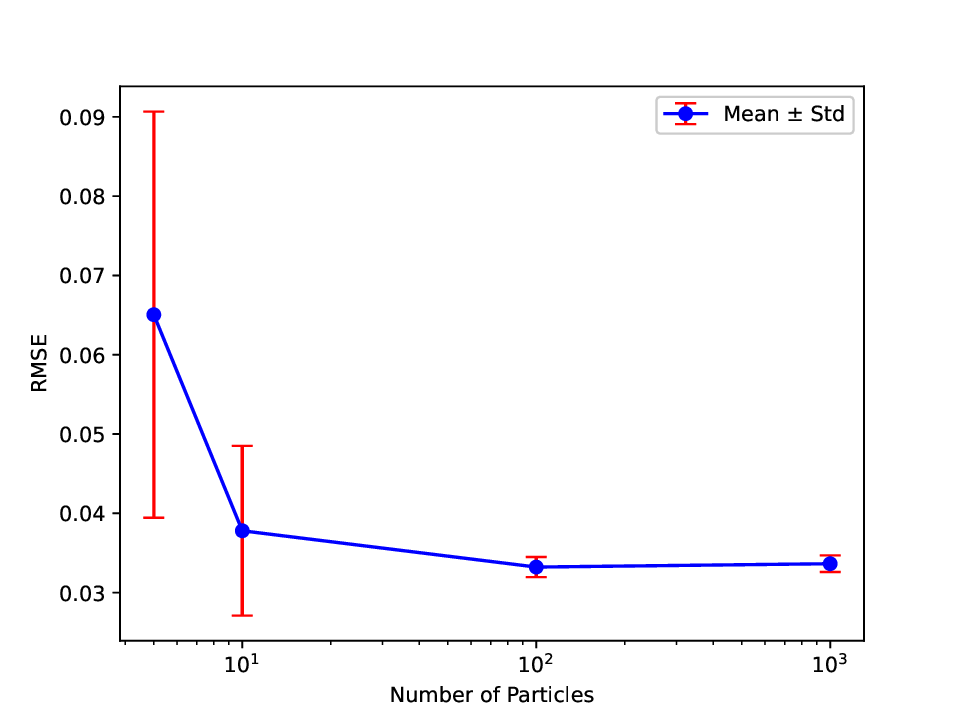}%
        \label{fig:pso-robust-particles}}
    \hfil
    \subfloat[]{\includegraphics[width=2.5in]{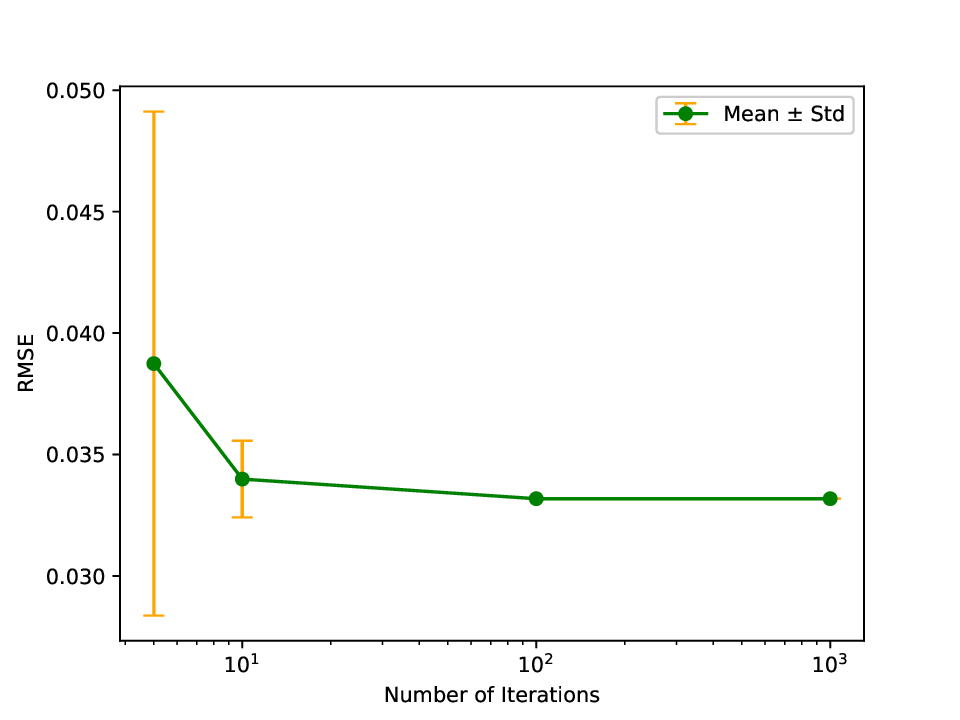}%
        \label{fig:pso-robust-iterations}}
    \caption{Robustness analysis of PSO. (a) Variation RMSE of estimated scaling factors with the number of particles. (b) Variation RMSE of estimated scaling factors with the number of iterations.}
    \label{fig:pso-robust}
\end{figure*}

%% file: Chapters/discussion.tex
\section{DISCUSSION}

The experimental results demonstrate that the proposed geometric projection method effectively mitigates large-scale, wavelength-independent spectral variability. By mapping dimensionally reduced pixel signatures onto an optimally estimated hyperplane, the algorithm corrects for scale distortions. This correction is critical because scale distortions, primarily driven by topography and illumination, often cause unmixing algorithms to misinterpret magnitude variations as changes in material abundance.

Regarding the scope of applicability under complex mixture conditions, our robustness analyses further validate that the proposed initialization and Particle Swarm Optimization (PSO) framework ensure stable scaling factor estimation even under conditions of high noise, moderate nonlinear mixing, and highly mixed pixels. Specifically, as demonstrated in Sections IV-F and IV-G, the estimation remains highly stable in the presence of bilinear nonlinearities and in low-purity environments lacking pure pixels. Consequently, applying this preprocessing step allows both traditional and deep learning unmixing algorithms to converge more accurately, as they are no longer penalized by dominant illumination artifacts.

While the current framework specifically models wavelength-independent scaling, real-world spectral variability also manifests as wavelength-dependent distortions due to complex atmospheric absorption or intrinsic material differences. It is important to note that in standard hyperspectral image processing pipelines, spectral bands heavily afflicted by strong atmospheric absorption (such as dense water vapor bands) are routinely removed prior to analysis. Applying the proposed preprocessing step after this standard band pruning effectively mitigates the impact of these specific atmospheric limitations on the scale estimation process.

To extend this framework to natively handle wavelength-dependent variability, the scalar parameter \(\mu_i\) could theoretically be expanded into a band-wise scaling matrix, or the spectrum could be segmented into piecewise regions where scaling remains approximately constant. However, natively estimating a distinct scaling factor for every spectral band or region within an unsupervised geometric projection is a highly underdetermined problem. It would necessitate enforcing strong spectral regularization or band-correlation priors. Therefore, such a setup would require specific domain knowledge or direct integration into the unmixing algorithm itself, rather than serving as a standalone preprocessing step.

Ultimately, a more immediate and practical application is to use the proposed algorithm as a complementary foundational step. By stripping away massive, dominant scale disparities, the scale-corrected outputs provide an ideally stabilized input. This allows downstream unmixing models to dedicate their expressive capacity entirely to resolving residual, fine-grained wavelength-dependent variations.

%% file: Chapters/Conclusion.tex
\section{Conclusion}

In this paper, we introduced a novel preprocessing algorithm designed to correct scale-induced spectral variability in hyperspectral images, a common artifact introduced by illumination, topography, and shadowing. By estimating and reversing these large-scale scalar distortions through a robust perspective-projection framework, the algorithm restores the fundamental geometric structure of the data. Extensive experimental validation demonstrates that this preprocessing step significantly improves the abundance estimation accuracy of a wide variety of state-of-the-art unmixing algorithms, yielding significant error reductions.
Future work could explore integrating the proposed preprocessing step directly into the optimization frameworks of existing unmixing algorithms, particularly deep learning–based methods, enabling joint learning of scale corrections and abundance estimation.

%% file: Chapters/appendix.tex
\appendix
This section will outline details of certain mathematical results which were omitted in the main article for the sake of brevity and clarity.
\subsection{Proof of the Hyperplane Equation}
\label{appendix:hyperplane-proof}
In this section, it will be shown that the hypothetical dimensionally reduced pixels without scaling factors (\(\yystar{i}\)) adhere to the hyperplane equation given by \eqref{eq:hyperplane}.

We can assume that the dimensionally reduced endmember signatures are Linearly Independent (otherwise one can be written as a mixture of others) so \(\MM\) is a full rank matrix. Thus, \(\exists\; \nnstar \in \real^K\) such that, \(\MM^T\nnstar = \onemat \) where \(\onemat = (1,1, \dots 1)^T \in \real^K\).

Let the \(i^\tth\) pixel be \(\yystar{i} = \MM \abdo_{i}\). Then \(\forall i\),
\begin{align*}
    \abdo_i^T\MM^T\nnstar          & = \abdo_i^T \onemat        \\
    \implies (\MM\abdo_i)^T\nnstar & = 1  \; \; \text{(by ASC)} \\
    \implies (\yystar{i})^T\nnstar & = 1  \; \;
\end{align*}

Now consider, a set of scalars \(\{\lambda_1,\lambda_2, \dots \lambda_N\}\) such that \(\sum_{i=1}^N \lambda_j = 1\). Then,

\begin{align*}
    \sum_{j=1}^N \left(\lambda_j(\yystar{j})^T\nnstar\right) & = \sum_{j=1}^N \lambda_j \\
    \left(\sum_{j=1}^N \lambda_j\yystar{j}\right)^T\nnstar   & = 1
\end{align*}
Thus, for any \(i\),
\begin{align*}
    (\yystar{i})^T\nnstar -
    \left(\sum_{j=1}^N \lambda_j\yystar{j}\right)^T\nnstar & = 1 -1 \\
    (\yystar{i})^T\nnstar -
    (\ccstar)^T\nnstar                                     & = 0    \\
    (\yystar{i} - \ccstar)^T\nnstar                        & =0
\end{align*}
with \(\ccstar\) satisfying \eqref{eq:point_formula} and \(\nnstar\) satisfying
\begin{align}
    \label{eq:nstarM-eq}
    M^T\nnstar = \onemat
\end{align}
By letting \(\lambda_i = \frac{\mu_i}{N}\) and using \(\mean{\mu} = \sum_{i=1}^N{\frac{\mu_i}{N}} = 1\) we get formula, \eqref{eq:cstar}
\subsection{Normal Candidate Generation Equation}
\label{appendix:candidate-normal}
For initializing the PSO algorithm, Equation \eqref{eq:nn-candidate} was used to calculate normals based on randomly selected pixels. This was done to generate normals in the neighborhood of the true normal. In this section, it will be shown that this equation will generate normals equal to the true normal if all the pixels that are randomly chosen had equal scaling factors.

Note that these randomly selected pixels need not be pure endmembers; they can be arbitrarily mixed. This derivation only uses the (approximate) equality of the scaling factors within the selected set and the existence of a well-defined normal (i.e., \(B\) is invertible).

Let \(\{\yy{1},\yy{2}, \dots \yy{K}\}\) be a \(K-\) set of pixels from the dataset. Assume that they share the same scaling factor \(\mu_1 = \mu_2 =\dots = \mu_K = \mu\). Then \(\yy{j} = \mu \MM\abdo_{j}\) for \(j \in {1,2,\dots K}\).
Then,
\begin{align*}
    \yy{i}^T\nnstar & = (\mu \MM\abdo_{j})^T \nnstar                                              \\
    \yy{i}^T\nnstar & =  \mu \abdo_{j}^T \MM^T\nnstar                                             \\
    \yy{i}^T\nnstar & =  \mu \abdo_{j}^T \onemat = \mu \;\; \text{(ASC and \eqref{eq:nstarM-eq})}
\end{align*}
Stacking these into one matrix equation with \(B \in \real^K, B = (\yy{1},\yy{2}, \dots, \yy{K})\) gives,
\begin{align*}
    B^T\nnstar & =  \mu \onemat
\end{align*}
Note that the scaling factor \(\mu \) only affects the magnitude of \(\nnstar\). Thus, it can be dropped since we are not interested in the magnitude of \(\nnstar\).

This means, \(\nnstar\) satisfies \eqref{eq:nn-candidate} under this condition.

In practice, we reject sampled sets that produce ill-conditioned \(B\) (near linear dependence) and normalize the resulting normal vector; this improves numerical stability without changing the derivation.

The motivation for this initialization strategy is that, among the randomly selected pixel sets, there is a chance of selecting one that is reasonably uniformly illuminated, resulting in \(\mu_i\) values that are close to each other. Such a candidate would lie near the true normal we aim to estimate.

\subsection{Proving the Optimality of the Estimator for the Hyperplane Normal}
\label{appendix:estimator-proof}

Let the minimizer of \(\zeta(\nn) = \ee{\Nsum{\norm{\yy{i}-\yyhat{i}}^2}}\) be \(\nhat\).

\begin{align}
    \zeta(\nn) & = \ee{\Nsum{\norm{\yy{i}-\yyhat{i}}^2}}    \\
    \nhat      & = \arg\min_{\nn} \left\{\zeta(\nn)\right\}
\end{align}
Let \(\nn \in \real^\KK\) be some normal, then it can be shown that,

\begin{align}
    {\norm{\yy{i}-\yyhat{i}}^2}       & = (\scalo_i - \gammai{i}{\nn})^2\norm{\yystar{i}}^2 \\
    \text{Where,\;\;} \gammai{i}{\nn} & = \frac{(\ccstar)^T \nn}{(\yystar{i})^T \nn}
\end{align}

Now, based on the assumptions made on the distribution of the scaling factors, it can be shown that,
\begin{align}
    \zeta(\nn) & = \Nsum{\mathcal{K}_i}+H(\nn)
\end{align}
Where \(\mathcal{K}_i\) is a quantity independent of \(\nn\) given by,
\begin{align*}
    \mathcal{K}_i & = \left(\text{Var}[\mu_i]+\frac{2}{N}\text{Var}[\mu_i]\right)\norm{\yystar{i}}^2
\end{align*}
and \(H(\nn)\) is the the part dependent on \(\nn\) given by,
\begin{align}
    \label{eq:H-n}
    H(\nn) & = \Nsum{\left(\text{Var}[\gammai{i}{n}]+(\ee{\gammai{i}{n}}-1)^2\right)\norm{\yystar{i}}^2}
\end{align}

Therefore, minimizer of \(\zeta(\nn)\)  is the same as that of \(H(\nn)\). Thus,
\begin{align*}
    \nhat & = \arg\min_{\nn} \left\{H(\nn)\right\}
\end{align*}

As mentioned previously, the variance of the possible scaling factors can vary significantly based on the abundances as textural properties of different compositions are different. Let the variance of the scaling factors for abundances with maximum scale variance and minimum scale variance be  \(\maxvar\) and \(\minvar\). i.e.

\(\maxvar = \max_i \{\text{Var}[\mu_i]\}\) and \(\minvar = \min_i \{\text{Var}[\mu_i]\}\)
\hfill

It can be shown using the assumed distributions on \(\mu_i\) that,
\begin{align*}
    \text{Var}[\gammai{i}{\nnstar}] \leq \frac{\maxvar}{N} \;\;\text{and}\;\;   \ee{\gammai{i}{\nnstar}} = 1
\end{align*}

Using the above result and substituting \(\nn=\nnstar\) for \(H(\nn)\),
\begin{align*}
    H(\nnstar)             & = \Nsum{\text{Var}[\gammai{i}{\nnstar}]}\norm{\yystar{i}}^2 \\
    \Rightarrow H(\nnstar) & \leq \maxvar\mean{\norm{\yystar{i}}^2}
\end{align*}

Because \(\nhat\) is the minimizer if \(H(\nn)\), \(H(\nhat) \leq H(\nnstar)\)
\begin{align}
    \label{eq:H-upperbound}
    \therefore H(\nhat) & \leq \maxvar\mean{\norm{\yystar{i}}^2}
\end{align}

Furthermore, it can be shown that, for any \(\nn\)
\begin{align*}
    \Nsum{\text{Var} [\gammai{i}{\nn}] {\norm{\yystar{i}}^2}}\geq \minvar {\mean{\norm{\yystar{i}}}}^2
\end{align*}

This result in conjunction with \eqref{eq:H-n} leads to the result, \(\forall \nn\)
\begin{align*}
    H(\nn) & \geq \minvar {\mean{\norm{\yystar{i}}}}^2+ \Nsum{\left(\ee{\gammai{i}{\nn}}-1\right)^2\norm{\yystar{i}}^2}
\end{align*}

By setting \(\nn = \nhat\) we get,
\begin{align}
    \label{eq:Hnhat-lowerbound}
    H(\nhat) & \geq \minvar {\mean{\norm{\yystar{i}}}}^2+ \Nsum{\left(\ee{\gammai{i}{\nhat}}-1\right)^2\norm{\yystar{i}}^2}
\end{align}

Equation \eqref{eq:H-upperbound} and Equation \eqref{eq:Hnhat-lowerbound} finally leads to the following result.

\begin{align}
    \maxvar\mean{\norm{\yystar{i}}^2} \geq \begin{aligned}[t] & \minvar {\mean{\norm{\yystar{i}}}}^2                                         \\
                   & + \Nsum{\left(\ee{\gammai{i}{\nhat}}-1\right)^2\norm{\yystar{i}}^2}\nonumber\end{aligned} \\
    \Nsum{(\ee{\gammai{i}{\nhat}}-1)^2}\norm{\yystar{i}}^2 \leq \begin{aligned}[t]
                                                                     & \maxvar\mean{\norm{\yystar{i}}^2}      \\
                                                                     & - \minvar {\mean{\norm{\yystar{i}}}}^2
                                                                \end{aligned}
\end{align}

Furthermore,

\begin{align}
    \ee{\gammai{i}{\nhat}} = \frac{\mean{(\yystar{})}^T\nhat}{(\yystar{i})^T\nhat}
\end{align}

Observe that \(\ee{\gammai{i}{\nhat}}\) is the scaling factor required to project \(\yystar{i}\) from the true hyperplane (where \(\unpixlo{i}\) reside) to the estimated hyperplane with the estimated normal \(\nhat\). In the ideal case where \(\nhat = \nnstar\), this should be 1. Furthermore, \((\ee{\gammai{i}{\nhat}}-1)^2\norm{\yystar{i}}^2 = (\ee{\gammai{i}{\nhat}}\norm{\yystar{i}}-\norm{\yystar{i}})^2\) represents the squared distance between the \(\yystar{i}\) in the the true hyperplane (corresponding to \(\nnstar\)) and the hyperplane predicted by \(\nhat\). (If \(\nhat\) happens to be \(\nnstar\) this will be zero.).

Therefore, we will represent this error distance for the \(i^\tth\) pixels by \(\delta_i\), which leads to the fact that,

\begin{align}
    \Nsum{\delta_i^2} & \leq \maxvar\mean{\norm{\yystar{i}}^2} - \minvar {\mean{\norm{\yystar{i}}}}^2
\end{align}

If the Root-Mean-Square (RMS) value of this distance error is given by \(\delta_{\text{rms}}\), then the error relative to the RMS pixel magnitude is given by Equation \eqref{eq:Upper-bound-on-the-Relative-RMSE-Error}.

\subsection{Useful Results for Appendix \ref{appendix:estimator-proof}}
This sections provide some results that are useful for following the derivations given in Appendix \ref{appendix:estimator-proof}.
\begin{align*}
    \mathbb{E}[\mean{\yy{}}^T \nn]
     & = \frac{1}{N} \sum_{j=1}^N E[\mu_j] (\yystar{j})^T \nn \\
     & = \frac{1}{N} \sum_{j=1}^N (\yystar{j})^T \nn
    = (\mean{\yystar{}})^T \nn                                \\
    \therefore\quad
    \mathbb{E}[\Gamma_i]
     & = \frac{(\mean{\yystar{}})^T \nn}{(\yystar{i})^T \nn}
\end{align*}

\begin{align*}
    \mathbb{E}[\mu_i\Gamma_i]
     & = \frac{\mathbb{E}\!\left[\mu_i\,\mean{\yy{}}^T \nn\right]}{(\yystar{i})^T \nn}           \\
     & = \frac{\mathbb{E}\!\left[\mu_i \frac{1}{N} \sum_{j=1}^N
    \mu_j (\yystar{j})^T \nn\right]}{(\yystar{i})^T \nn}                                         \\
     & = \frac{\frac{1}{N} \sum_{j\neq i}
    \mathbb{E}[\mu_i]\,\mathbb{E}[\mu_j]\,(\yystar{j})^T \nn}{(\yystar{i})^T \nn}                \\
     & \quad + \frac{\frac{1}{N} \mathbb{E}[\mu_i^2]\,(\yystar{i})^T \nn}{(\yystar{i})^T \nn}    \\
     & = \frac{(\mean{\yystar{}})^T \nn}{(\yystar{i})^T \nn} + \frac{\mathrm{Var}[\mu_i] + 1}{N} \\
     & = \mathbb{E}[\Gamma_i] + \frac{\mathrm{Var}[\mu_i] + 1}{N}                                \\
    \therefore\quad
    \mathrm{Cov}[\mu_i, \Gamma_i]
     & = \mathbb{E}[\mu_i\Gamma_i] - \mathbb{E}[\mu_i]\,\mathbb{E}[\Gamma_i]
    = \frac{\mathrm{Var}[\mu_i] + 1}{N}
\end{align*}

\begin{align*}
    \Var{\Gamma_i(\nnstar)}
     & = \Var{\frac{\mean{\yy{}}^T \nnstar}{(\yystar{i})^T \nnstar}}
    = \frac{1}{\big((\yystar{i})^T \nnstar\big)^2}
    \Var{\mean{\yy{}}^T \nnstar}                                     \\[4pt]
     & = \frac{1}{\big((\yystar{i})^T \nnstar\big)^2}
    \Var{\frac{1}{N}\sum_{j=1}^N \mu_j (\yystar{j})^T \nnstar}       \\[4pt]
     & = \frac{1}{\big((\yystar{i})^T \nnstar\big)^2}
    \frac{1}{N^2}\sum_{j=1}^N \Var{\mu_j} \big((\yystar{j})^T \nnstar\big)^2
\end{align*}

\begin{equation*}
    \begin{split}
        \implies \frac{\sigma_{\min}^2\sum_{j=1}^N \big((\yystar{j})^T \nnstar\big)^2}
        {N^2\big((\yystar{i})^T \nnstar\big)^2}
        \le{} & \Var{\Gamma_i(\nnstar)}                                               \\
        \le{} & \frac{\sigma_{\max}^2\sum_{j=1}^N \big((\yystar{j})^T \nnstar\big)^2}
        {N^2\big((\yystar{i})^T \nnstar\big)^2}\,.
    \end{split}
\end{equation*}

When \(\nn=\nnstar\) (the true normal) every unscaled pixel satisfies
\((\yystar{j})^T \nn = (\yystar{i})^T \nn\) = \(\kappa\) for all \(i,j\)

\begin{align*}
    \Var{\Gamma_i(\nnstar)}
     & \le \frac{\sigma_{\max}^2}{N^2}\;\frac{\sum_{j=1}^N \kappa^2}{\kappa^2}
    = \frac{\sigma_{\max}^2}{N}.
\end{align*}

Let \(s_j = (\yystar{j})^T \nn\).
From the earlier bound,
\begin{align*}
    \Var{\Gamma_i} & \ge \sigma_{\min}^2 \,\frac{\sum_{j=1}^N s_j^2}{N^2 s_i^2}.
\end{align*}
Multiplying by \(\|\yystar{i}\|^2\) and summing over \(i\) gives
\begin{align*}
    \sum_{i=1}^N \Var{\Gamma_i}\,\|\yystar{i}\|^2
     & \ge
    \sigma_{\min}^2\,\frac{\sum_{j=1}^N s_j^2}{N^2}
    \sum_{i=1}^N\frac{\|\yystar{i}\|^2}{s_i^2}.
\end{align*}

Applying the Cauchy-Schwarz inequality to the vectors \\
\(\bigl(s_j\bigr)_{j=1}^N\) and \(\bigl(\|\yystar{j}\|/s_j\bigr)_{j=1}^N\), we have
\begin{align*}
    \left(\sum_{j=1}^N s_j^2\right)
    \left(\sum_{j=1}^N \frac{\|\yystar{j}\|^2}{s_j^2}\right)
     & \ge
    \left(\sum_{j=1}^N \|\yystar{j}\|\right)^2.
\end{align*}

Combining yields
\begin{align*}
    \sum_{i=1}^N \Var{\Gamma_i}\,\|\yystar{i}\|^2
     & \ge
    \sigma_{\min}^2 \left(\frac{\sum_{j=1}^N \|\yystar{j}\|}{N}\right)^2 \\
    \sum_{i=1}^N \Var{\Gamma_i}\,\|\yystar{i}\|^2
     & \ge
    \sigma_{\min}^2\left(\mean{\|\yystar{}\|}\right)^2.
\end{align*}

\subsection{Non-negativity of \(\scalo_i\) and effect of \(\noiso_i\)}
\label{appendix:non-negativity-of-scaling-factors}
Under ideal conditions where \(\mathbb{T}(\noiso_i)\) is negligible, if the predicted \(\nhat\) is sufficiently close \(\nnstar\) then \(\muhat{i}\) can be shown to be positive.

Note that,

\begin{align*}
    \ccstar & = \frac{1}{N}\sum_{j=1}^N \pixlo{j} = \frac{1}{N}\sum_{j=1}^N \scalo_i\MM\abdo_j                              \\
            & = \MM \left(\frac{1}{N}\sum_{j=1}^N \scalo_i \abdo_j\right)                                                   \\
            & = \MM \mathbf{\bar{a}} \; ; \left(\mathbf{\bar{a}} \triangleq \frac{1}{N}\sum_{j=1}^N \scalo_i \abdo_j\right)
\end{align*}

Since \(\abdo_j\) are non-negative and \(\scalo_i\) are positive, \(\mathbf{\bar{a}}\) is a non-negative vector. Because \(\abdo_j\) satisfy the ASC, then at least one of the entries of \(\mathbf{\bar{a}}\) is positive.

From Equation \eqref{eq:muhat},

\begin{align*}
    \muhat{i} & = \frac{\scalo_i \abdo_i^T \MM^T \nhat}{\mathbf{\bar{a}}^T \MM^T \nhat} \\
\end{align*}

Now, \(\MM^T \nnstar = \onemat\), and as the optimization procedure attempts to find \(\nhat\) close to \(\nnstar\), if it is sufficiently close, \(\MM^T \nhat =\mathbf{v}\) will be close to \(\onemat\). Thus, \(\mathbf{v}\) will be quite close to \(\onemat\) and therefore will be a positive vector. Therefore,

\begin{align*}
    \muhat{i} & = \frac{\scalo_i \abdo_i^T \mathbf{v}}{\mathbf{\bar{a}}^T \mathbf{v}} \\
\end{align*}
Since \(\scalo_i\) is positive, \(\abdo_i\) is non-negative and \(\mathbf{v}\) is positive, the numerator is positive. Since \(\mathbf{\bar{a}}\) is non-negative with at least one positive entry and \(\mathbf{v}\) is positive, the denominator is also positive. Therefore, \(\muhat{i}\) will be positive.

Let, \(\mathbb{T}(\noiso_i)=\tilde{\noiso}_i\), also as it is always possible to possible to incorporate the mean of the noise terms into the LMM portion of the model we may take \(\mean{\tilde{\noiso}_i}=0\), then the predicted scaling factor is given by ,
\begin{align*}
    \muhat{i} & = \frac{\scalo_i \abdo_i^T \MM^T \nhat + \tilde{\noiso}_i^T\nhat}{\mathbf{\bar{a}}^T \MM^T \nhat + \mean{\tilde{\noiso}_i}^T\nhat} \\
              & = \frac{\scalo_i \abdo_i^T \MM^T \nhat + \tilde{\noiso}_i^T\nhat}{\mathbf{\bar{a}}^T \MM^T \nhat}
\end{align*}

For simplicty consider the case where \(\nhat \rightarrow \nnstar\) and therefore \(\MM^T \nhat = \onemat\). Then,

\begin{align*}
    \muhat{i} & = \scalo_i + \tilde{\noiso}_i^T\nnstar
\end{align*}

The first term is positive, but the second term can be positive or negative. The second term can be thought of as the error in the predicted scaling factor due to noise and modelling errors. If the noise is sufficiently small, then \(\muhat{i}\) will be positive. In fact since \(\norm{\tilde{\noiso}_i}\norm{\nnstar} \geq \tilde{\noiso}_i^T\nnstar\) a sufficient condition for \(\muhat{i}\) to be positive is for \(\norm{\tilde{\noiso}_i} < \scalo_i\) indicating that pixels with small scaling factors are more likely to have negative \(\muhat{i}\) values due to noise and modelling errors. This is consistent with the intuition that pixels with small scaling factors are more likely to be significantly affected by noise and modelling errors, which can lead to negative predicted scaling factors.

In general, if \(\noiso_i\) is small enough, then the SVD supress these components making the effect of \(\tilde{\noiso}_i\) negligible. However, when it is large , the SVD esimtation itself is impacted meaning that larger component of \(\noiso_i\) will be present in \(\tilde{\noiso}_i\) increasing its effect on the predicted scaling factors. This is consistent with the observation that the performance of the proposed method gracefully degrades at higher noise levels as well as higher levels of nonlinear mixing effects.

%% file: Chapters/Author_Biography.tex
\begin{IEEEbiography}[{\includegraphics[width=1in,height=1in,clip,keepaspectratio]{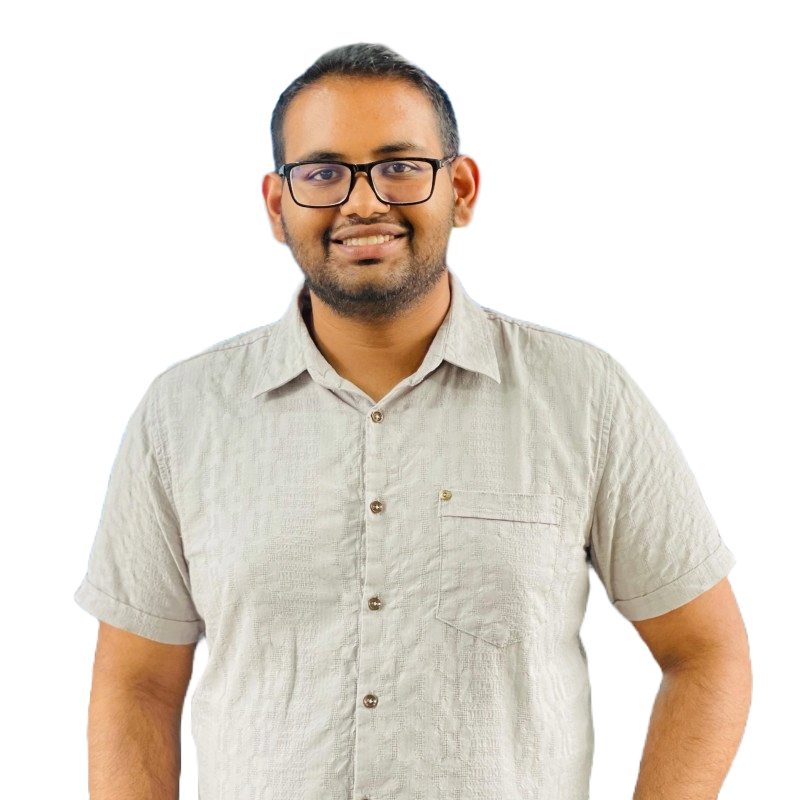}}]{Praveen Sumanasekara}
	is an undergraduate student in Electrical and Electronic Engineering at the University of Peradeniya, Sri Lanka. He has achieved top national and strong global rankings in IEEEXtreme and is a Huawei Seeds for the Future award recipient. His interests include  remote sensing, signal processing and theoretical foundations of machine learning.
\end{IEEEbiography}

\begin{IEEEbiography}[{\includegraphics[width=1in,height=1in,clip,keepaspectratio]{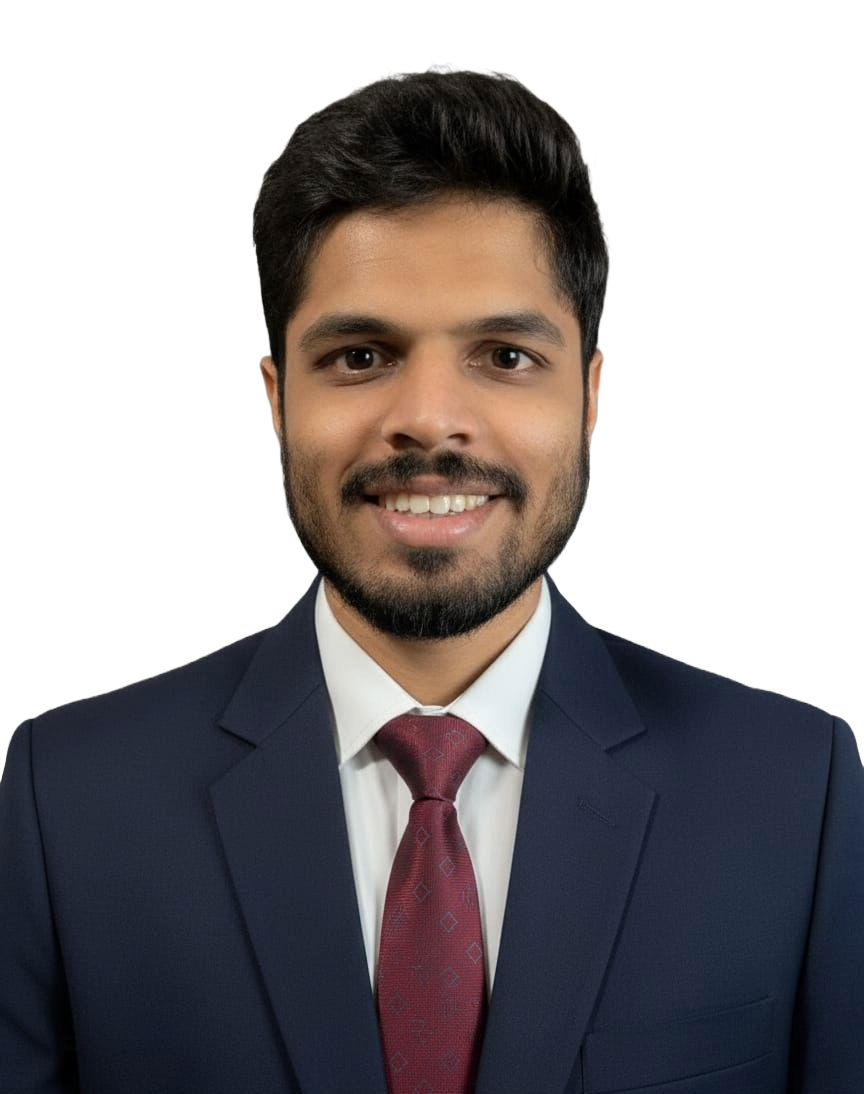}}]{Athulya Ratnayake}
	is an undergraduate student in Electrical and Electronic Engineering at the University of Peradeniya, Sri Lanka. His interests include applied mathematics, signal processing, and machine learning for remote sensing and Earth observation, and he has achieved top national and global results in IEEEXtreme.
\end{IEEEbiography}

\begin{IEEEbiography}[{\includegraphics[width=1in,height=1in,clip,keepaspectratio]{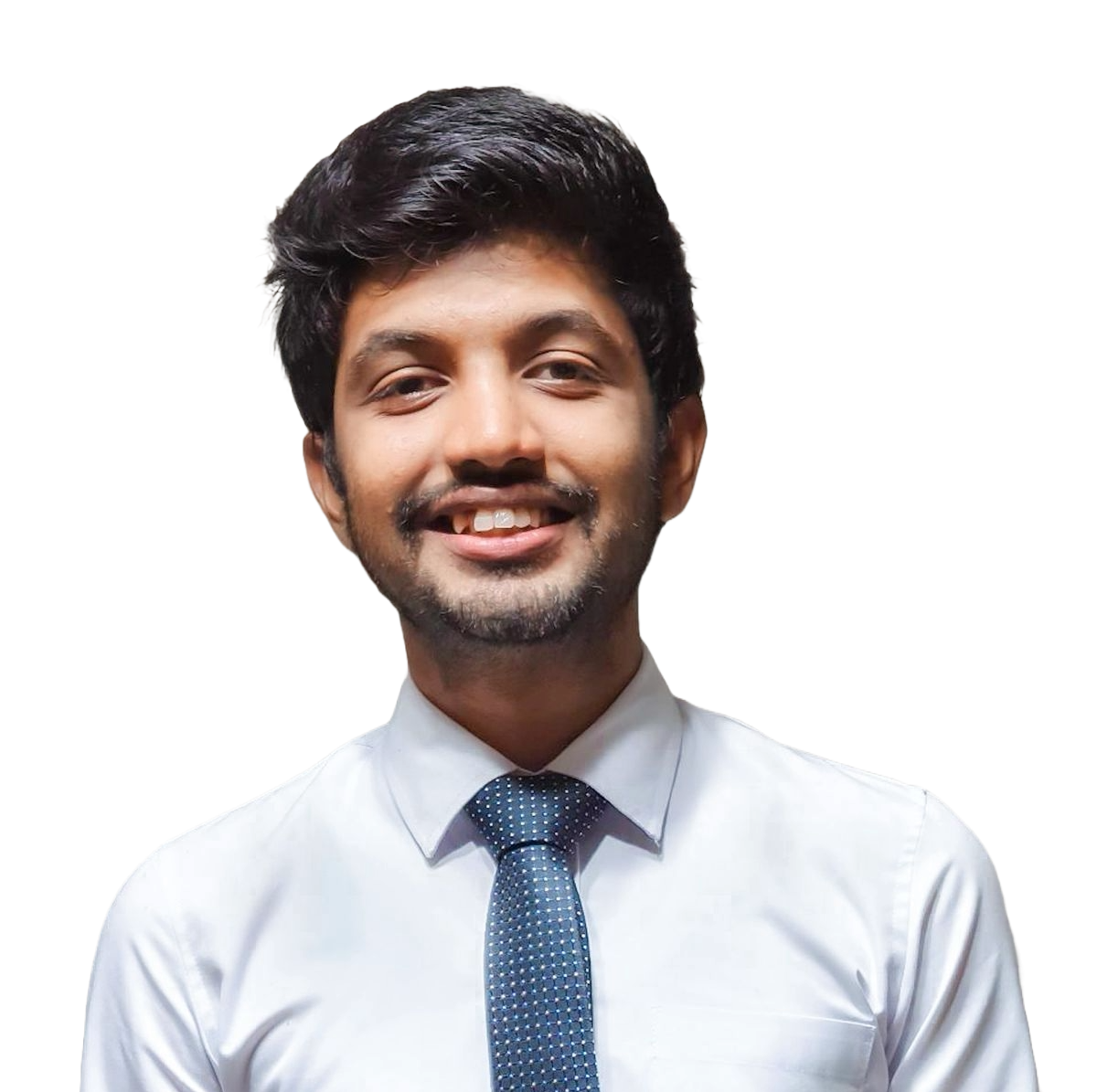}}]{Buddhi Wijenayake}
	is an undergraduate student in Electrical and Electronic Engineering at the University of Peradeniya, Sri Lanka. He has achieved top results in the IEEEXtreme Global Programming Challenge and has served as Chair of the IEEE MTT-S Student Branch Chapter. His research interests include remote sensing, signal processing, and computer vision.
\end{IEEEbiography}

\begin{IEEEbiography}[{\includegraphics[width=1in,height=1in,clip,keepaspectratio]{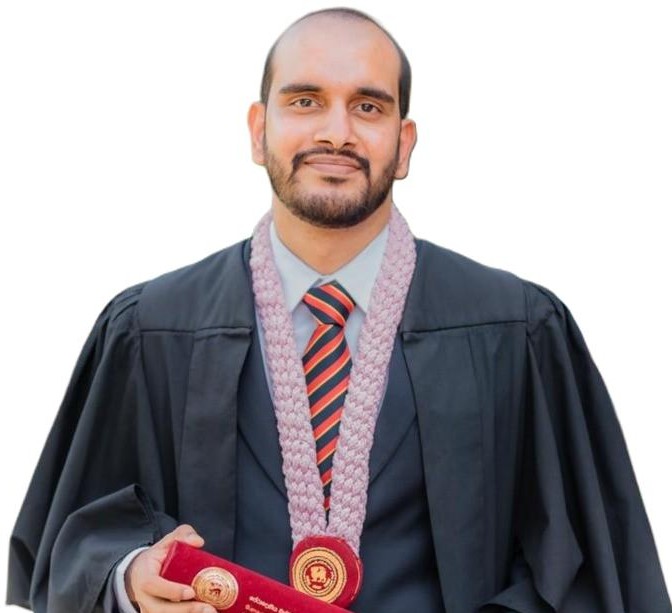}}]{Keshawa~Ratnayake}
	received the B.Sc. Eng. degree in Electrical and Electronic Engineering from the University of Peradeniya. He is currently a graduate student pursuing his PhD in Electrical and Computer Engineering at Purdue University College of Engineering. His interest include Remote Sensing, Signal Processing and Computer Vision.
\end{IEEEbiography}

\begin{IEEEbiography}[{\includegraphics[width=1in,height=1in,clip,keepaspectratio]{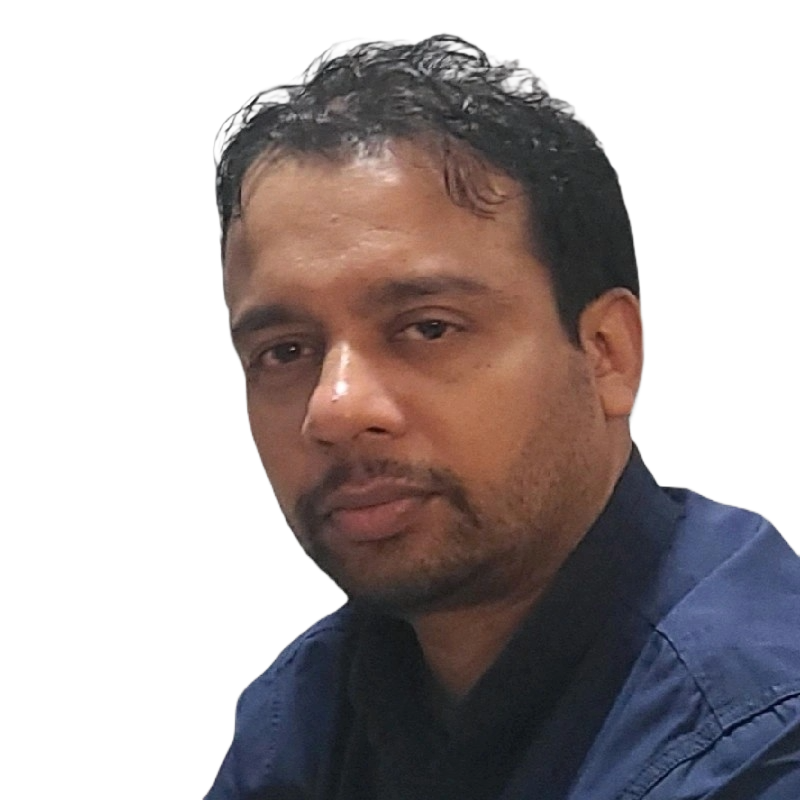}}]{Roshan Godaliyadda}
	is a Professor in Electrical and Electronic Engineering at the University of Peradeniya, Sri Lanka. He received the Ph.D. degree from the National University of Singapore and works in signal/image processing, computer vision, remote sensing, smart grids, and AI. He is a Senior Member of the IEEE.
\end{IEEEbiography}

\begin{IEEEbiography}[{\includegraphics[width=1in,height=1in,clip,keepaspectratio]{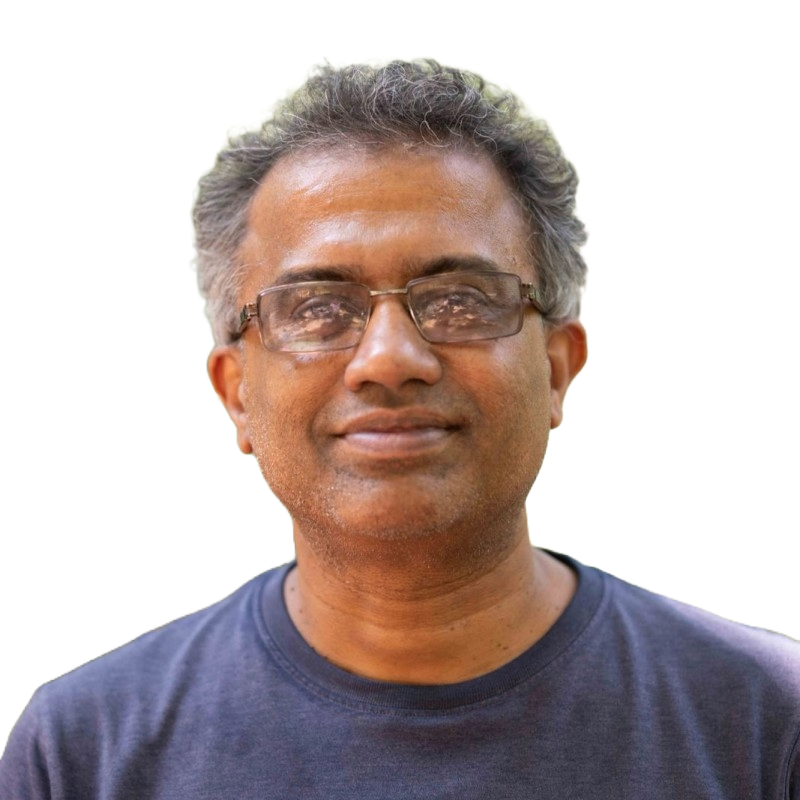}}]{Vijitha Herath}
	is a Professor and Head of the Department of Electrical and Electronic Engineering at the University of Peradeniya, Sri Lanka. He received the Ph.D. degree from the University of Paderborn, Germany. His interests include remote sensing, multispectral imaging, AI/ML, and optical wireless communications.He is a Senior Member of the IEEE.
\end{IEEEbiography}

\begin{IEEEbiography}[{\includegraphics[width=1in,height=1in,clip,keepaspectratio]{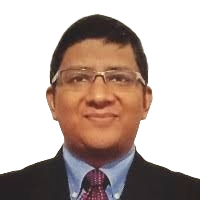}}]{Parakrama Ekanayake}
	is a Professor in Electrical and Electronic Engineering at the University of Peradeniya, Sri Lanka. He received the Ph.D. degree in Applied Mathematics from Texas Tech University, USA. His interests include signal/image processing, AI/ML, control, and computational modelling.He is a Senior Member of the IEEE.
\end{IEEEbiography}